\documentclass[preprint,showpacs,preprintnumbers,amsmath,amssymb]{revtex4-2}
\usepackage{graphicx}
\usepackage{bm}
\usepackage{color}
\usepackage{xcolor}
\usepackage{caption}
\usepackage{subcaption}
\usepackage{amssymb}
\usepackage{epsfig}

\usepackage{amsmath}
\usepackage{array,multirow}
\DeclareGraphicsExtensions{.png,.pdf}
\usepackage[colorlinks=true, allcolors=blue]{hyperref}

\begin{document}



\title{
Edge of entanglement in non-ergodic states: a complexity parameter formulation
}
\author{Devanshu Shekhar and Pragya Shukla}
\affiliation{ Department of Physics, Indian Institute of Technology, Kharagpur-721302, West Bengal, India }

\date{\today}

\widetext

\begin{abstract}

We analyse the subsystem size scaling of the entanglement entropy of a non-ergodic pure state that can be described by a multi-parametric Gaussian ensemble of complex matrices in a bipartite basis.  Our analysis indicates,  for a given set of global constraints,   the existence of infinite number of universality classes of local complexity,  characterized by the complexity parameter,  for which the entanglement entropy reveals a universal scaling with subsystem size.  A rescaling of the complexity  parameter  helps us  to identify the critical regime for the entanglement entropy of a broad range of pure non-ergodic states.


\end{abstract}

\maketitle

\section{Introduction} \label{intro}

Random quantum states, engineered or that of a physical observable,  belong to an important class that are increasingly making their robust appearance in many domains \cite{nielson-book,ajsfc, wpgcrsl, nechita, collins}.
For example, with quantum entanglement used as a resource, the engineering of quantum states to achieve specific objectives for quantum information processing  has become an active area of research \cite{nielson-book,ajsfc}.
Contrary to quantum states of a physical observable, e.g., Hamiltonian, the states used in quantum information can consist of components determined by their proposed function/utility (subject to normalization condition), and evolve under unitary gate operations.  Indeed a typical processing unit  in quantum information  and communication studies can consist of single or multiple layers of quantum gates/ channels (e.g., a quantum circuit) and often use the initial state as the engineered random state of many interacting qubits.
As indicated by many studies in past, it is often advantageous to consider random gates (e.g., for quantum error correction \cite{brown2013short,choi2020quantum}). 
This  however leads to a randomization of the incoming state $|\Psi_{in} \rangle$, with nature and type of the randomness dependent on the choice of the unitary operator $U$ as well as $|\Psi_{in} \rangle$: $|\Psi_{out} \rangle = U |\Psi_{in} \rangle$. Due to presence of many gates usually, a typical component of the outgoing state consists of a sum over many random variables and by invoking central limit theorem, its distribution can often be predicted as a Gaussian. But the distribution parameters of the components can in general differ; this in turn leads to a representation of the state by a multiparametric Gaussian density. Even in case of non-random gates, a state used for realistic communication purposes has to transmit through noisy channels; this in turn randomizes the components. For cases, where the effect of noise leaves the information only about the first two moments, the maximum entropy hypothesis (MEP) permits the components to be described by  independent  (need not be identical though) Gaussian distributions subject to normalization condition. This motivates the present work, with a primary focus to numerically analyse the bipartite entanglement aspects of a specific class of such states, namely, pure states with Gaussian components. 

The manifestation of randomness in a natural quantum state (e.g of a physical observable) can intuitively be explained as follows. A quantum state often reflects the complexity of the system,  with its dynamics expected,  in general,  to be sensitive to many system parameters,  quantifying the local constraints, e.g., disorder and global constraints, e.g., symmetry, conservation laws etc. Although inherently probabilistic in nature, its components in a Hilbert space basis can be determined to a reasonable accuracy, e.g., if the eigenstates and eigenvalues of an operator, e.g., Hamiltonian can be determined exactly. A lack of detailed information due to complexity renders an exact  determination of the components in a physically motivated basis technically difficult, often leaving them best described by a distribution even in absence of disorder. The nature and type of the randomness can again be determined by invoking maximum entropy hypothesis under known system constraints, e.g., symmetry, conservation laws, disorder, dimensionality etc.  

While a consideration of independent Gaussian distributed components for realistic quantum states may seem to have limited applications, however, as revealed by previous studies of the Hermitian ensembles representing a wide range of complex systems, this seems to be a good approximation. For example, almost all components of a typical state of a quantum chaotic system can be described by independent, identical Gaussian distributions subjected to normalization condition (as corroborated by the success of well-known Porter Thomas distribution of local intensity for quantum chaotic states) \cite{zycz5} (also discussed in \cite{haak}). We recall that, notwithstanding the normalization condition subjecting the sum of the components to a fixed value, the components  can still be statistically uncorrelated and therefore described by independent distributions.  A Gaussian behaviour of the eigenfunctions components is also indicated by the study \cite{bogo1} for  Rosenzweig-Porter ensemble (matrix elements Gaussian distributed with zero mean and a fixed ratio of diagonal to off-diagonal variances); the latter is believed to be a good model for many body localization $\to$ delocalization transition, if subjected to an additional constraint of fixed diagonals. In absence of the latter, the components remain independent although their behaviour in the bulk is only approximately Gaussian (see eq.(20) of \cite{bogo1}). Indeed the observed Gaussian behaviour of the components numerically motivated consideration of the Hilbert-Schmidt ensembles for generic quantum states and in the investigation of any quantum information processing task, it is always desirable to make use of generic Gaussian states to benchmark the performance \cite{wei1}. For example, some of the well known quantum states of light, playing important role in optical communication are fermionic Gaussian states which also consist of Gaussian components \cite{bhk1, hw}.

Based on prevailing system conditions,  a quantum system can in general be in an ergodic or non-ergodic state.  
While an ergodic state is typically attributed to strong quantum correlations, e.g., high degree of entanglement \cite{page1993average}, the latter can vary among different non-ergodic states of a given system.  The changing system conditions, e.g.,  variation of disorder strength and/or many-body interactions, can also affect the correlations leading to a crossover from  localized $\to$ non-ergodic extended $\to$ delocalized state, or separable $\to$ entangled $\to$ maximally entangled states in the bipartite basis.  More specifically,  the entanglement entropy can evolve with changing system conditions, from zero (separable) to a maximum possible limit, with the rate of evolution in general dependent on many system parameters.  As mathematical derivations are often based on system-specific approximations, this renders a theoretical formulation of the entanglement entropy of non-ergodic states, applicable under all possible system conditions,  a technically challenging task. It is therefore desirable if possible to identify the classes of non-ergodic states for which a common mathematical formulation of the entanglement measures can be derived.  An important theoretical insight in this context was given in  \cite{Shekhar_2023},  implying that a controlled variation of the entanglement of a specific class of non-ergodic random states (those with multiparametric Gaussian distributed components),  between zero and a maximum limit, can be achieved by varying just a single  functional of the system parameters,  hereafter referred as the complexity parameter. 
The primary objective of the present work is to pursue the insight further, numerically analyse its implication and establish that a size based  rescaling of the complexity parameter not only helps in identifying the universality classes of the ensemble averaged entanglement entropy but also reveals its critical regime. We note that the scaling of entanglement entropies \cite{Horodecki2009} of a quantum state with sub-system size is a topic of immense interest in various many-body physics contexts, e.g., seeking criticality in local quantum systems \cite{PhysRevLett.90.227902}, understanding the topological properties of ground states \cite{PhysRevLett.96.110404}, pinpointing quantum phase transition point in systems driven by a quenched disorder parameter \cite{Nandkishore2015, Luitz2015,Zhang2022, Skinner2019}, and for establishing connection between quantum mechanics and gravity \cite{Qi2018}. While the scaling behaviour in case of ergodic states is believed to be well-understood \cite{page1993average} (with the entanglement entropy scales as the volume of the subsystem), many aspects of the non-ergodic pure states are still not properly understood. The criticality of entanglement too has many ramifications, e.g., in the  study of phase transitions of many-body systems.

The paper is organized as follows.
As our search for the above mentioned universality is based on the complexity parameter formulation \cite{Shekhar_2023}, it is briefly reviewed in section \ref{comparam} to keep the paper self-contained. The section also specifies the specific class of non-ergodic quantum states, to which our theoretical formulation is applicable. An important aspect of the formulation i.e the role of constants of evolution,  not well-emphasized in \cite{Shekhar_2023}, is also now elaborated. A few misprints appeared in \cite{Shekhar_2023} are also now corrected. Section \ref{evolution} and \ref{fss} describe the numerical analysis of the universality and criticality of the average entanglement entropy for a few  state matrix ensembles with different variance structures but for two subsystems of equal sizes. To understand the influence of subsystem size on universality, section \ref{cut} analyses the case with different subsystems sizes. We conclude in section \ref{conclusion} with main results and open questions.

\section{Complexity parameter formulation of the evolution: a brief review} \label{comparam}

Consider a pure bipartite quantum state $|\Psi \rangle = \sum C_{kl} \; | a_k \rangle \; | b_l \rangle$  with coefficients $C_{kl}$ (complex or real) as a measure of  the correlations between the orthogonal subspaces of its sub parts $A$ and $B$, consisting of basis vectors $|a_k \rangle$ and $| b_l \rangle$, $k=1 \to N_A, l=1 \to N_B$ respectively. The standard entanglement measures for a pure bipartite state, viz., the von Neumann $R_1$ and other R\'enyi entropies $R_n$ with $n >1$, are functions of the eigenvalues  of  the $N_A \times N_A$ reduced density matrix for $A$ as $W = C \cdot C^{\dagger}$ (also known as Schmidt eigenvalues). For notational ease, hereafter we use $N_A=N$ and $N_B=N+\nu_0$.

A real physical system is in general subject to many global constraints, e.g., symmetries and conservation laws. In presence of exact discrete symmetries and if the quantum state preserves them, the physically motivated basis to study the dynamics is the one that preserves the symmetry. The eigenstate in a symmetry resolved basis, can be written as products of the eigenfunctions of different symmetry blocks. 
The state matrix $C$ in such a basis  will however appear in a block diagonal form, with each block  symmetry resolved, i.e., related to a symmetry related quantum number. As the latter are independent, it is appropriate to confine the entanglement analysis to one of the symmetry resolved blocks. The Schmidt eigenvalues for one such block can only have accidental degeneracy, also known as avoided crossings. (This is discussed in detail in \cite{mehta, haak}  in context of Hermitian matrices in general). Besides a consideration of the state matrix with no symmetries is relevant in context of an engineered quantum state too. For clarity of presentation of our ideas, here we assume that the state $|\Psi \rangle$ is not subjected to any  discrete symmetry constraints.

\subsection{Ensemble Density}

As mentioned in section \ref{intro}, a lack of exact determination of the components due to underlying complexity leaves their statistical description as the best option. A  quantum state with random components is best described by an ensemble of exact replicas of the state matrix. The nature and type of the ensemble representing a random  quantum state depends on the nature of its dynamics in the chosen basis.  While an ergodic state can be well-represented by a basis-independent ensemble, the non-ergodicity requires the ensemble to be basis dependent.  Here we consider an ensemble of state matrices representing a wide range of non-ergodic states.
Referring the joint probability density function (JPDF) of the components $C_{kl}$  as ${\rho_{cn}}(C; h,b)$ (also known as the ensemble density), it can be expressed as
\begin{eqnarray}
{\rho_{cn}}(C; h,b) = {\rho_c}(C; h,b) \; \delta \left(\sum |C_{kl}|^2 -1 \right) 
\label{rhon}
\end{eqnarray}
where $ {\rho_c}(C; h,b)$ describes the JPDF of the state without normalization constraint. 

To determine the average behaviour of $R_n$ from eq.(\ref{rhon}), we proceed as follows. The JPDF $\rho_w(W)$ of the reduced density matrix $W$ can be expressed as
\begin{eqnarray}
\rho_w(W) &=& \int \delta(W - C \cdot C^{\dagger}) \; {\rho_{cn}}(C; h,b) \; {\rm D}C \\
&=& \delta({\rm Tr}\, W -1) \; \int \delta(W - C \cdot C^{\dagger}) \; {\rho_c}(C; h,b) \; {\rm D}C 
\label{w1}
\end{eqnarray}

As a consequence of randomized  $W$, the Schmidt eigenvalues  and thereby entanglement measures can fluctuate over the ensemble (from one sample to another) and it is necessary to consider their distribution over the ensemble. Defining the eigenvalues of $W$ as $e_1, \ldots, e_N$, the JPDF $P_c(\lambda_1, \ldots, \lambda_N)$ of the $n^{th}$ eigenvalue to lie between $\lambda_n \to \lambda_n+d\lambda_n$  (for $n=1 \to N$) can be given as 
\begin{eqnarray}
P_c(\lambda_1, \ldots, \lambda_N) &=& \int \prod_{n=1}^N \delta(\lambda_n- e_n) \; \rho_w (W) \; {\rm D}W \\
&=& C_{hs} \;  \delta\left(\sum_n \lambda_n -1\right) \; P_{\lambda}(\lambda_1, \ldots, \lambda_N) \label{pndf}
\end{eqnarray}
with $P_{\lambda}(\lambda_1, \ldots, \lambda_N)=\int \prod_{n=1}^N \delta(\lambda_n- e_n) \; \delta(W - C \cdot C^{\dagger}) \; \rho_c(C; h,b) \; {\rm D}C  \; {\rm D}W$. Here $C_{hs}$ is a normalization constant: choosing $\int \delta\left(\sum_n \lambda_n -1\right) \; P_{\lambda}(\lambda_1, \ldots, \lambda_N) \; {\rm D}\lambda ={1/C_{hs}}$ renders $ P_c$ normalized to unity: $\int P_c \; {\rm D}\lambda =1$ with ${\rm D}\lambda \equiv \prod_{n=1}^N {\rm d}\lambda_n$.

While, similar to a non-random state,   the entanglement entropy of a single sample of  a  random state  too does not depend on the basis, but its ensemble average can in general depend on the ensemble parameters. This can be seen from following expression for average entanglement entropy \cite{Shekhar_2023}, 
\begin{eqnarray}
\langle R_n \rangle &=& \int  R_n(\lambda_n) \;\delta\left(S_1-\sum_n \lambda_n\right) \; P_{\lambda} \; {\rm D} \lambda
\label{rndf}
\end{eqnarray}
with $R_1(\lambda_n)= -\sum_n \lambda_n \; \log \lambda_n$, $R_2 = - \log \left(\sum_n \lambda_n^2 \right)$ and $S_1$ is an arbitrary constant constraint on the trace, later to be set as $S_1=1$.


As the components $C_{kl}$ are measures of the complicated correlations between the two sub-parts, their determination for generic system conditions can often be technically non-trivial and may involve statistical errors. Besides even if the components of the initial state say $|\Psi(0) \rangle$ are known exactly, a transmission through a quantum channel in presence of an uncontrollable environment may randomize the state: 
$|\Psi(t) \rangle = U(t,0) |\Psi(0) \rangle$ with $U$ as the random unitary  operator describing the time-evolution of the state. This gives 
$C_{kl}(t)= \sum_{ij} U_{kl; ij} C_{ij}(0)$ where $U_{kl; ij} \equiv \langle a_k b_l \mid U \mid a_i b_j \rangle$ are random variables (subject to unitary constraint). Thus, for non-random  components at time $t=0$, their coupling with $U$ randomizes them for later times; invoking central limit theorem then predicts them to be Gaussian distributed if e.g., $U_{kl; ij}$ are random variables with finite second moments. This motivates us to consider, in present study,  the case in which $C_{kl}$  are best known up to their mean and variances only and subjected to normalization condition  $\sum_{k,l} |C_{kl}|^2 = 1$. Based on maximum entropy hypothesis,  the joint probability density function (JPDF) of the components can then be given by eq.(\ref{rhon}) with 
\begin{eqnarray}
    {\rho_c}(C; h,b) \propto {\rm exp}\left[-\sum_{k,l,s} \;\frac{1}{2h_{kl;s}} \left( C_{kl;s} -b_{kl;s} \right)^2 \right] 
    \label{jpdfMulti}
\end{eqnarray}
with $C_{kl}=C_{kl,1} + i \, C_{kl;2}$ with $k=1 \to N, l=1 \to N+\nu_0$ and $s=1,2$. An important point worth emphasizing here is as follows:  while $C_{kl}$ are not independent due to normalization condition, their distributions can still be uncorrelated; this is indeed the idea used in the theory of random ergodic states (see \cite{zycz5})  or Berry's random wave conjecture of \cite{be77, be772}.

Eq.(\ref{rhon}) along with eq.(\ref{jpdfMulti}) describes a multiparametric Gaussian ensemble of state  matrices representing a normalized pure state  in a bipartite basis. 
As the ensemble parameters are a measure of the fluctuations of the components (e.g., due to estimation error or in presence of disorder) 
different choices of  parametric  matrices $h\equiv \left[h_{kl,s} \right]$ and $b\equiv \left[b_{kl,s} \right]$ can correspond to different pure states. For example, using limit of a Gaussian as a $\delta$-function, we have $\lim_{h \to 0} \rho_c(C) \to \delta(C -b)$ (with  $\lim_{h \to 0}$ implying $h_{kl;s} \to 0$  $\forall k,l,s$); the limit therefore leads to a non-random state. Similarly a choice of $h_{kl;s} \to \; \alpha_k \;\delta_{l1}$ and $b_{kl;s} \to \; 0$  $\forall k,l,s$ gives a typical state of the ensemble separable in the chosen bipartite basis: $|\Psi \rangle = |\Phi_A\rangle \; | B_1 \rangle$ where $|\Phi_A\rangle \sim \sum_{k=1}^{N_A}\alpha_k \; | \alpha_k \rangle)$ is a state in $A$-subspace.  (We note that a  separable state in general corresponds  to each component written as a product i.e., $C_{kl}=\alpha_k \beta_l$. Assuming both $\alpha_k$ as well as $\beta_l$ as non random, the pdf of $C_{kl}$ can be expressed as  $\delta(C_{kl}- \alpha_k \beta_l) =\lim_{v \to 0} {\rm e}^{-(C_{kl}- \alpha_k \beta_l)^2\over 2 v}$.  Alternatively, a Gaussian distribution for $C_{kl}$, for the case in which $\alpha_k$ and 
$\beta_l$ are randomly distributed real variables with their PDFs as  $\rho$ and $\sigma$, can arise  if following condition is satisfied: $ \sigma (\beta_l) = \int {\rm exp}\left[- \;\frac{\alpha_k^2 \, \beta_k^2}{2 h_{kl}} \right] \; \rho(\alpha_k)  \; {\rm d}\alpha_k$; here $C_{kl}$ is assumed real  with its mean zero to simplify explanation without any loss of generality).

 In general, a non-ergodic random state can consist of components non-Gaussian, non-identically distributed (e.g., both Gaussian and non-Gaussian as well as a few of them non-random too) or even correlated.  A theoretical analysis of a generic non-ergodic random  states including all cases is however not only  technically complicated but also time consuming. While eq.(\ref{rhon}) represents only a class of   non-ergodic random states, it still represents their wide range  and its analysis can help us to gain insights for more generic cases. This motivates us to keep  the present analysis confined to only cases represented by the ensemble density of the above type.

\subsection{Complexity Parameter Formulation}

The contact with the environment subjects a physical system to many perturbations, often causing a variation of system parameters with time. The latter in general affects the state of the system and, thereby matrix elements $C_{kl}$,  resulting in an evolution of $\rho_c$ in matrix space. But, for $\rho_c$ to represent one of the  quantum states, it is necessary that the ensemble parameters take  into account all system-specifics, thereby rendering them to be  functions of the system parameters. Consequently, they also vary, leading to an evolution of $\rho_c$ in the $\{h_{kl}, b_{kl}\}$ space too.  For the evolving ensemble to continue representing the system,  it is desirable that the evolution of $\rho_c$ in matrix space  is exactly mimicked by that in ensemble parameter space.  As discussed in detail in \cite{Shekhar_2023, Shukla_2017} (also see {\it appendix \ref{comparamApp}}),  indeed a constant diffusion of $\rho_c$ in the matrix space, with  a drift under harmonic potential, can be exactly described by  a specific combination of the first order variation  of the ensemble parameters.

The multiparametric evolution can further be simplified by a transformation from $M$ dimensional space $b_{kl;s}, h_{kl;s}$ to another space $Y_1, \ldots, Y_{M}$ that maps the JPDF in eq.(\ref{jpdfMulti})   to  $\rho(H; Y_1, Y_2, \ldots, Y_{M})$ and leads to its single parametric  evolution \cite{Shekhar_2023}, 

\begin{equation} 
 \frac{\partial \rho}{\partial Y_1} =  L \rho, \hspace{0.2in} \frac{\partial \rho}{\partial Y_{\alpha}}=0  \hspace{0.1in} (\alpha > 1)
 \label{evo1}
\end{equation}  
with $L=\sum_{k,l,s} \frac{\partial}{\partial C_{kl;s}}\left[ \frac{\partial  }{\partial C_{kl;s}}+\gamma \; C_{kl;s} \;  \right]$ and   $\rho = {\mathcal N}_0 \; \rho_c$ with ${\mathcal N}_0$ as a constant  \cite{Shekhar_2023}.  Here $\gamma$ is an arbitrary non-zero parameter, related to variance of  $C_{kl}$ in equilibrium limit. We note $\gamma=0$ corresponds to an evolution with no equilibrium limit.

As discussed in {\it appendix \ref{comparamApp}}, $Y_1$ can be obtained by solving a set of characteristic equations, 
$\frac{d h_{kk;s}}{ f_{kk;s}} = \ldots = \frac{d h_{kl;s}}{f_{kl;s}} = \frac{db_{kl;s}}{b_{kl;s}} = \frac{dY_{\alpha}}{\delta_{\alpha 1}}$ with $f_{kl;s} \equiv | 2(1- \gamma h_{kl;s})|$; the solution can be given as 
 \begin{equation} 
Y_1=-\frac{1}{2 M \, \gamma} \; \sum_{k,l,s}'  (\ln f_{kl;s} + \gamma \ln |b_{kl;s}|^2 )+{\rm const.,} 
  \label{Yparam}
\end{equation}
where the notation $\sum_{k,l}'$ implies a sum over $h_{kl;s} \not= \gamma^{-1}$ and $b_{kl;s} \not=0$ only \cite{Shekhar_2023}.
 We note that for the evolution occurring in the neighbourhood of $b_{kl}=0$, or $h_{kl;s} \to \gamma^{-1}$ the corresponding parametric derivative does not contribute to $Y_1$ 
(discussed in more detail in {\it appendix} \ref{comparamApp}).

With $\frac{\partial \rho}{\partial Y_{\alpha}}=0$ for $n >1$, Eq.(\ref{evo1}) implies an evolution of the JPDF $\rho(C)$  in terms of  $Y_1$ subjected to $M-1$ constants of evolution, namely, $Y_2, \ldots, Y_{M}$. As discussed in {\it appendix} \ref{comparamApp}, $Y_{\alpha}$, for $\alpha >1$ can be obtained by solving the eq.(\ref{ltna}).

As discussed in detail in \cite{Shekhar_2023}, eq.(\ref{evo1}) along with eq.(\ref{w1}) and eq.(\ref{pndf}) can further be used to derive the $Y$-governed evolution of the reduced density matrix $\rho_A \equiv C \cdot C^{\dagger}$ and thereby Schmidt eigenvalues,
\begin{eqnarray}
\frac{\partial P_{\lambda}}{\partial Y_1}= {\mathcal L}_{\lambda} \, P_{\lambda},  \hspace{0.5in}  \frac{\partial P_{\lambda}}{\partial Y_{\alpha}}=0, \hspace{0.1in} ({\alpha} > 1)
\label{pdl1}
\end{eqnarray}
where 
${\mathcal L}_{\lambda} \equiv \sum_{n=1}^N \left[\frac{\partial^2 (\lambda_n \; P_{\lambda})}{\partial \lambda_n^2} - \frac{\partial}{\partial \lambda_n}\left( \sum_{m=1}^N \frac{\beta \lambda_n}{\lambda_n- \lambda_m} + \frac{\beta \,(\nu_0+1)}{2} - 2 \gamma \lambda_n  \right)  P_{\lambda}\right]$ and $\nu_0= N_B-N_A$. Here $\beta=2$ for $C$ as a complex matrix; (as discussed in \cite{Shekhar_2023}, $\beta=1$ for $C$ as a real matrix but the numerical analysis in present study is confined to $\beta=2$ case only).  We note that $Y_1$ was replaced by $Y$ in \cite{Shekhar_2023} for clarity purposes. As discussed in {\it appendix \ref{lentR12}}, the level repulsion term  (also known as avoided crossing) in the above equation plays an important role in the evolution of Schmidt eigenvalues. Indeed, in separability limit,  all eigenvalues except one of them are zero; in maximally entangled limit of the ensemble, eigenvalues are separated by a mean level spacing. The evolution of a typical eigenvalue between the two limits is significantly influenced by the level repulsion term.

Eq.(\ref{rndf}) along with the above equation leads to the complexity parameter governed evolution equation for $\langle R_n \rangle(\Lambda, Y_2, \ldots, Y_M, S_1)$  (discussed in \cite{Shekhar_2023} and referred hereafter as $\langle R_n \rangle(\Lambda, S_1)$ for brevity). Although technically complicated for finite $N$ case, the equation can be simplified in limit $N \to \infty$, leading to

\begin{equation}
    {2\over \beta}  \frac{\partial \langle R_n \rangle}{\partial \Lambda} = g_n(\Lambda, S_1) - \frac{\partial \langle R_n \rangle}{\partial S_1}, \hspace{0.5in}
 \frac{\partial \langle R_n \rangle}{\partial Y_{\alpha}} = 0,   \hspace{0.1in} 
 (\alpha > 1)
    \label{vnd1}
  \end{equation}
with $S_1 = {\rm Tr}(C \cdot C^{\dagger}) = \sum_{n=1}^N \lambda_n$ with $S_1=1$ refers to unit trace constant and 
\begin{eqnarray}
\Lambda =N_A \, N_B \, (Y_1-Y_{0})
\label{alm}
\end{eqnarray}
 as the rescaled ensemble complexity parameter with $N_A N_B$ as the size of the Hilbert space in which state $| \Psi\rangle$ resides, and
 
\begin{eqnarray}
g_1(\Lambda, S_1) \equiv \left( {\langle R_0 \rangle\over N}- q_0 \; J\right), \hspace{0.2in} g_2(\Lambda, S_1) \equiv - {2 \beta \, S_1\over N} {\left\langle {1\over S_2} \right\rangle}
\label{gn}
\end{eqnarray}
with $R_0 = -\sum_n \log \lambda_n$ and $S_2=\sum_{n=1}^N \lambda_n^2$ and  $q_0 =\frac{2 N+ \nu_0 -1}{N+\nu_0}$. Here $J$ is the normalization constant set to unity in the present work: $J \equiv J(S_1) =  \int  {\rm D} \lambda \; P_c =1$
 We note, for latter reference, 
$g_1(\infty,1) \approx \frac{\langle R_0(\infty, 1) \rangle}{N} \approx \log N $ and $g_2(\infty, S_1) \equiv - {2 \beta \over N} {\langle {1\over S_2} \rangle} \sim \log N$ with $\langle R_1 (\infty, 1 )\rangle = \left(\log N - \frac{N}{2 (N+\nu_0)}\right)$ and $\langle R_2 (\infty, 1 )\rangle  =\log \frac{N \,(N+\nu_0)} {2 N +\nu_0 - 1}$. For brevity,  large $N$ limit of $\langle R_n (\infty, 1 )\rangle$ will also be referred as $R_{n, \infty}$.

Eq.(\ref{vnd1}) can be solved by method of characteristics leading to a general solution $\langle R_n \rangle(\Lambda, S_1)$ for arbitrary trace $S_1$. As our interest is in $S_1=1$ and for a separable initial state chosen at $Y_1 = Y_0$ i.e. $\Lambda=0$ with $\langle R_n (0, 1) \rangle=0$,  the solution becomes ({\it appendix} \ref{appAvgRn}),


\begin{equation}
    \langle R_n (\Lambda, 1) \rangle \approx {(-1)^{n-1}\over n} \, g_n(\Lambda, 1) \, \left[1 - L^{- \tau \left(\frac{\beta \,\Lambda}{2}-1\right)}\right].
    \label{avgRnye}
\end{equation}
Here $\tau$ is an arbitrary function independent of $\Lambda$ and is introduced to ensure correct limiting behaviour for  $\langle R_n \rangle$ in the limits $\Lambda \to 0$ and $\infty$; this requires $\tau \to 0$ in large $N$ limit but $\tau \Lambda \to \infty$ if $\Lambda \to \infty$. (As discussed later in section III.B, $\tau=1/D_n$ where $D_n$ is defined in eq.(\ref{yeLa})). 
The above exponential form for $ \langle R_n \rangle$ is also supported by the solution obtained by an alternative route  discussed in \cite{Shekhar_2023} (given by eq.(43) therein) as well as numerical analysis. (We recall that   $\langle R_n (\Lambda, 1 ) \rangle$  theoretically predicted in \cite{Shekhar_2023} is of form similar to eq.(\ref{avgRnye}) but with $\tau=1$ and  $g_n(\Lambda, 1)$ approximated as $R_{n, \infty}$; the latter approximation was based on ignoring the $\Lambda$-variation of $\langle R_0 \rangle(\Lambda,1) $ with respect to $\langle R_n\rangle(\Lambda,1) $).

Eq.(\ref{avgRnye}) depends on $g_n(\Lambda, 1)$ and thereby requires information about $\Lambda$-dependence of ${\langle R_0 \rangle}$ and ${\langle {1\over S_2} \rangle}$. Proceeding as in case of $R_1, R_2$, the $\Lambda$ dependent behaviour of ${\langle R_0 \rangle}$ and $\langle Q \rangle \equiv \langle{1\over S_2}\rangle$ can be given as (details in {\it appendices \ref{solR0}, \ref{solQ}} respectively)
\begin{eqnarray}
 \langle R_0\rangle (\Lambda, 1) &=& \langle R_0 \rangle(\infty, 1)  +   {N^2 \over  2 \gamma \tau (\Lambda-1)}
 \label{rt9d} \\
\langle Q\rangle(\Lambda, 1) &=& {\rm e}^{\tau \,\left(1-(\beta \,\Lambda/2)\ \right)} + {\beta \over 2 N}  \left\langle Q^2\right\rangle \left(1- \; {\rm e}^{\tau \,\left(1-(\beta \,\Lambda/2) \right)} \right) 
\label{qt9d} 
\end{eqnarray}


For brevity,  a function $\langle F \rangle (\Lambda,1)$ will hereafter be denoted as $\langle F \rangle (\Lambda)$.

\subsection{Determination of rescaling parameter $\tau$}

The unknown $\tau$, appearing as a rescaling parameter of $\Lambda$ in eqs.(\ref{avgRnye}, \ref{rt9d}, \ref{qt9d}),  has an important role: it is introduced to satisfy the desired limiting behaviour of the function at $\Lambda=0, \infty$. To determine $\tau$, we proceed as follows. 

We note that the first term in the right side of eq.(\ref{pdl1}) 
smaller by $O(N)$ as compared to others term and can be neglected in large $N$ limit; this can be seen by a rescaling of Schmidt eigenvalues $\lambda_n \to r_n D$, with $D \sim N^{b}$ ($b \sim 1$ based on numerics discussed in section III). With 
$P_{\lambda}(\lambda_1,\ldots, \lambda_N) \, {\rm D}\lambda =   P_r(r_1,\ldots, r_N) \, {\rm D}r $, eq.(\ref{pdl1}) in terms of rescaled variables  becomes 

\begin{eqnarray}
 \frac{\partial P_{r}}{\partial Y_1} \approx  - \sum_{n=1}^N  \frac{\partial}{\partial r_n}\left( {1\over D} \sum_{m=1}^N \frac{\beta r_n}{r_n- r_m}  - 
 2 \gamma r_n  \right)  P_{r}
\label{pdl1a}
\end{eqnarray}
%
The constraint $\sum_n \lambda_n =1$ on the eigenvalues 
$\lambda_n$ now becomes $\sum_n r_n ={1\over D}$. As clear from the above, the repulsion term contribute significantly for $D \sim \Delta^{-1}$ where $\Delta$ is the average level spacing between Schmidt eigenvalues, and, ${\mathcal J}$ gives the normalization of JPDF $P_r$:
\begin{eqnarray}
 {\mathcal J}  = \int  \delta \left({\mathcal S}_1-\sum_k r_k \right) \; P_r \; {\rm D} r.
\label{sdfa}
\end{eqnarray}
The choice of $J=1$ then gives normalization constant $C_{hs}$ for $P_c$ as  $C_{hs} = { D \over {\mathcal J}}$. We also define
\begin{eqnarray}
\langle {\mathcal F} \rangle = {\mathcal J}^{-1} \; \int {\mathcal F}(r) \;\delta \left({\mathcal S}_1-\sum_k r_k \right) \; P_r \; {\rm D} r
\label{r1df1a}
\end{eqnarray}
A substitution of  ${\mathcal R}_1(r)=- \sum_k r_k \log r_n$ and ${\mathcal R}_2(r) = -\log (\sum_k r_k^2 )$ in the above equation gives ${\langle {\mathcal R}_1 \rangle}$, 
${\langle {\mathcal R}_2\rangle}$ respectively. Similarly a substitution of ${\mathcal R}_0(r)=- \sum_k \log r_n$ and ${\mathcal Q}(r)=- \frac{1}{\sum_k r_n^2}$  also gives $\langle {\mathcal R}_0\rangle$ and $\langle {\mathcal Q} \rangle $.

Using the definition in eq.(\ref{rndf}) and $S_1 = D \, {\mathcal S}_1 $, where ${\mathcal S}_1 \equiv \sum_n r_n$, we now have 
\begin{eqnarray}
\langle R_1(\Lambda,S_1) \rangle &=& 
D \, {\langle {\mathcal R}_1(\Lambda,{\mathcal S}_1) \rangle} -  {\log D}      \label{ax1} \\
\langle R_2(\Lambda,S_1) \rangle &=& 
\langle {\mathcal R}_2(\Lambda,{\mathcal S}_1) \rangle -  
 2 \, \log D   \label{ax2}
\end{eqnarray}    
Similarly    
\begin{eqnarray}   
\langle R_0 \rangle(\Lambda,S_1) &=& \langle {\mathcal R}_0(\Lambda,{\mathcal S}_1) \rangle -   N \, \log D  
    \label{ax0} \\
\langle Q\rangle(\Lambda,S_1) &=& \langle {\mathcal Q} \rangle (\Lambda,{\mathcal S}_1) \over D^2 
    \label{axq}
\end{eqnarray}

As given by eq.(\ref{r1df8aa}) in {\it appendix D}, the mathematical forms of ${\mathcal R}_n$ satisfying required boundary conditions can be determined without any unknowns. A substitution of eq.(\ref{r1df8aa}) in eq.(\ref{ax1}) gives

\begin{eqnarray}
    \langle R_1(\Lambda, 1) \rangle 
&=& {\langle R_0 \rangle\over N}  \left(1- L_A^{\left({1\over D}-{\frac{\chi_1 \beta \Lambda}{2 D}}\right)} \right) 
\label{ax3}
\end{eqnarray}

Similarly substitution of eq.(\ref{r1df8aa}) in eq.(\ref{ax2}) gives

\begin{eqnarray}
\langle R_2(\Lambda, 1) \rangle 
&=& {\chi_2 \over \chi_1} \, {4 \langle Q \rangle \over \beta N}  \left(1- L_A^{\left({\chi_2 \over \chi_1 D}-{\frac{\chi_2 \beta \Lambda}{2 D}}\right)} \right) 
\label{ax4} 
\end{eqnarray}
with $\chi_1= \frac{N+2\nu D-1}{N+2 \nu-1}$, $\chi_2= \frac{N+\nu D-1}{N+2 \nu-1}$. With $\nu = (N_B-N_A +1)/2$ and $N_A=N$, we note that $\chi_1, \chi_2$ become significant for cases with $N_B \gg N_A$.

In large $N$-limit, and for cases with $\chi_2 \sim \chi_1$, eq.(\ref{ax3}) and eq.(\ref{ax4}) can further be approximated as 

\begin{eqnarray}
\langle R_n(\Lambda) \rangle \approx 
{(-1)^{n-1}\over n} \; g_n(\Lambda) \;  ( 1- L_A^{-\frac{\beta \Lambda}{2 D_n}})     \hspace{1in}  n=1,2
\label{avgRnye1}
\end{eqnarray}   
where $g_n(\Lambda)$ is given by eq.(\ref{gn}) (with $\langle R_0 \rangle$ and $\langle Q \rangle$ by eq.(\ref{rt9d}) and eq.(\ref{qt9d}) respectively) and $D_n = \frac{D}{ \chi_n}$ .

Intuitively the rescaling of $\Lambda$ in eq.(\ref{rlamda})  can be explained as follows:
$\langle R_n (\Lambda ) \rangle$ is a function of Schmidt eigenvalues $\lambda_1, \ldots, \lambda_N$ of the  matrix $C \cdot C^{\dagger}$. While in separability limit, a typical eigenvalue is zero (with only one of them equal to $1$), it is $\sim 1/N$ in maximum entanglement limit or the ergodic limit of the ensemble (stationary Wishart limit). Further the typical mean level spacing between eigenvalues in the former limit is $0$, it is $1/N$ in the latter limit. Based on the second order perturbation theory of the eigenvalues of Hermitian matrices, a change from one limit to the other is therefore expected to take place for $(Y_1-Y_0) \sim o(1/N)$ (i.e., the parameter value at which the repulsion becomes significant enough). The appropriate parameter governing the evolution in this regime is therefore $\sim N (Y_1-Y_0)$ equivalently $\Lambda/ D$. (As confirmed by numerics too, the entropy shows a rapid change in the regime $Y_1-Y_0 \sim 1/N$). Indeed this is equivalent to  a rescaling of the Schmidt eigenvalues so as to capture the contribution from the eigenvalue repulsion term in eq.(\ref{pdl1}) to $\langle R_n \rangle$.

\subsection{Role of constants of evolution}

The theoretical study in \cite{Shekhar_2023} indicated that the evolutionary paths for $\langle R_n \rangle(\Lambda)$  for different ensembles, with initial condition as a separable state for each of them, are governed by a single parameter $\Lambda$ but did not elaborate on the role of the constants $Y_2, \ldots, Y_M$. The numerical study in \cite{Shekhar_2023} 
however indicated almost analogy of the evolutionary paths in terms of $\Lambda$ (an almost collapse onto the same curve in terms of $\Lambda$) thus implying either no role of constants or same constants for all ensembles considered for numerics. The results in \cite{Shekhar_2023} were however based on small-$N$ numerics, and for balanced condition $N_A=N_B=N$. As the previous study left many queries left unanswered, it is relevant  to pursue a detailed numerical analysis in large $N_A$ limit and verify the improved theoretical prediction given by eq.(\ref{avgRnye}).

The transformation $\{h_{kl}, b_{kl} \} \to Y_1, \ldots, Y_M$     also implies the transformation of the ensemble density in eq.(\ref{rhon}) to $\rho(C; Y_1,\ldots Y_M)$ . The ensembles with different sets of $\{h_{kl}, b_{kl} \}$   can have different values of $Y_1, \ldots, Y_M$, the transformed ensemble density   is therefore represented by different points in complexity parameter space.

In general different states will evolve along different paths in $Y$-space.  It is however possible that two different Gaussian states, represented in two different basis, still have same sets of constants $Y_2, \ldots, Y_M$ for both of them. Such states will evolve, with $Y_1$ as the evolution parameter, along same path in $Y$-space  but that does not ensure  same entanglement entropy for them. As discussed in section III, the latter requires analogy of a rescaled  $Y_1$.

A relevant query in the above context is whether the constants of evolution  $Y_2, \ldots, Y_M$ can always be chosen same for a set of states represented by eq.(\ref{rhon}) if they are subjected to same set of global constraints e.g symmetry and conservation laws?  Intuitively the answer is in affirmative and can be justified as follows:   a  physically motivated basis to represent  a state is usually based on its global constraints.  Different states with a common set of global constraints can then be represented  in same/ equivalent basis. A typical matrix in general has  many basis constants \cite{basis} which can then be chosen as the constants of evolution $Y_2, \ldots, Y_M$. This is later explained through three simple examples (section III.A).

Eq.(\ref{avgRnye1})  suggest a deep universality underlying the pure states represented by the multiparametric Gaussian ensembles. This can be elucidated as follows. 
Different states are in general represented by different ensemble parameters ($\{h, b \}$ sets) in eq.(\ref{jpdfMulti}) and a variation of these parameters is in  general expected to lead to their different evolutions. 
But, following from eq.(\ref{avgRnye1}), the evolution is not governed by the details of $\{h, b \}$ sets but only by a specific function $\Lambda_{ent} (\{h, b \}, N_A)$ and a set of constants of evolution $Y_2, \ldots, Y_M$. In general, if the latter are different, notwithstanding same $\Lambda_{ent}$ for the two states, the paths of their evolution are expected to differ. But, as explained above, the latter can be chosen as the functions of basis constants. As a matrix has a large number of basis constants, the constants $Y_2, \ldots, Y_M$ can be chosen same for different ensembles represented in a fixed basis \cite{basis}. This in turn leaves the evolution characterized only by $\Lambda_{ent}$.  
At some intermediate stage of evolution, it may happen that two different states (i.e., represented by different $\{h, b \}$ sets in eq.(\ref{jpdfMulti}))  may correspond to same $\Lambda_{ent}$ value; Eq.(\ref{avgRnye1}) then predicts their entanglement entropies to be equal at that point if their constants of evolution are also same. 
More clearly, different states with same $\Lambda_{ent}$ value  are predicted to share same entanglement entropy, irrespective of the details about their $\{h,b\}$ sets and therefore form a universality class characterized by $\Lambda_{ent}$. As for finite $N$, the latter can vary continuously between $0 \to \infty$, the above  predicts infinite number of universality classes of the entanglement statistics among pure Gaussian states characterized by $\Lambda_{ent}$. 
  
Any change in system conditions with time however may change the system parameters and thereby the state. This in turn is expected to reflect on the parameters of the state matrix ensemble (if it is a suitable representation of the state). Consequently the two evolving states need not retain the analogy of $Y_1$ and thereby $\Lambda_{ent}$ afterwards.

We emphasize that the universality reported here is different and indeed much deeper  from the one discussed previously, e.g., for excited states \cite{vidmar2018volume},  ground  states \cite{audenaert2002entanglement, osterloh2002scaling, PhysRevA.66.032110, PhysRevLett.90.227902,ding2008subarea, wolf2006violation, gioev2006entanglement, gioev2006entanglement, li2006scaling, cramer2007statistics},  correspondence between criticality and logarithmic  entanglement entropy scaling \cite{PhysRevLett.90.227902, hastings2007area, calabrese2004entanglement, fradkin2006entanglement, kallin2011anomalies, metlitski2011entanglement},  states with topological properties \cite{PhysRevLett.96.110404, PhysRevLett.96.110405, haque2007entanglement}, etc.

\section{Numerical analysis} \label{numerics}

Our primary focus in this work is to numerically analyse the states with Gaussian distributed components and seek 
(i) the validity of eq.(\ref{avgRnye}), i.e, whether a universality of entanglement entropy can be achieved  for the state matrix ensembles described by eq.(\ref{rhon}) with different sets of $h_{kl}, b_{kl}$ in eq.(\ref{jpdfMulti}), 
(ii) a search for a critical regime in which the entanglement entropy of a non-ergodic state does not change with system size, 
(iii)  the sub-system size scaling of the entanglement entropy for a fixed complexity parameter and to understand robustness of universality to sub system-size variation,

\subsection{Details of the ensembles} \label{defEnsembles}

To verify the above claims, we consider four different ensembles with  independent Gaussian distributed matrix elements with zero mean but different decaying-types of variances; the latter imply different decay rates for the typical correlations  between  two orthogonal subspaces of the bipartite state.  The specific details for each ensemble can be given as follows (noting that the row index  refers to orthogonal base of subsystem A and the column index to that of B).

\textbf{\textit{Brownian ensemble (BE):}}  with variance $ h_{k1} =1, \; l=1; \hspace{0.1in} h_{kl}  =  (1+\mu)^{-1},  \;  l > 1$ and mean  $b_{kl;s} = 0 (\forall \; k,l)$ in Eqn. (\ref{jpdfMulti}), typically we have $C_{kl} \sim (1+\mu)^{-1/2}$ for $l >1$ in this case. 
 The latter describes a state with typically same order of overlap between  basis state pairs of $A$ with $B$ except for pairs $| a_k b_1 \rangle$.  With initial ensemble chosen with $\mu \to \infty$,  Eqn. \eqref{alm} gives

\begin{eqnarray}
\Lambda=- \frac{N_A \, (N_B-1)}{2\gamma}\; {\rm ln} \left(1-\frac{2\gamma}{(1+\mu)} \right).
\label{yrp}
\end{eqnarray}

\textbf{\textit{Power law decay ensembles (PE):}} { with $h_{kl} = \left(1+\frac{k (l-1)}{\mu}\right)^{-1}, \;  b_{kl}  =  0 \quad \forall \;\; k, l$, $C_{kl}$ here typically decays as a power law along rows as well as columns. Contrary to BE case, this implies a variation of typical  contributions from the  basis state-pairs $\mid a_k \rangle \mid b_l \rangle$ for $k \not=l$. 
With initial ensemble  parameter $\mu \rightarrow 0$,  Eqn. \eqref{alm} gives }
 
\begin{eqnarray}
\Lambda=-\frac{1}{2\gamma}\; \sum_{r_1=1}^{N_A} \sum_{r_2=1}^{N_B-1} {\rm ln} \left(1-\frac{2\gamma}{1+\frac{r_1 \, r_2}{\mu}}\right)
\label{ype}
\end{eqnarray}

\textbf{\textit{Exponential decay ensemble (EE):}} with $h_{kl} = {\rm exp}\left(-\frac{k |l-1|}{\mu}\right),\;  b_{kl} = 0  \;  \forall \; k, l$; this case is similar to  that of  PE  except for an exponential decay of $C_{kl}$  for sub-bases pairs with higher indices.  Here again with initial ensemble  parameter $\mu \rightarrow 0$,  Eqn. \eqref{alm} gives

\begin{eqnarray}
\Lambda=- \frac{1}{2\gamma}\; \sum_{r_1=1}^{N_A} \sum_{r_2=1}^{N_B-1} \; {\rm ln} \left[ 1- {2\gamma}\; {\rm exp}\left(-\frac{r_1 \; r_2}{\mu}\right)\right].
\label{yee}
\end{eqnarray}

\textbf{\textit{Sparse Ensemble (SE):}} 
Contrary to previous three ensembles, some of the matrix elements in this case are exactly zero, with sparsity decreasing as $n$ increases.  Here again we choose $b_{kl} = 0$ for all entries, and,
\begin{equation}
    h_{kl} = \begin{cases}
        \exp\left[-\left(\frac{k(l-1)}{w^2}\right)\right] \hspace{1cm} d(k, l) = 0, \\
        \exp\left[-\left(\frac{k(l-1)}{w_s^2}\right)\right] \hspace{1cm} 0 < d(k, l) \leq n,
    \end{cases}
\end{equation}
where, $d(k, l)$ is the \textit{hamming distance} between the bases $k$ and $l$, defined as the number of bit flips required to make $k = l$, which has a practical significance to quantum error correction \cite{preskill1998lecture}. In contrast to the ensembles mentioned above, this case has more free parameters: $w$, $w_s$, and $n$. The  initial ensemble in this case corresponds to the choice $w, w_s \to 0$ which gives
\begin{equation}
    \Lambda = -\frac{1}{2\gamma} \left[\sum_{r=1}^{N_A} \; {\rm ln} \left(1- {2\gamma}\; {\rm exp}\left(-\frac{r (r-1)}{w^2}\right)\right) + \sum_{r_1, r_2}^{'}{\rm ln} \left(1- {2\gamma}\; {\rm exp}\left(-\frac{r_1 (r_2-1)}{w_s^2}\right)\right)\right],
    \label{yse}
\end{equation}
where $\sum_{r_1, r_2}$ is over $r_1$ and $r_2$ satisfying $0 < d(r_1, r_2) \leq n$.

We note that each ensemble mentioned above  correspond to zero centred Gaussian entries; the latter implies the typical values of the components of the state (i.e., typical entries  in a typical state matrix of the ensemble)  of the order of variance. Further BE, PE and EE are same as used in our previous study \cite{Shekhar_2023} of entanglement entropy with an important difference: previously  the subsystem sizes $N_A, N_B$ were kept fixed  with only  parameter $\mu$ varying. In the present analysis, $c, \alpha$ and $N_A$  act as free parameters ($\mu \equiv c \, N_A^{\alpha}$; see section \ref{fss}), thereby facilitating the analysis of subsystem size effect on  the universality of the evolutionary path from separability to maximum entanglement.  The choice of the same ensembles helps us in  a direct comparison with our previous results too.

Another point worth noting is the choice of the initial ensemble for the evolution: in each case it corresponds to an ensemble of matrices $C$ with non-zero elements (distributed as zero centred Gaussians and typically of the order of the variance)  only in the first column, i.e., $C_{kl} = 0$ if $l \not=1$.  A typical matrix $C \cdot C^{\dagger}$ in the initial ensemble corresponds to zero entanglement in following sense:  as variances of the higher columns also go to zero (except first one), this leaves state matrix as rank one and therefore  state as separable: $\Psi =  \sum_k C_{k1} |a_k\rangle |b_1\rangle$.  While a separable state can be represented by alternative $C$ matrix-forms, the one mentioned above however has an additional advantage: with initial state chosen in the above form, the evolution of entanglement manifests directly through the increase in elements in higher columns,  with almost all columns becoming of the same order in maximum entanglement limit. 

As mentioned in section II, the parameter $\gamma$ corresponds to the variance of the matrix elements of the ensemble at the end of crossover. For the ensembles mentioned above, the Wishart limit is reached as $h_{kl} \approx h_{kk} =1$. This gives $\gamma=1$ for our numerical analysis of the evolution.

As mentioned above, the condition ${\partial \rho\over \partial Y_{\alpha}}=0$ for $\alpha > 1,$  implies $Y_{\alpha}$ as the constants of evolutions; the latter can be determined by solving the characteristic equation (for the case $b_{kl;s} \not=0$)
$ \frac{d h_{kk;s}}{2 (1-\gamma \;  h_{kl;s})} = \ldots = \frac{d h_{kl;s}}{2 (1-\gamma \;  h_{kl;s})} = \frac{db_{kl;s}}{\gamma b_{kl;s}} = \frac{d Y_{\alpha}}{0}$.
A general solution of the above equation is $F(Y_{\alpha}, \alpha_1, \ldots, \alpha_{M-1})=0$ where $\alpha_n$ are the constant surfaces obtained by solving the above set of $M$ differentials. 
For example, for the matrix of mean values $b= \{b_{kl;s} \}=0$, the choice of an arbitrary pair of differentials gives  $\frac{d h_{mn;s}}{1-\gamma \;  h_{mn;s}} = \frac{d h_{ij;s}}{1-\gamma \;  h_{ij;s}}$ with the corresponding solution as  $\log\left(\frac{1-\gamma \;  h_{mn;s}}{1-\gamma \;  h_{ij;s}}\right) = {\rm constant}$;  the constant so obtained can then be chosen as one of the $Y_{\alpha}$'s for $\alpha >1$. 
Indeed, for BE, the above condition can be satisfied by any two pairs of indices $m,n$ and $i,j$ giving $\log\left(\frac{1-\gamma \;  h_{mn;s}}{1-\gamma \;  h_{ij;s}}\right) = 0$; as many such pairs can be formed, we have $Y_{\alpha}=\left(\frac{1-\gamma \;  h_{mn;s}}{1-\gamma \;  h_{ij;s}}\right)=1$ for all $\alpha >1$. For PE and EE, the condition is satisfied by a specific combination of pairs, i.e., $m,n$ and $n-1,m+1$: 
$ d\log({1-\gamma \;  h_{mn;s}}) - d\log({1-\gamma \;  h_{n-1,m+1;s}})=0$ giving $\log\left(\frac{1-\gamma \;  h_{n-1,m+1;s}}{1-\gamma \;  h_{mn;s}}\right)   = 0$; $Y_k$ can then be chosen as $Y_{\alpha} = \frac{1-\gamma \;  h_{n-1,m+1;s}}{1-\gamma \;  h_{mn;s}}=1$. Proceeding similarly in SE case, pairs of $h_{mn;s}$ can be identified which lead to $Y_k=1$. As the above indicates, it is possible to choose same set of constants $Y_{\alpha}$ for BE, PE and EE, namely $Y_{\alpha}=1$.

We note that the existence of a solution of the type mentioned above requires $h_{mn;s} \propto h_{ij;s}$ (satisfied by all pairs in BE case and for specific pairs in PE and EE case) and need not exist for all ensembles. The alternative solutions leading to constants $Y_{\alpha}$ can then be obtained, e.g., by considering  combinations of more than a pair of  differentials (e.g ${\rm d}Y_{\alpha} \equiv \sum_{mn;s} {q_{mn;s} \;  {\rm d}h_{mn;s}\over 1-\gamma h_{mn;s}} =0$ with $\sum_{mn;s} q_{mn;s}=0$ with $q_{mn;s}$ arbitrary functions).
Indeed as indicated by the study \cite{basis}, a matrix of $N \times N$ has many basis constants  which can be chosen  as the constants of evolution $Y_{\alpha}$; this implies that the transformation $h,b \to Y$-space can always be defined for a given basis.

\subsection{Universality  of evolutionary route} \label{evolution}

A  bipartite quantum state represented by $\rho_{cn}$ (eq.(\ref{rhon})) with $\rho_c$ given by eq.(\ref{jpdfMulti}) corresponds to a point in $h, b$-space.  The $\{h, b\} \to Y $-space mapping gives the location of the state in $Y$-space, labelled by $Y_1, \ldots, Y_{M}$. Due to any change in system conditions, different state ensembles and thereby their entanglement entropies $\langle R_n \rangle$ would then evolve along different invariant curves (in $M$-dimensional space with $M-1$ invariants) marked by the corresponding set of constants $Y_2, \ldots, Y_M$.  (The different ensembles in this paper only refer to those with different means and variances in eq.(\ref{jpdfMulti}).)  The paths of evolution however would be analogous if the constants of evolutions are same for all ensembles under consideration. The entanglement entropies of two such ensembles, say ``1" and ``2" would also be same if $\Lambda_{ent,1} = \Lambda_{ent,2}$.

In \cite{Shekhar_2023}, We  compared $\langle R_1 \rangle$ and $\langle R_2 \rangle$ numerically for each one of the three ensembles mentioned above and found an analogous evolution of the two entropies in terms of the parameter $\Lambda_{ent}$ for ``balanced condition" $N_A=N_B$. The analogy was  consistent with our theoretical prediction in \cite{Shekhar_2023} with $g_{n} \approx R_{n, \infty}$  and $\tau = 1$; here the notation $R_{n, \infty}$ implies $\lim_{\Lambda \to \infty} \langle R_n \rangle(\Lambda)$.
This was indeed expected as the constants of evolution are same for both entropies of a given ensemble and $R_{1, \infty} = R_{2, \infty}$.

\begin{figure}
    \centering
     \begin{subfigure}[b]{0.45\textwidth}
         {\includegraphics[width = \textwidth]{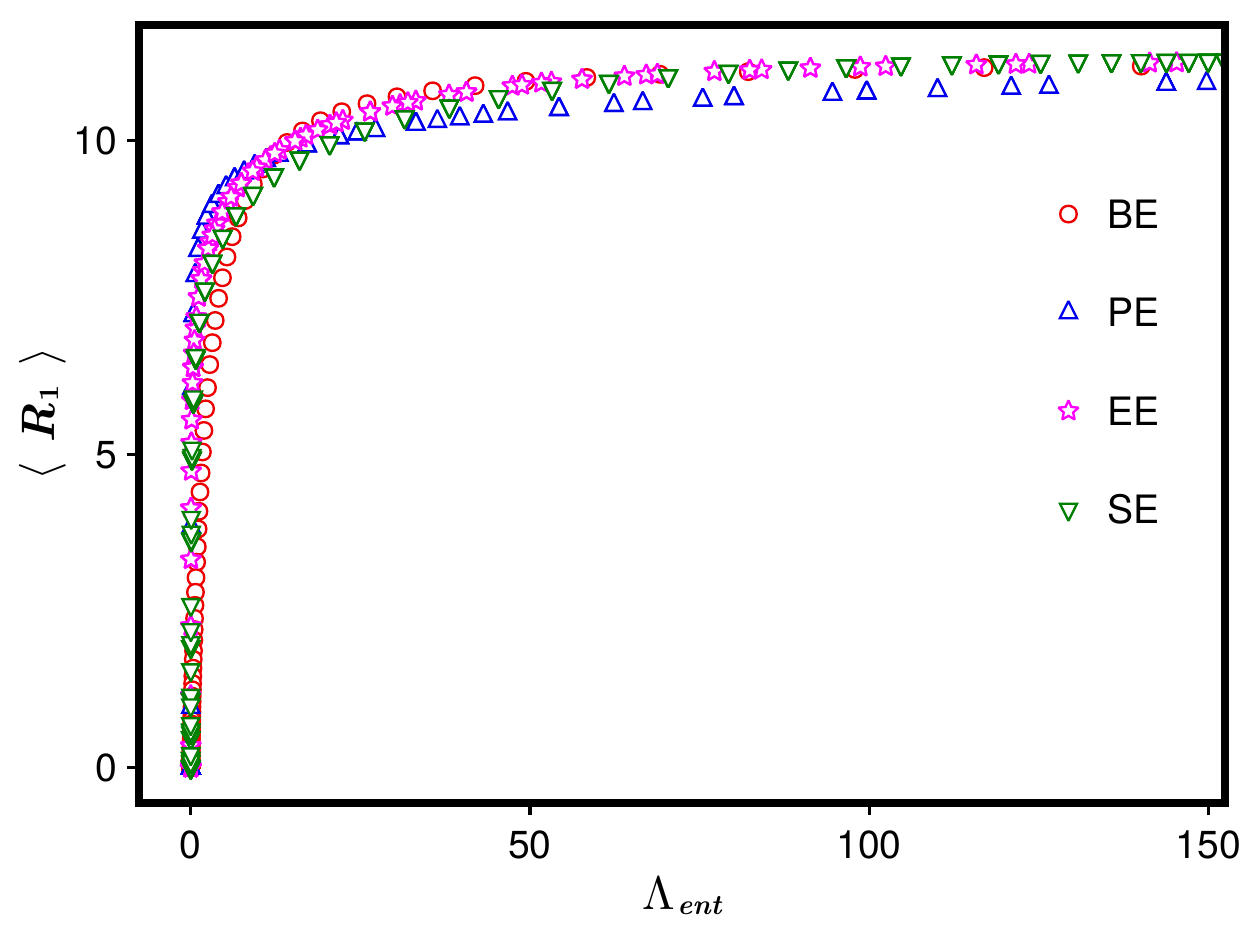}}
     \end{subfigure}
     \begin{subfigure}[b]{0.45\textwidth}
         {\includegraphics[width = \textwidth]{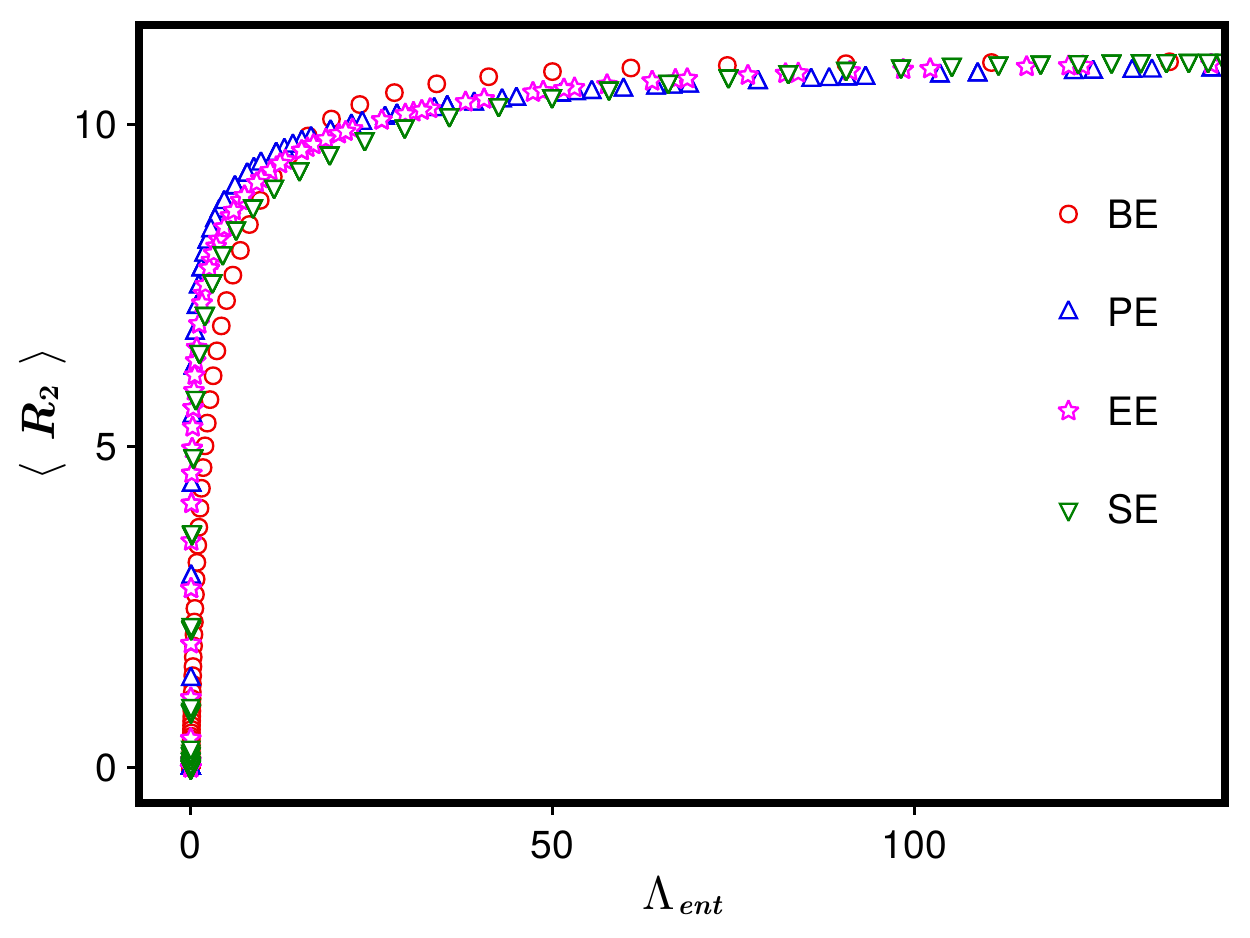}}
     \end{subfigure}
     \caption{Comparison of entanglement dynamics for various ensembles, \textit{e.g.,} ensemble whose variance is constant (BE) Eqn. \eqref{yrp}, decays as power-law (PE) Eqn. \eqref{ype}, decays exponentially (EE) Eqn. \eqref{yee}, with respect to the first column, and the sparse ensemble (SE) Eqn. \eqref{yse} is shown for (a) $\langle R_1 \rangle$ and (b) $\langle R_2 \rangle$, with the \textit{complexity-parameter} ($\Lambda$) rescaled with the respective parameters $D_1$ and $D_2$. The partition in this case is balanced, and the parameter $D_1$ and $D_2$ for BE, PE, EE and SE for both the measures can be interpolated from figs. \ref{Lambda_e} for $L_A = 12 $.}
     \label{tauYye}
 \end{figure}

The study \cite{Shekhar_2023} also suggested that the evolutionary paths for $ \langle R_n  \rangle$ for different ensembles, with initial condition as a separable state for each of them, are almost analogous in terms of $\Lambda$.   
The study was however based on small-$N$ numerics as well as balanced condition $N_A=N_B=N$ and indicated only an almost collapse onto the same curve in terms of $\Lambda$. 
We pursue the investigation further in the present study by a numerical comparison of the dynamics of $\langle R_1 \rangle $ and $\langle R_2 \rangle$ for large size ensembles and for unbalanced condition $N_A \not= N_B$. Here  $N_A= q^{L_A}$ with $q$ as the size of the local Hilbert space (i.e., the one for a  basic unit of which subsystems A and B consist of). For clarity of presentation, here we consider the basic units as the qubits, thus implying $q=2$. For numerical analysis, the Schmidt eigenvalues are derived by an exact diagonalization of the  ensembles of $N_A \times N_A$ matrices $W=C \cdot C^{\dagger}$, with $C$ taken from a BE, PE, EE or SE. Invoking the relation $R_1=-\sum_{k=1}^{N_A} \lambda_k \log \lambda_k$ and $R_2=-\log \left(\sum_{k=1}^{N_A} \lambda_k^2 \right)$ now gives $R_1$ and $R_2$ for each matrix; the average over an ensemble for fixed ensemble parameters then leads to $\langle R_1 \rangle(\Lambda), \langle R_2 \rangle(\Lambda)$ for a given $\Lambda$-value. Repeating the procedure for each ensemble for many  values of the ensemble parameters thereby leads to the results describing evolution of $\langle R_n \rangle$ as a function of $\Lambda$; the above procedure is later referred as the {\it numerical diagonalization route}. As shown in Fig. \ref{tauYye}  for each case (with $L_A$ fixed),  we find the analogy only if $ \Lambda$ is further rescaled by an ensemble specific parameter ($D_n$) leading to 
\begin{eqnarray}
\Lambda_{ent} = \frac{ \Lambda}{D_n}.
\label{rlamda}
\end{eqnarray}

\begin{figure}
     \begin{subfigure}[b]{0.25\textwidth}
         \centering
             {\includegraphics[width=\linewidth]{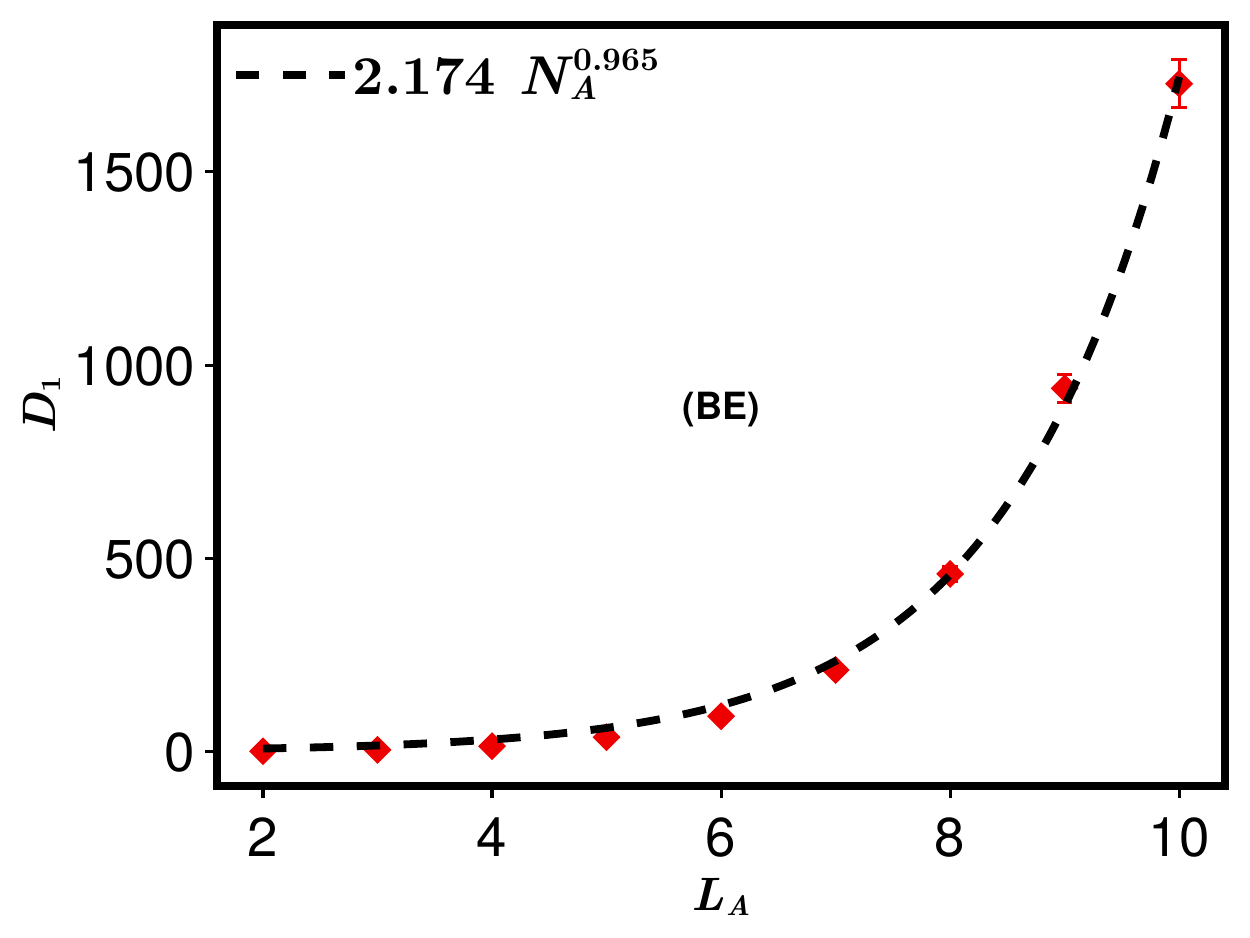}}
         \end{subfigure}
         \hfill
     \begin{subfigure}[b]{0.25\textwidth}
         {\includegraphics[width=\linewidth]{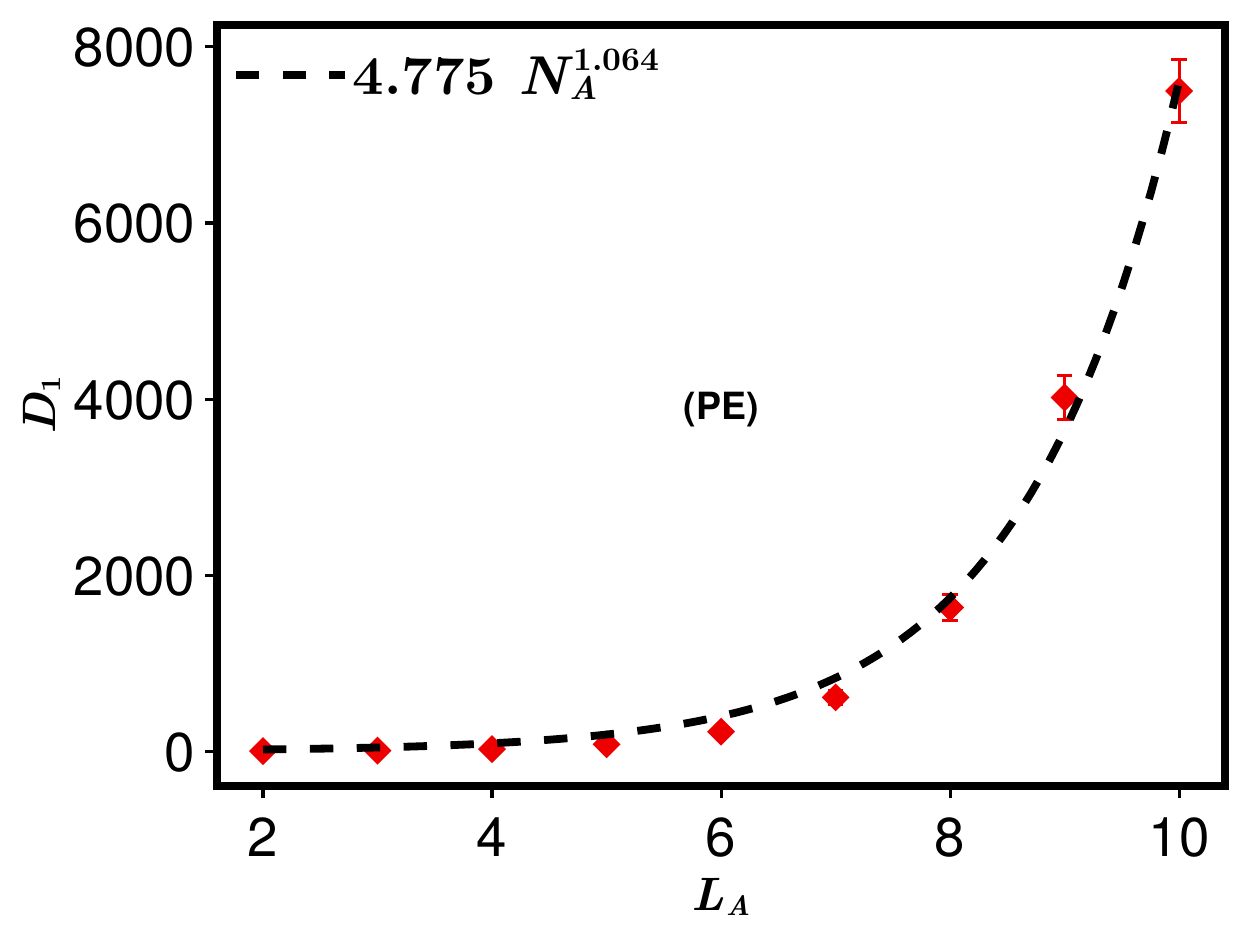}}
     \end{subfigure}
     \hfill
     \begin{subfigure}[b]{0.25\textwidth}
        {\includegraphics[width=\linewidth]{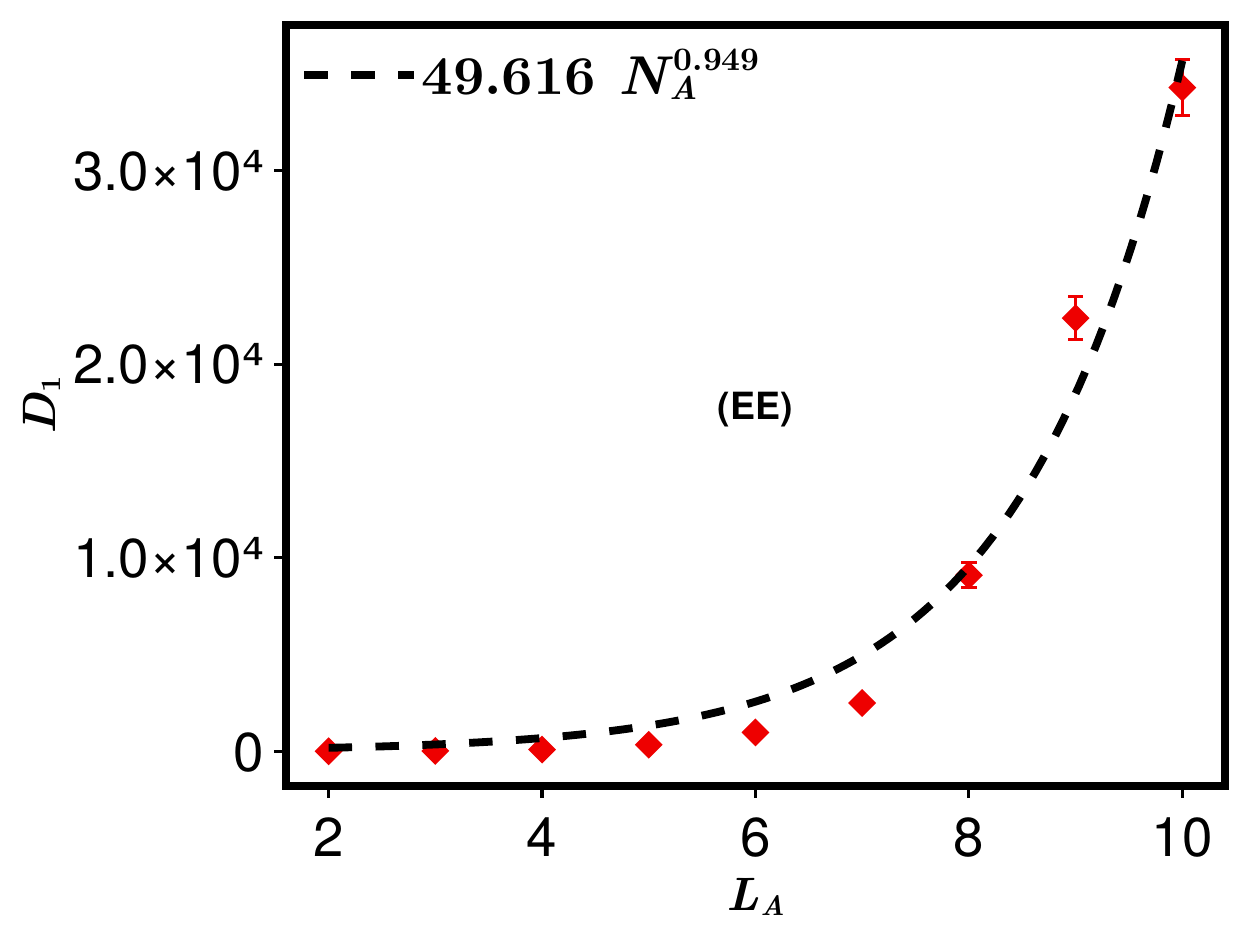}}
     \end{subfigure}
     \hfill
     \begin{subfigure}[b]{0.25\textwidth}
         {\includegraphics[width=\linewidth]{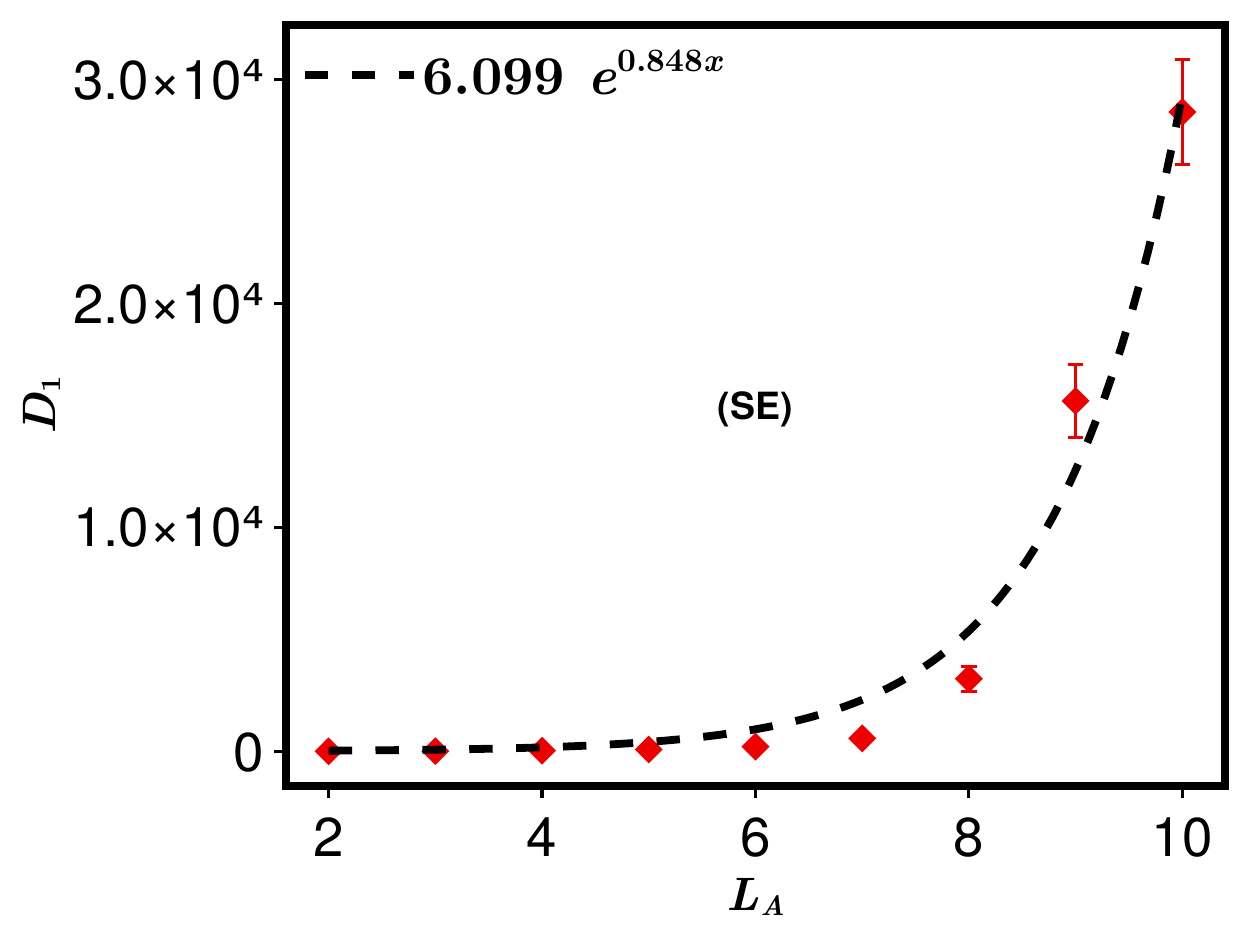}}
     \end{subfigure}
     \hfill
     \begin{subfigure}[b]{0.25\textwidth}
         \centering
             {\includegraphics[width=\linewidth]{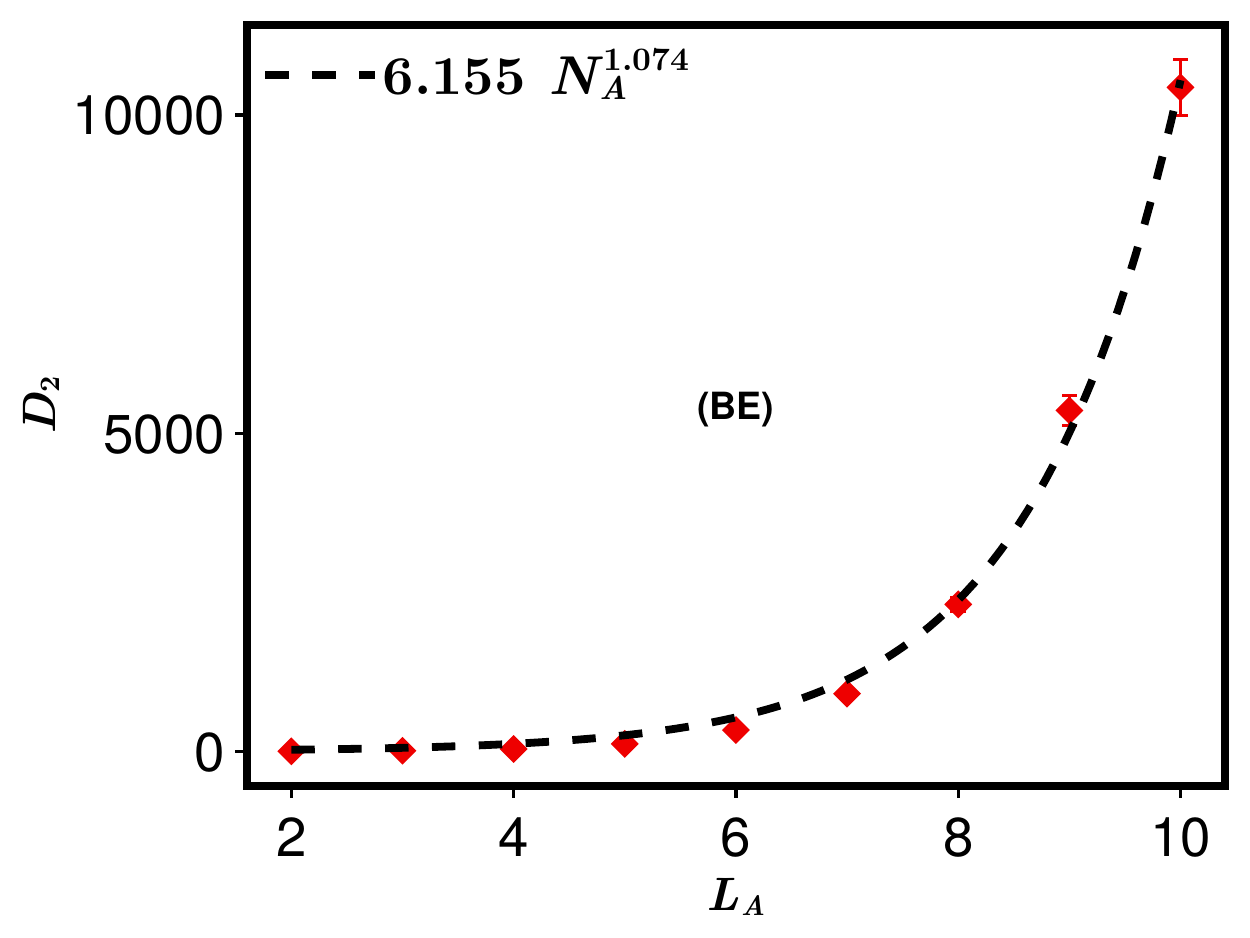}}
         \end{subfigure}
         \hfill         
     \begin{subfigure}[b]{0.25\textwidth}
       {\includegraphics[width=\linewidth]{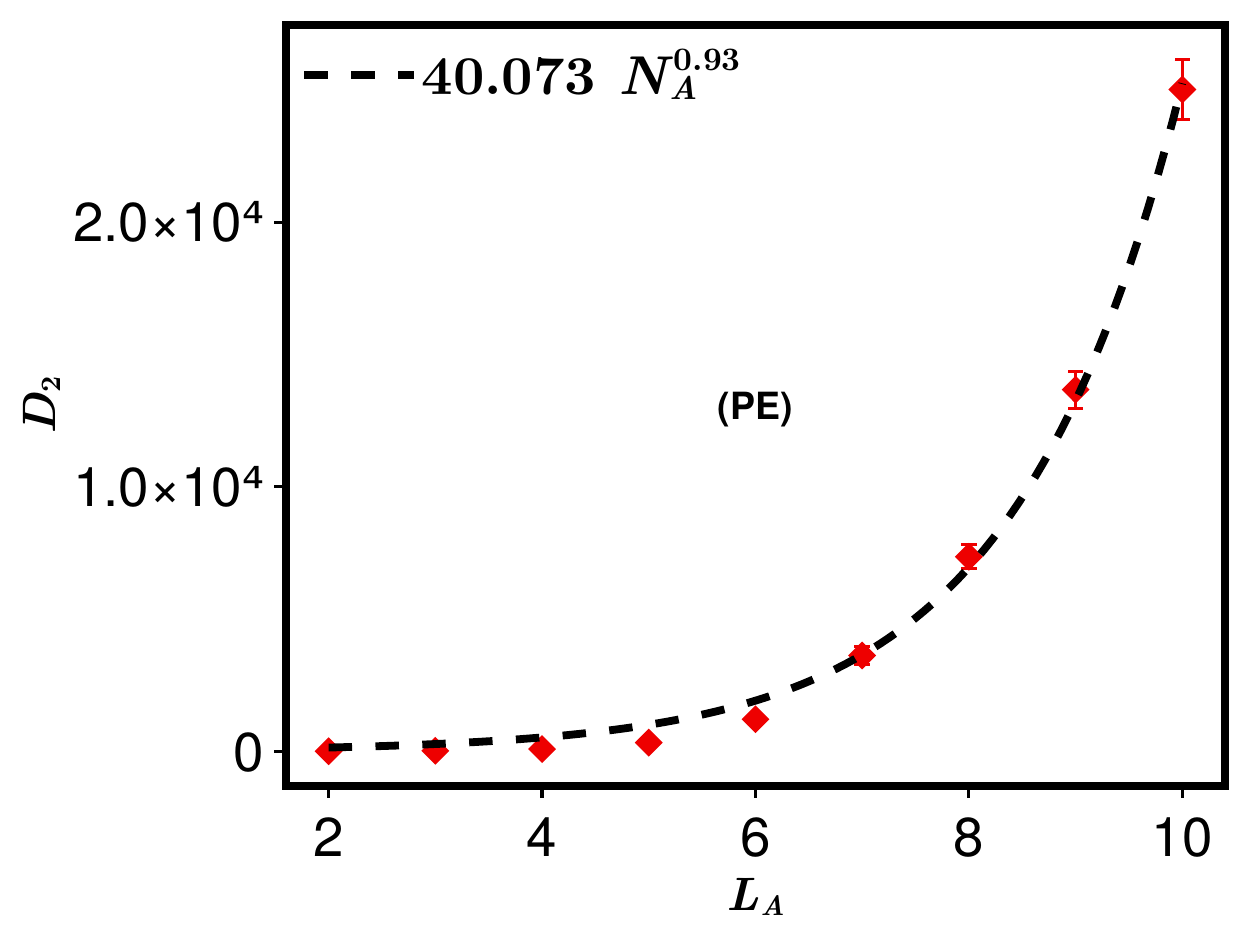}}
     \end{subfigure}
     \hfill
     \begin{subfigure}[b]{0.25\textwidth}
        {\includegraphics[width=\linewidth]{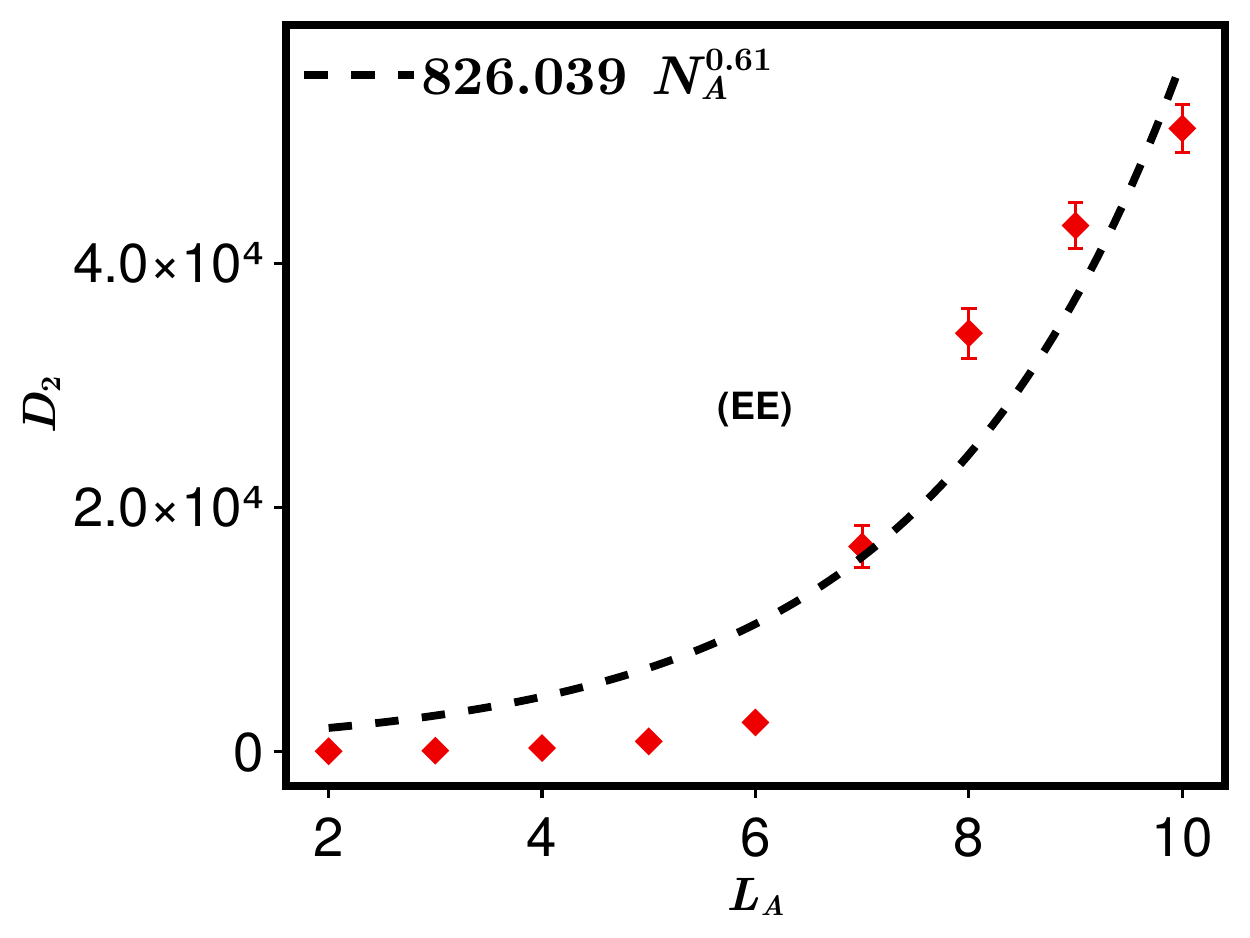}}
         \end{subfigure}
         \hfill
         \begin{subfigure}[b]{0.25\textwidth}
         {\includegraphics[width=\linewidth]{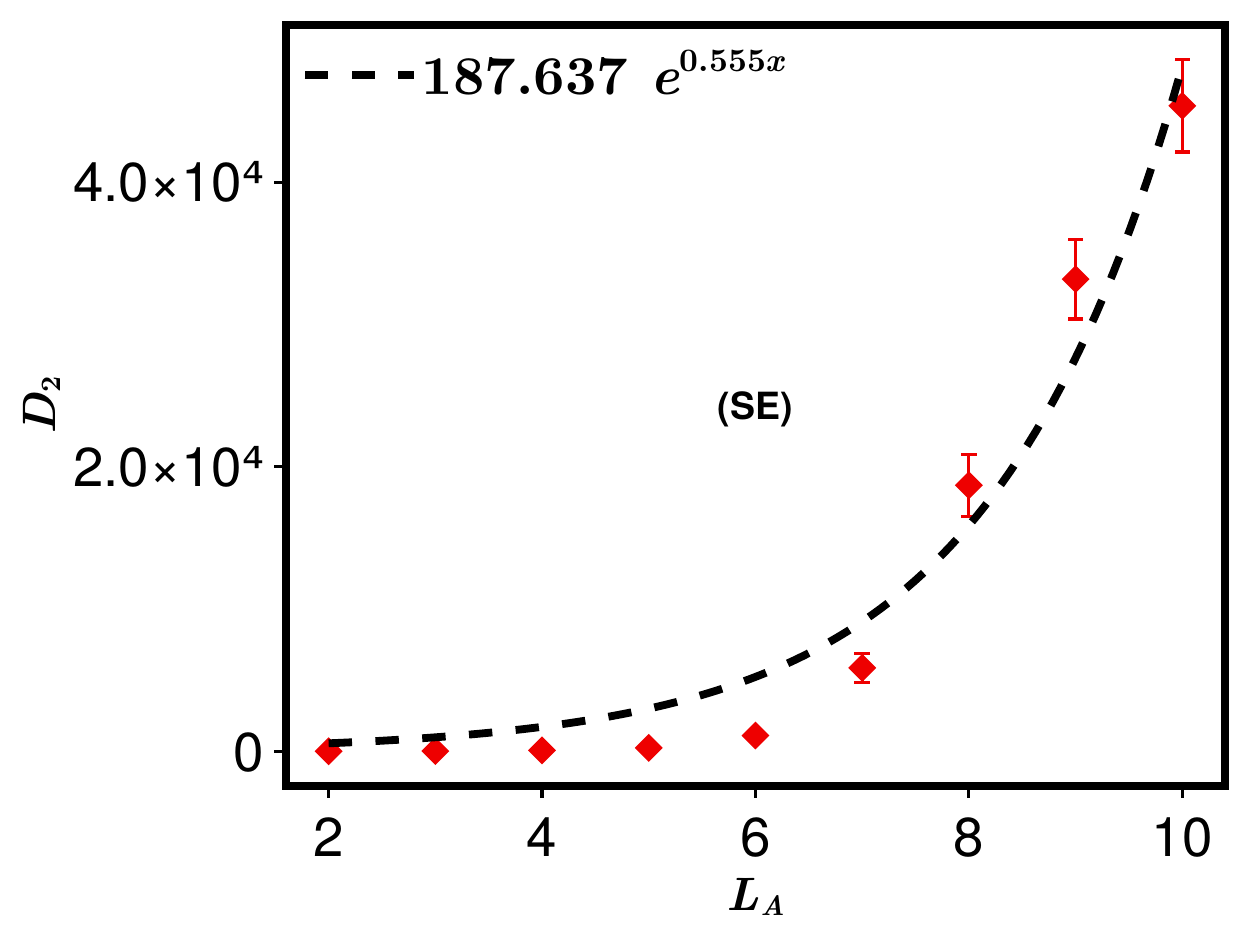}}
         \end{subfigure}
         \caption{$D_1$ and $D_2$ are numerically obtained as a function of sub-system size ($L_A$) for the three different ensembles: BE, PE, EE and SE from (a)-(d) respectively for $R_1$ and (e)-(h) for $R_2$. We analyse the entanglement dynamics for various ``cut'' and later fit the analytic form Eqn. \eqref{avgRnye1} for $R_1$ and $R_2$. We find $D_n$ scaling exponentially with the sub-system size to be the optimum fit, whose goodness has been quantified by its $R^2$ value. The function, along with the fit parameters, is shown in the legend ($N_A = 2^{L_A}$).}
         \label{Lambda_e}
 \end{figure}

To gain insight in  the ensemble-dependence of $D_n$, we numerically analyse  $\langle R_n (\Lambda ) \rangle$ for the four ensembles for all possible bipartitions  (i.e., for a fixed system size $L$ but varying subsystem size $L_A$) and extract $D_n$ values by fitting to Eqn. (\ref{avgRnye}). Based on numerical insights, we conjecture 
\begin{equation}
    D_{1,2}(N_A)  = a \; N_A^b  = a \; 2^{b \, L_A} 
 \label{yeLa}
\end{equation}
with $b \sim 1$ and $a$ dependent on the nature of the ensemble e.g.,  nature of  the decay of the higher column variances (with respect to first one).  The numerical data obtained from four different ensembles, displayed in figs. \ref{Lambda_e}  supports the above conjecture (more discussion in next section).  Further the numerical insight about universality in terms of a rescaled $\Lambda$ (eq.(\ref{rlamda})) is consistent with eq.(\ref{avgRnye1}).

\subsection{Finite Size Scaling and Criticality of the entanglement} \label{fss}

For clarity of presentation, here we consider $R_1$ only and for equal subsystem sizes  $N_A=N_B= 2^{L_A} =2^{L_B}$ i.e. balanced condition $L_A=L_B$ (with $ L_A= \log_2 N_A$).

Based on previous studies, $R_1$ for a generic state of a one dimensional system is believed to scale with subsystem size $L_A$ as 
\begin{eqnarray}
R_1 = q_1 \; L_A + q_2 \; \log L_A + constant
\label{stan}
\end{eqnarray}
with $q_1,\, q_2$ as constants. Typically, $q_1 \neq 0, q_2=0$ for ergodic states \cite{vidmar2018volume, bianchi2022volume}, while $q_1,  q_2=0$ for some non-ergodic states and $q_1=0, q_2 \neq 0$ for states in critical regime \cite{eisert2010colloquium}.   

It is natural to query whether a similar behaviour can be seen for $R_1$ averaged over the ensembles of states given by eq.(\ref{rhon}).  Eqn. (\ref{avgRnye1}) for the latter implies 

(i) $\langle R_1 (\Lambda) \rangle \to  L_A$ and thereby approaching volume law  for $\Lambda  \gg D_1$ and in large $L_A$ limit,  

(ii) $ \langle R_1 (\Lambda) \rangle \to 0$ and thereby approaching separability for $\Lambda \ll  D_1$, 

(iii) For $\Lambda \sim {D_1 \over \log L_A}$, however, the critical behaviour can result  if $g_1(\Lambda)$ and $L^{-\Lambda/ D_1}$ conspire together to give  $\langle R_1 \rangle  \sim \log L_A$ with $g_1(\Lambda) \approx \langle R_0 \rangle(\Lambda)/N$.
From eq.(\ref{rt9d}),  $\langle R_0 \rangle(\Lambda)$ in this can be approximated by  $\langle R_0(\infty) \rangle$.

Further, with both $\Lambda$ and $D_1$  dependent on $L_A$, the above suggests a $L_A$-scaling of $\langle R_1\rangle$  near  the critical point $\Lambda_{ent}^* =\frac{\Lambda^*}{D_1}$. To verify this aspect, we apply a finite size scaling approach as follows.  Assuming  $\Lambda = \Lambda(\alpha L_A)$,  $\Lambda^* = \Lambda(\alpha^* L_A)$ as its value at the  critical point $\Lambda_{ent}^*$ and $\alpha$ as a variable used for scaling analysis,  we  write $\langle R_{1}(\Lambda_{ent}) \rangle$ a function of  the scaling variable $(\alpha-\alpha^*)  L_A^{1 \over \nu}$,
 \begin{equation}
 \langle R_{1}(\Lambda_{ent}) \rangle =  F\left( \left(\alpha-\alpha^*\right) L^{1/\nu}\right)
    \label{xalm0}
\end{equation}
with $\alpha^*$ as the critical point of the transition in  $\langle R_{1}(\Lambda_{ent}) \rangle$ with a critical exponent  $\nu$ and $F(x)$ as an arbitrary function of $x$. To analyse the scaling of $\langle R_{1}(\Lambda_{ent}) \rangle$ near the critical point $\alpha^*$,  we consider  Eqn. (\ref{xalm0}) with $\langle R_{n}(\Lambda^*_{ent}) \rangle=F(0)$.

In order to seek the scaling behaviour and whether  a critical ensemble parameter $\alpha^*$ can indeed exist for the non-ergodic states represented by eq.(\ref{rhon}), we consider two routes: (i) the {\it numerical diagonalization route} mentioned in section \ref{evolution}, and, (ii) as   $\langle R_1 \rangle$ is theoretically predicted by Eqn. (\ref{avgRnye1}),  it is natural to query whether it displays the same scaling behaviour as by the previous route  (referred below as {\it theory} route). We note that our critical point analysis here is confined to BE, PE and EE only with required $\Lambda_{ent}$  obtained from  Eqn.(\ref{rlamda}) along with  Eqn.(\ref{yrp}, \ref{ype}, \ref{yee}) for $\Lambda$ and Eqn.(\ref{yeLa}) for $D_1$.
Indeed the determination of the critical parametric value for SE requires a separate analysis, without providing any additional insights.

 For BE, we consider the case with $\mu = c_1 \, N_A^{\alpha} = c_1 \, 2^{\alpha \; L_A}$ with $c_1$  a non-zero, finite constant; Eqn. (\ref{yrp}) gives 
\begin{eqnarray}
\Lambda =  c \; 2^{(2-\alpha) L_A}.
\label{lamnum}
\end{eqnarray}
with $c=1/c_1$.
 For PE and EE with   $\mu=c_1 \, N_A^{2-\alpha} = c_1 \, 2^{(2-\alpha) L_A} $, Eqn. (\ref{ype}) and Eqn. (\ref{yee}) again give eq.(\ref{lamnum}) but now $c = c_1 \; H_{N_A} H_{N_B-1}$, for PE with $H_N$ as the $N^{th}$ Harmonic number, and, $c =c_1 \; \Gamma \left[0, (c_1 \, 2^{(2-\alpha^*) L_A})^{-1} \right]$ for EE, with $\Gamma(a,x)$ as the incomplete Gamma function.  Eqn.(\ref{lamnum}) then  gives  $\Lambda^* =  c \; 2^{(2-\alpha^*) L_A}$,  this along with  Eqn. (\ref{rlamda}) and Eqn.(\ref{yeLa}) give  $\Lambda_{ent}$-value  at the critical point as $\Lambda_{ent}^* = {\Lambda^* \over D_n} = {c\over a} \; 2^{(2-b-\alpha^*) L_A}$. We note that while $c$ for BE is a constant,  $c$ for PE and EE retains, albeit weak, dependence on $L_A$.

\begin{figure}[!h]
    \centering
    \includegraphics[scale=0.55]{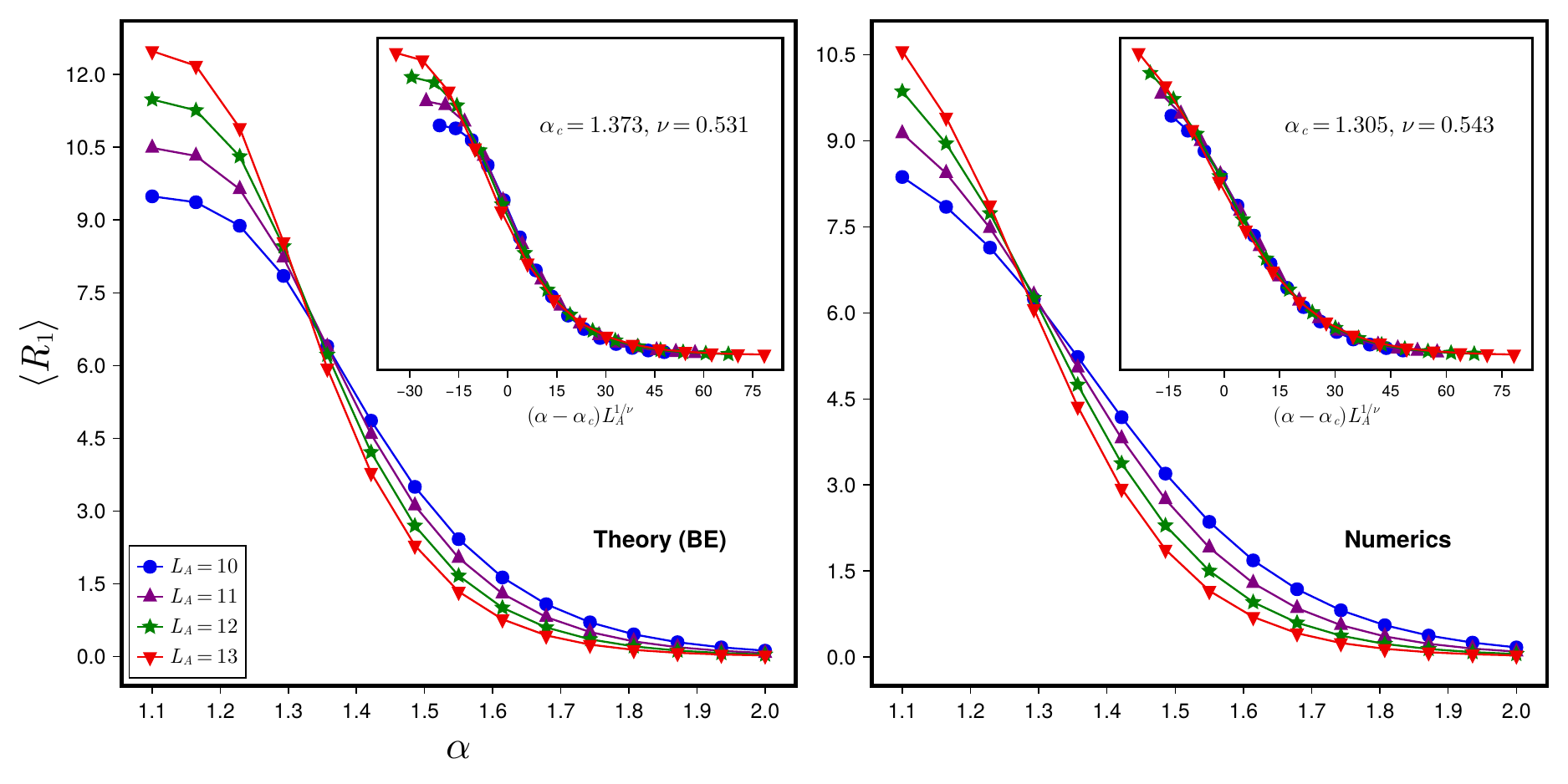}
    \caption{(Right) The scaling of $\langle R_1 \rangle$ with $\alpha$ for different sub-system sizes $L_A = L/2$, for $L = 20,22,24,26$ using the exact diagonalization route is shown, with a scaling collapse, using finite-size scaling, shown in the inset for BE. (Left) Eq. \eqref{avgRnye1} is plotted for a theoretical comparison. The critical point ($\alpha^*$) and exponent ($\nu$) for each ensemble are also displayed in the figures. A close analogy between theory and numerics can also be observed. Similar results for PE and EE are also shown in figs. \ref{scalingCollpasePE} and \ref{scalingCollpaseEE} respectively.}
    \label{scalingCollpaseBE}
\end{figure}

\begin{figure}[!h]
    \includegraphics[scale=0.55]{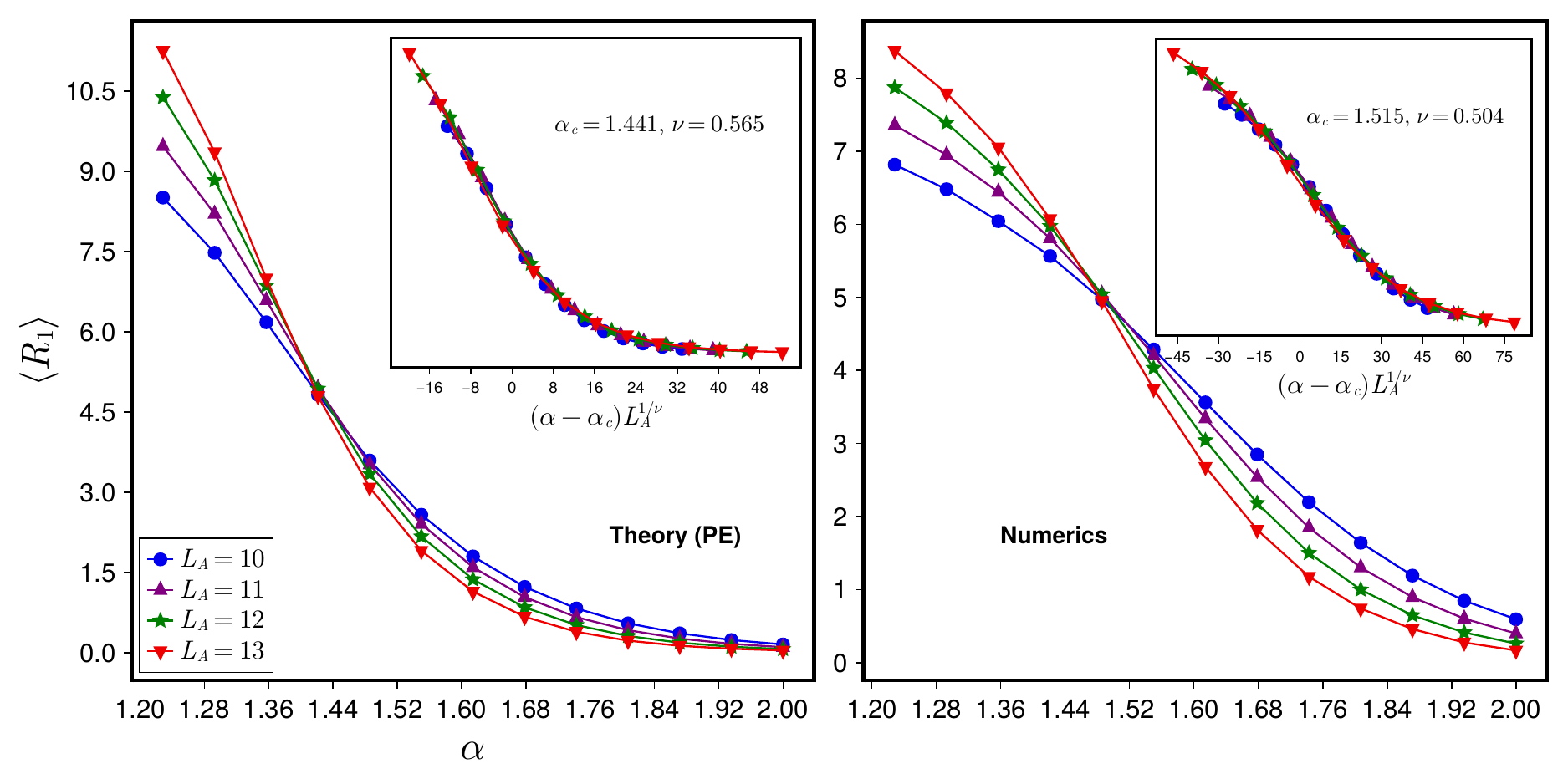}
    \caption{The calculations as in fig. \ref{scalingCollpaseBE} is repeated here for the PE.}
    \label{scalingCollpasePE}
\end{figure}

\begin{figure}[!h]
    \includegraphics[scale=0.55]{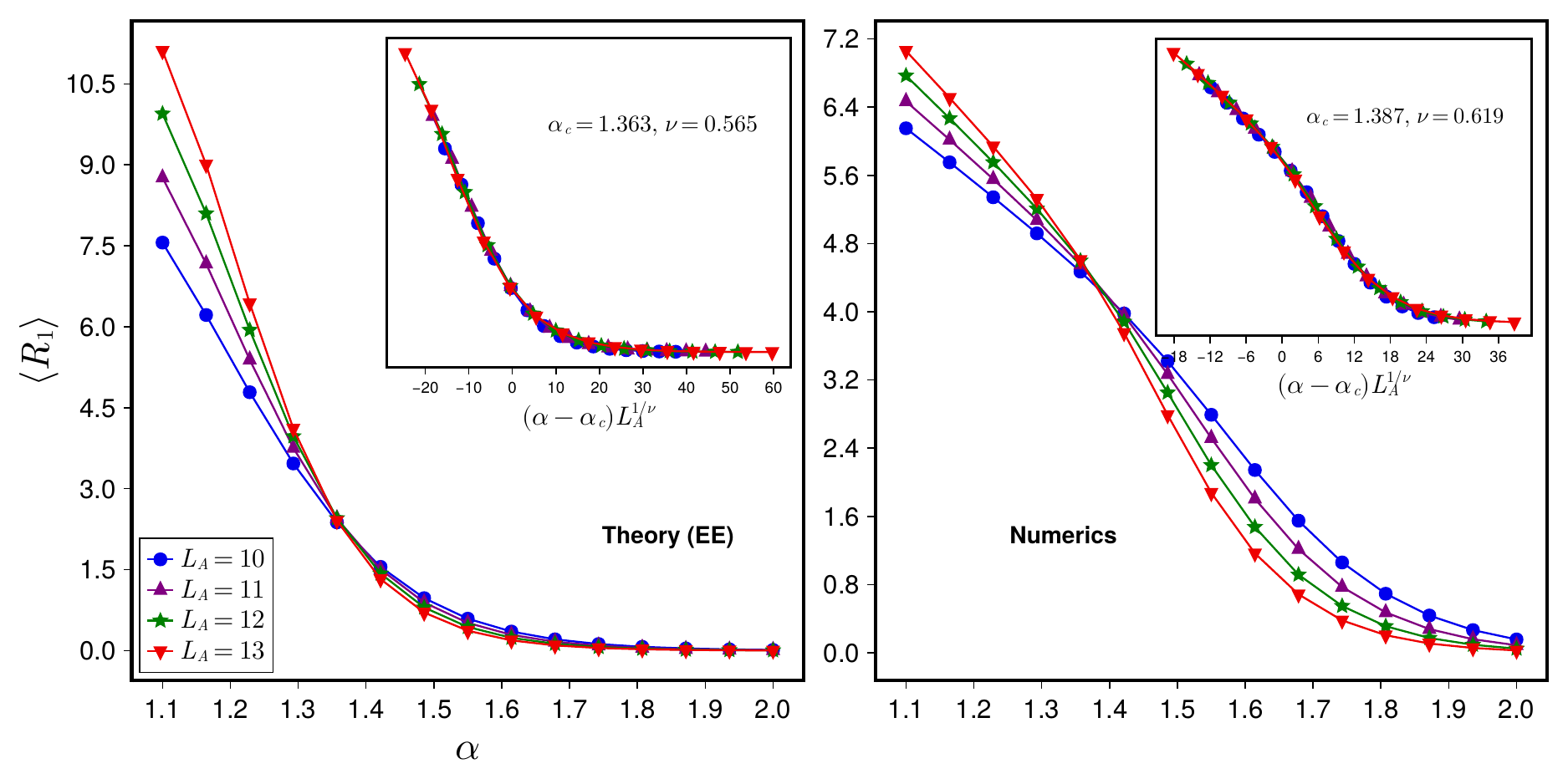}
    \caption{The calculations as in figs. \ref{scalingCollpaseBE} and \ref{scalingCollpasePE} is repeated here for the EE.}
    \label{scalingCollpaseEE}
\end{figure}

Eq.(\ref{rlamda}) along with eq.(\ref{lamnum}) gives $\Lambda_{ent} = {c \over a} \; 2^{(2-b-\alpha) L_A} = 2^{(\alpha^*-\alpha) L_A} \; \Lambda_{ent}^* $.
%
%
For $\alpha > \alpha^*$,  $\lim_{L_A \to \infty}  \Lambda_{ent} \to 0$,  $\langle R_1 \rangle  \to 0$; from eq.(\ref{avgRnye1}), the state is therefore expected to approach the separability limit. For $\alpha <  \alpha^*$,  $\lim_{L_A \to \infty}  \Lambda_{ent} \to \infty$, $\langle R_1 \rangle  \to {\langle R_0(\infty) \rangle \over N} \approx L_A$  and the state therefore approaches the maximum entanglement limit. To seek the critical value $\alpha^*$,  we analyse $\alpha$-dependence of $\langle R_1 \rangle$ for the BE, PE and EE cases for a fixed value of $c_1$ (ref. Table \ref{critTable} for details). The results for each ensemble, obtained by  numerical as well as  theoretical route, are displayed in Fig. \ref{scalingCollpaseBE}, \ref{scalingCollpasePE}, \ref{scalingCollpaseEE} respectively. Here the required theoretical value for $\langle R_0(\Lambda^*_{ent}) \rangle$ to plot theoretical prediction in eq.(\ref{avgRnye}) is determined by eq.(\ref{rt9d}). 
As the figures indicate, $\langle R_1 \rangle$ vs $\alpha$ curves for different $L_A$ values indeed intersect each other at a common point. The corresponding scaling behaviour using $\langle R_1(\Lambda^*_{ent}) \rangle=F(0)$ are also displayed in these figures (with the left displaying theoretical and the right numerical results). As can be seen from the  figures,  $\langle R_1\rangle$ behaves as a sigmoid  function of $(\alpha -\alpha^*) \, L^{1\over \nu}$. (We note that, in large $N$ limit, eq.(\ref{avgRnye1}) can be approximated by a sigmoid form $\langle R_1 \rangle \approx {g_1 \; {\rm e}^{x} \over 1+ {\rm e}^{x}}$ with $x = \Lambda_{ent}^* \log L_A$; the latter form is also consistent  by the displays in the right parts of Figs.\ref{scalingCollpaseBE}, \ref{scalingCollpasePE}, \ref{scalingCollpaseEE}. Table \ref{critTable} displays the values of $\alpha^*, \nu$ and $R^* \equiv \langle R_1(\Lambda_{ent}^*)$ as well as $a, b$  for  each ensemble (for arbitrarily chosen $\mu$, equivalently $c$-value).   A small difference between the critical values obtained from the two routes in Figs. \ref{scalingCollpaseBE} - \ref{scalingCollpaseEE} can be attributed to numerical errors in estimation of Schmidt eigenvalues through exact diagonalization as well as  $D_n$; we recall that the latter is again a numerical prediction.

In general,  the above suggests a classification of the  non-ergodic states described by Eqn. (\ref{jpdfMulti}) in three main classes as $L_A \to \infty$: maximum entanglement, separability, and the third 
a critical regime with $\langle R_1\rangle \propto \log L_A$. 
%
%
Based on BE, PE and EE analysis,  we find  that the critical behaviour 
$\langle R_1 \rangle \sim \log L_A$ occurs only for a specific scaling of  $\Lambda$  with $D_1$. Based on eq.(\ref{avgRnye1}), this is expected to be valid for any ensemble described by eq.(\ref{jpdfMulti}). As both $\Lambda$ as well as $D_1$ depend on a combination of the ensemble parameters, a search for the appropriate combinations leading to critical behaviour  for the SE case is technically time consuming; we expect to pursue it as  a separate analysis.

\begin{table}[h]
    \centering
    \begin{tabular}{ |c|m{3.5cm}|c|c|c|c|c|c|c|c|c|c|c| }
     \hline
     \multirow{2}{*}{\textbf{Ensemble}} & \multirow{2}{*}{\textbf{Variance ($h_{kl}$)}} & \multicolumn{2}{|c|}{$D_1$} & \multicolumn{2}{|c|}{$D_2$} & \multirow{2}{*}{\textbf{c}} & \multicolumn{3}{|c|}{\textbf{Theory}} & \multicolumn{3}{|c|}{\textbf{Numerics}}\\\cline{3-6}\cline{8-13}
      & & $a$ & $b$ & $a$ & $b$ & & $R^*$ & $\alpha^*$ & $\nu$ & $R^*$ & $\alpha^*$ & $\nu$ \\
     \hline \hline
     \textbf{BE} & 
     \[
        \begin{cases}
        1, \; l=1 \\
        \frac{1}{1 + \mu}, \; l \neq 1
     \end{cases}
     \]
     & 2.174 & 0.965 & 6.155 & 1.074 & 0.1 & 7 & 1.373 & 0.531 & 6.5 & 1.305 & 0.543 \\ \hline
     \textbf{PE} & 
     $$\frac{1}{1 + \frac{k \, (l-1)}{\mu}}$$
     & 4.775 & 1.064 & 40.073 & 0.93 & 1 & 4.75 & 1.441 & 0.565 & 5 & 1.515 & 0.504 \\ \hline
     \textbf{EE} & 
     $$\exp \left[-\left(\frac{k \, (l-1)}{\mu}\right)^2\right]$$ & 49.616 & 0.949 & 826.039 & 0.61 & 7 & 2.4 & 1.363 & 0.565 & 4.2 & 1.387 & 0.619 \\
     \hline
    \end{tabular}
    \caption{\label{critTable} Here, $\mu = c\,N_A^{\alpha}$, with $c$ given in the seventh column above for each of the ensembles and $N_A = 2^{L_A}$.}
\end{table}

For further insight, it is instructive to rewrite Eqn. (\ref{xalm0}) as  $\langle R_1 \rangle   =  F\left(\left({L_A\over \xi}\right)^{1\over \nu}\right)$ with $\xi =|\alpha-\alpha^*|^{- \nu}$ (for $\nu >0$).  
We note that $\xi$ has a role in separability $\to$ entanglement transition akin to that of a wavefunction correlation length  in localization $\to$ delocalization transition (e.g. similar to metal $\to$ insulator transition,  $\xi$ diverges at $\alpha=\alpha^*$); this motivates us to refer  $\xi$ as the {\it entanglement length}.  Further, as $\alpha$ refers to an ensemble parameter reflecting complexity of bipartite correlations in the quantum state,  and,  as the ensemble represents a quantum state with maximum entanglement for $\alpha <\alpha^*$, separability for $\alpha > \alpha^*$ and a critical statistics at $\alpha=\alpha^*$,  we refer $\alpha^*$ as marking the {\it edge of entanglement}.

The analysis presented above describes the search for the critical point of $\langle R_1 \rangle$.  Following similar steps, the critical point of $\langle R_2 \rangle$ can also be identified; the analysis is omitted here as it does not give any new insights about critical statistics.

\subsection{Universality and role of ``cut" of Bipartition (relative subsystem size)} \label{cut}

Previous section analysed the role of $\Lambda_{ent}$ in entanglement entropy growth under ``balanced constraint" on subsystem sizes i.e., $L_A = L_B$: section \ref{evolution} confirmed the universality of the growth in terms of $\Lambda_{ent}$ for different ensembles if their constants of evolutions have same values and have statistically analogous initial condition, and, section \ref{fss} established $\Lambda_{ent}$ as the parameter to characterize criticality.

With $L_A$ a system parameter too, it is natural to query as to how the entanglement entropies evolve if  the balanced constraint is removed and  $L_A$ is varied while keeping total system size $L=L_A+L_B$ as well as  $\Lambda$ fixed? We note that $\Lambda_{ent}$  still changes due to  variation of $D_n$ with $L_A$  and therefore $\langle R_n\rangle$ is expected to vary too.  But, from Eqn.(\ref{yeLa}), $D_n$ depends on the ensemble specific constants $a, b$.   It is therefore not obvious whether  an analogous evolutionary path of entanglement entropy, with $L_A$ as the governing parameter,  can still exist for ensembles with same $\Lambda$ as well as constants of evolution?

To seek the answer,  we numerically analyse the evolution of  $\langle R_1\rangle$ and $\langle R_2 \rangle$ for  BE,  PE, EE and SE as the ``cut" of the bi-partition, i.e., $L_A$ is varied at different  fixed levels of complexity quantified by $\Lambda$ (again using {\it numerical diagonalization route} mentioned in section \ref{evolution}). Here a fixed $\Lambda$ for each ensemble is obtained by invoking following condition
\begin{eqnarray}
\Lambda= \Lambda_{BE} = \Lambda_{PE} = \Lambda_{EE} = \Lambda_{SE}
\label{lamnum2}
\end{eqnarray}
Following from eq.(\ref{yrp}), eq.(\ref{ype}), eq.(\ref{yee}) and eq.(\ref{yse}), this leads to
\begin{eqnarray}
    \Lambda= c_b \; 2^{(L-\alpha_b L_A)} =  c_p \; 2^{(L-\alpha_p L_A)} = c_e \; 2^{(L-\alpha_e L_A)} = \Lambda_{SE}
    \label{lamnum3}
    \end{eqnarray}
with $c_b, c_p, c_e$ and $\alpha_b, \alpha_p, \alpha_e$ as specific $c$ and $\alpha$ values for BE, PE and EE which satisfy the above condition. For SE,  the free parameters $w, w_s$ and $n$ are varied to match $\Lambda$. Clearly the same $\Lambda$ value can occur for many combinations of $(c_b, \alpha_b), (c_p, \alpha_p), (c_e, \alpha_e),$ and $(w, w_s, n)$.

\begin{figure}[h]
    \begin{subfigure}[b]{0.3\textwidth}
    \centering
        {\includegraphics[width=\linewidth]{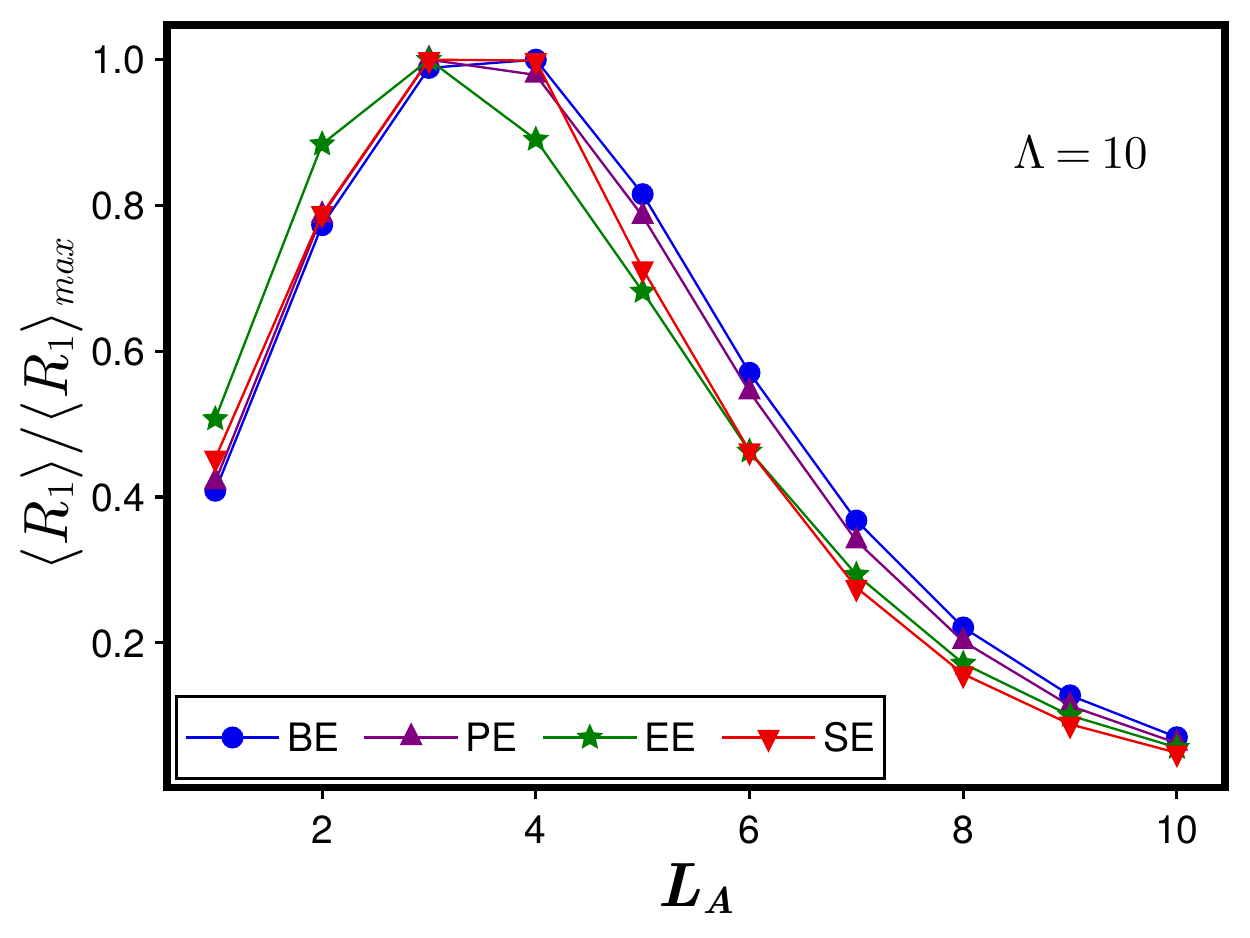}}
        \label{y1r1}
    \end{subfigure}
    \hfill
    \begin{subfigure}[b]{0.3\textwidth}
        {\includegraphics[width=\linewidth]{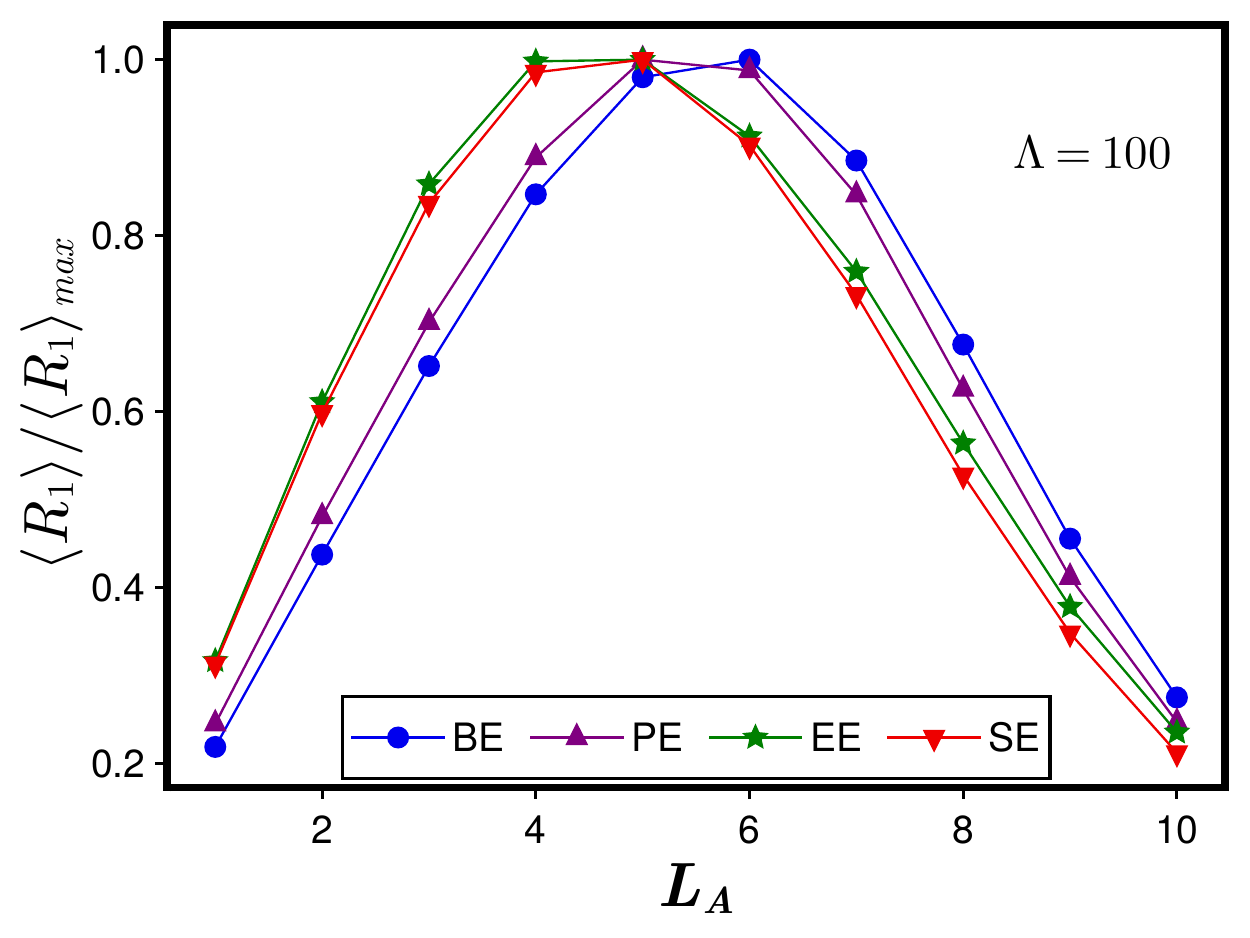}}
        \label{y2r1}
    \end{subfigure}
    \hfill
    \begin{subfigure}[b]{0.3\textwidth}
        {\includegraphics[width=\linewidth]{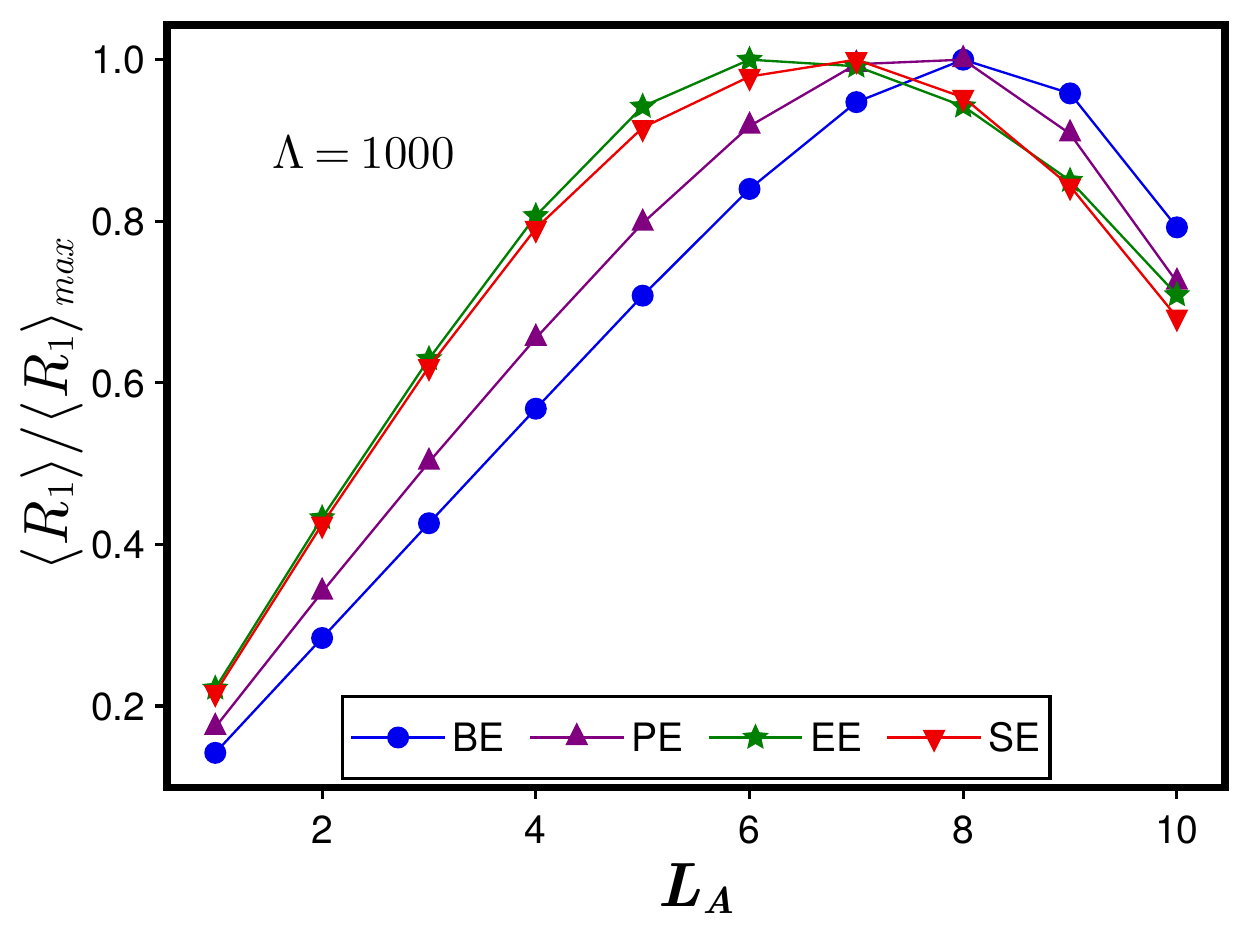}}
        \label{y3r1}
    \end{subfigure}
    \hfill
    \begin{subfigure}[b]{0.3\textwidth}
        {\includegraphics[width=\linewidth]{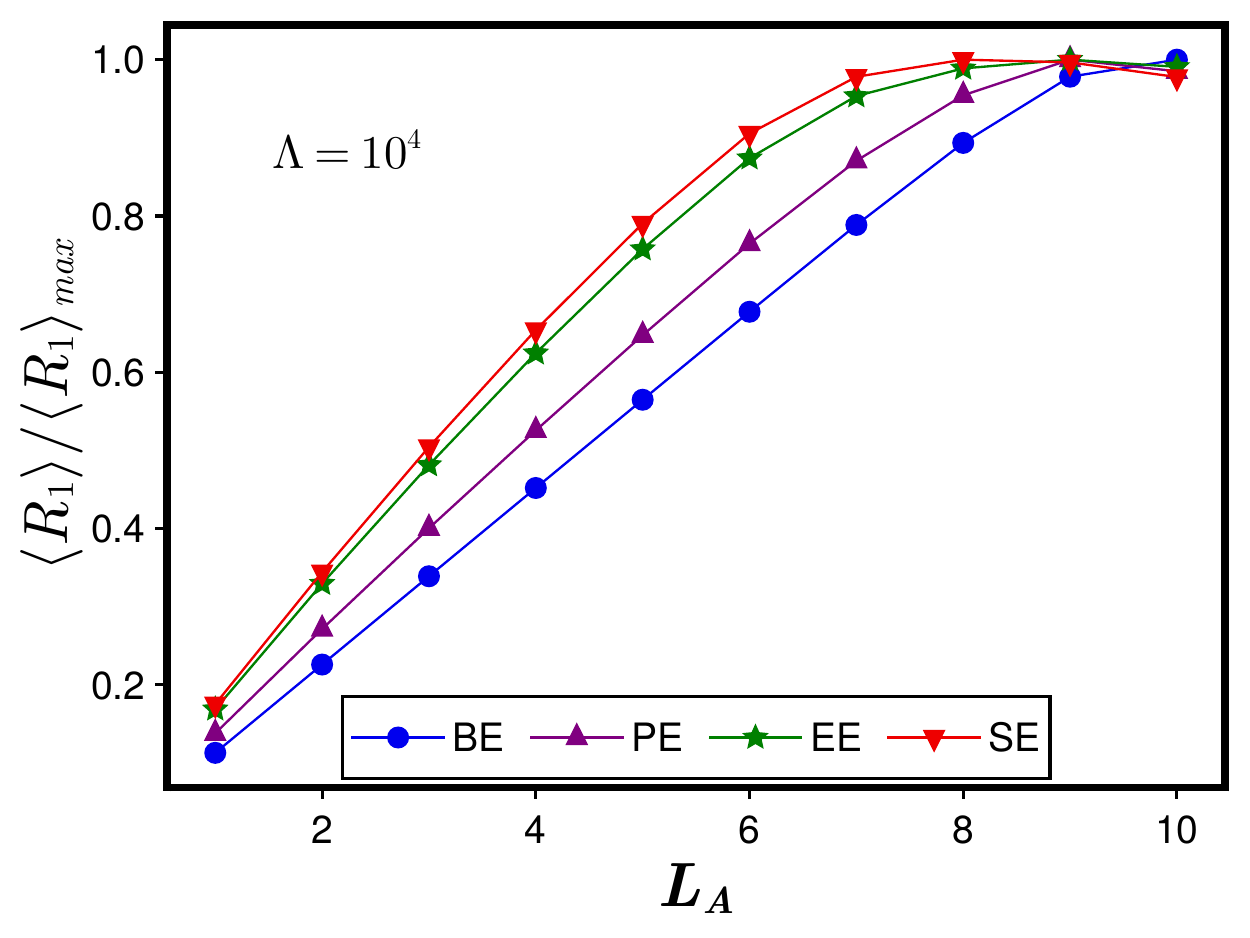}}
        \label{y4r1}
    \end{subfigure}
    \hfill
    \begin{subfigure}[b]{0.3\textwidth}
        {\includegraphics[width=\linewidth]{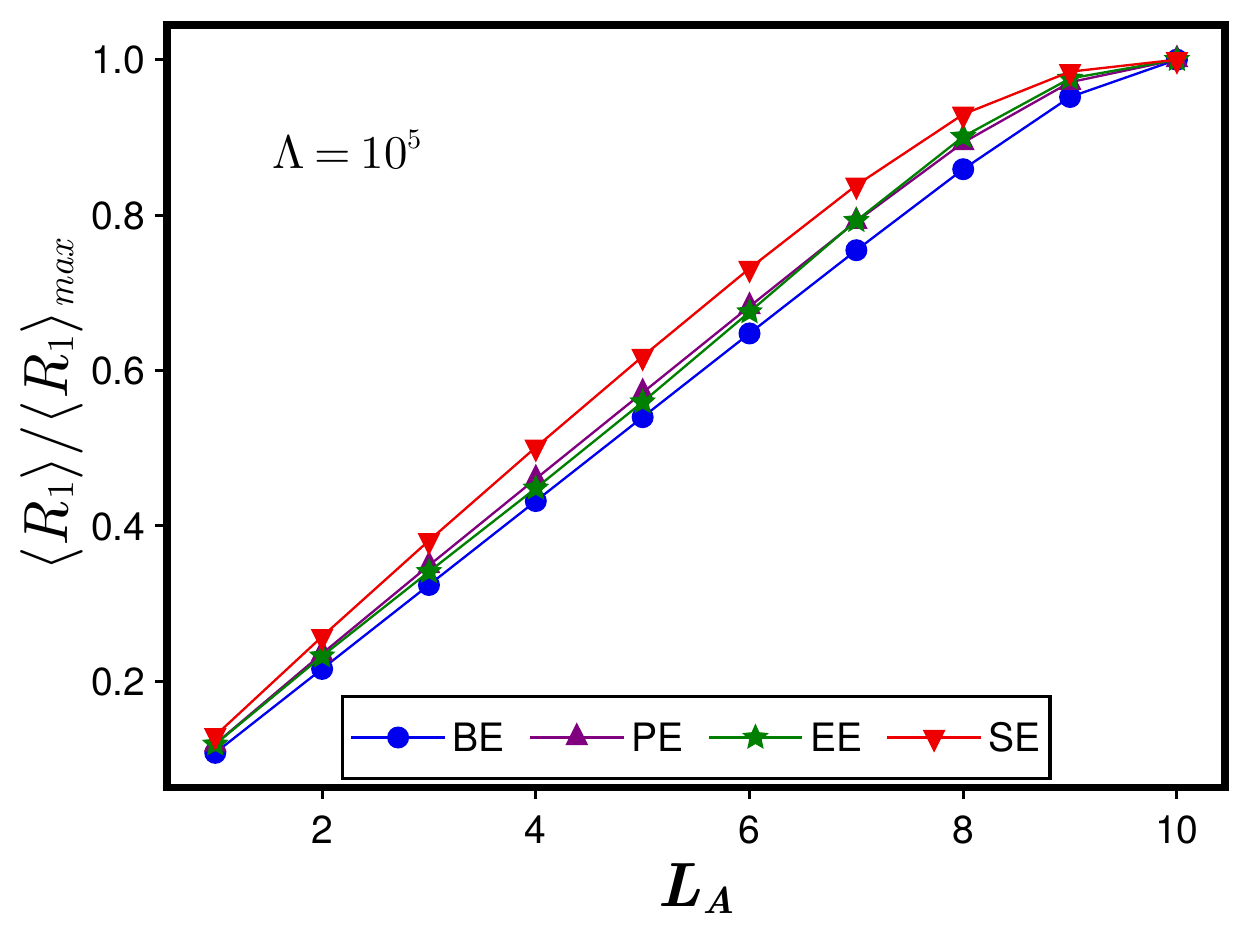}}
        \label{y5r1}
    \end{subfigure}
    \hfill
    \begin{subfigure}[b]{0.3\textwidth}
        {\includegraphics[width=\linewidth]{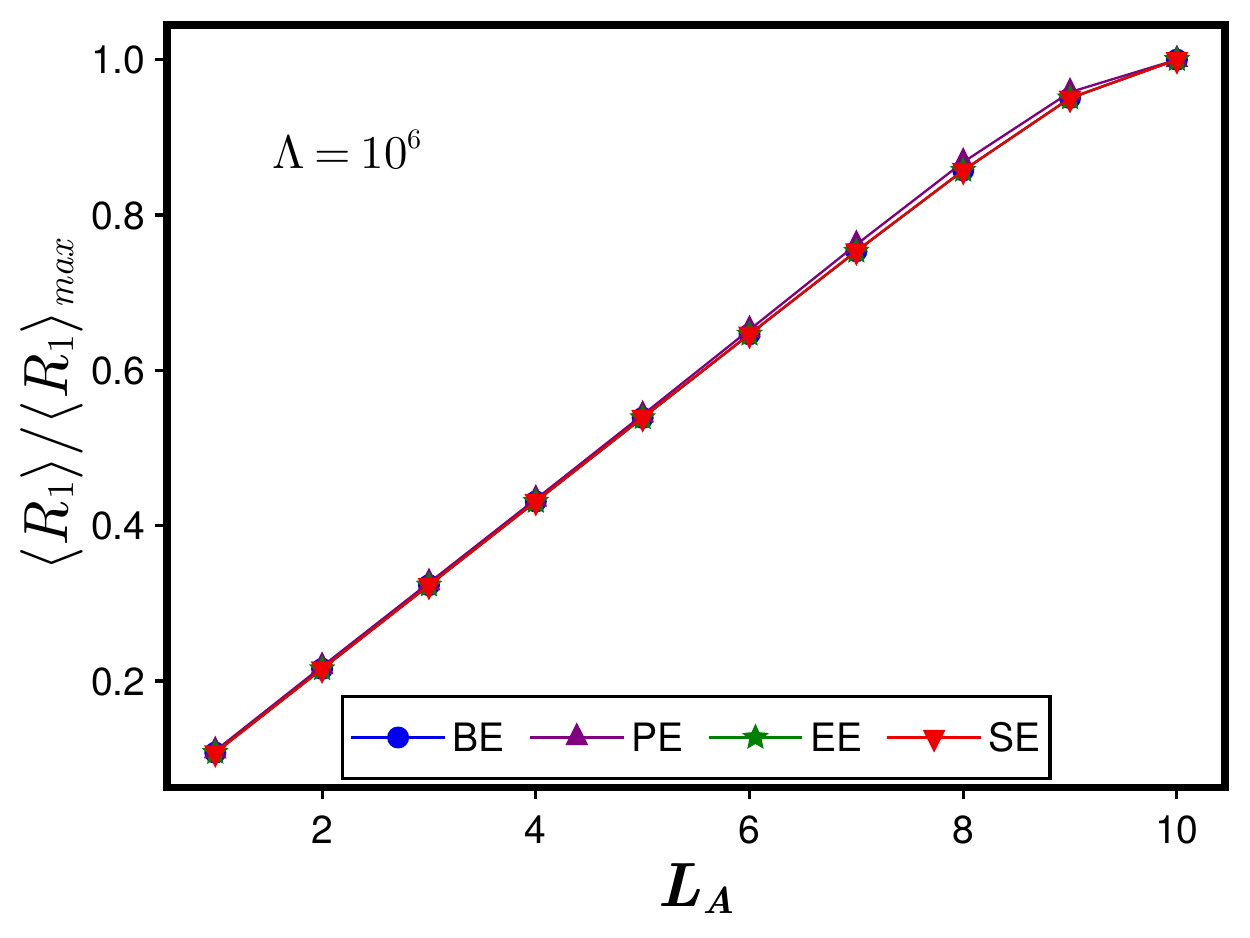}}
        \label{y6r1}
    \end{subfigure}
    
    \caption{The scaling of normalized $R_1$ with sub-system size $L_A$, ranging from $1 \to 10 (=\frac{L}{2})$, is shown for increasing ergodicity (increasing $\Lambda$) in the quantum state from (a) $\Lambda \sim 10$ to (f) $\Lambda \sim 10^6$. The qualitative similarity among the ensembles signifies the universality of $\Lambda$.}
    \label{r1scaling}
\end{figure}

\begin{figure}[h]
    \begin{subfigure}[b]{0.3\textwidth}
    \centering
       {\includegraphics[width=\linewidth]{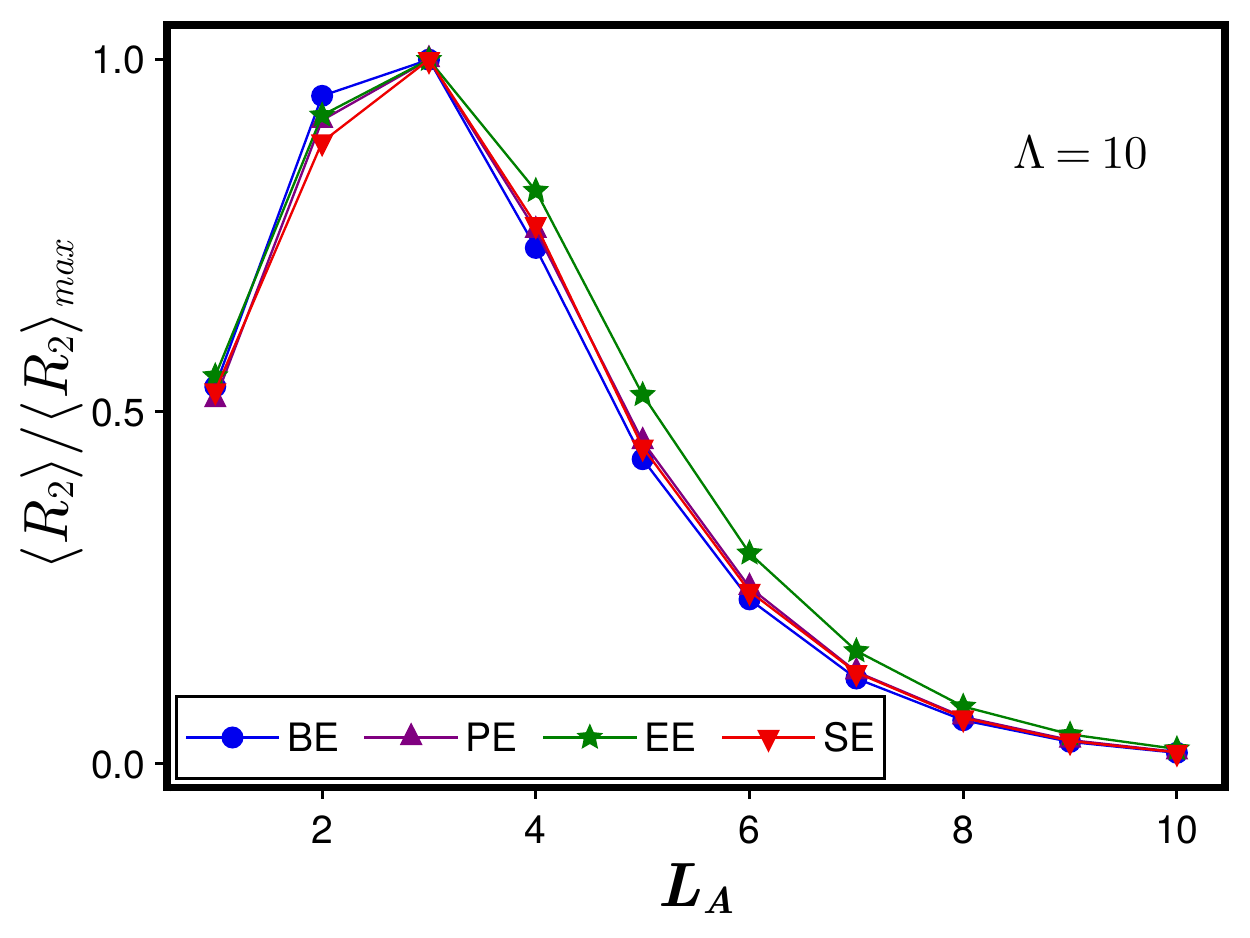}}
        \label{y1r2}
    \end{subfigure}
    \hfill
    \begin{subfigure}[b]{0.3\textwidth}
        {\includegraphics[width=\linewidth]{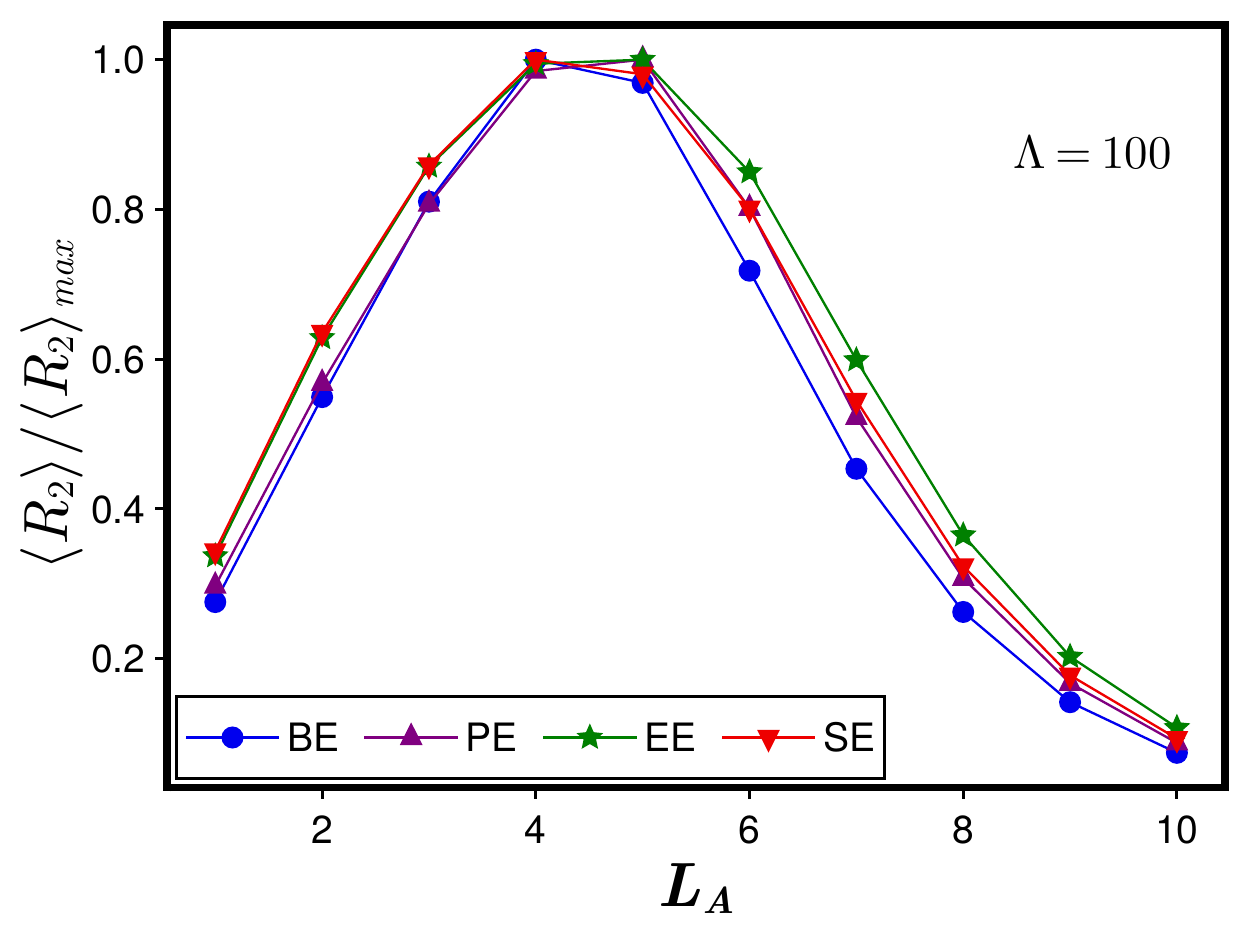}}
        \label{y2r2}
    \end{subfigure}
    \hfill
    \begin{subfigure}[b]{0.3\textwidth}
        {\includegraphics[width=\linewidth]{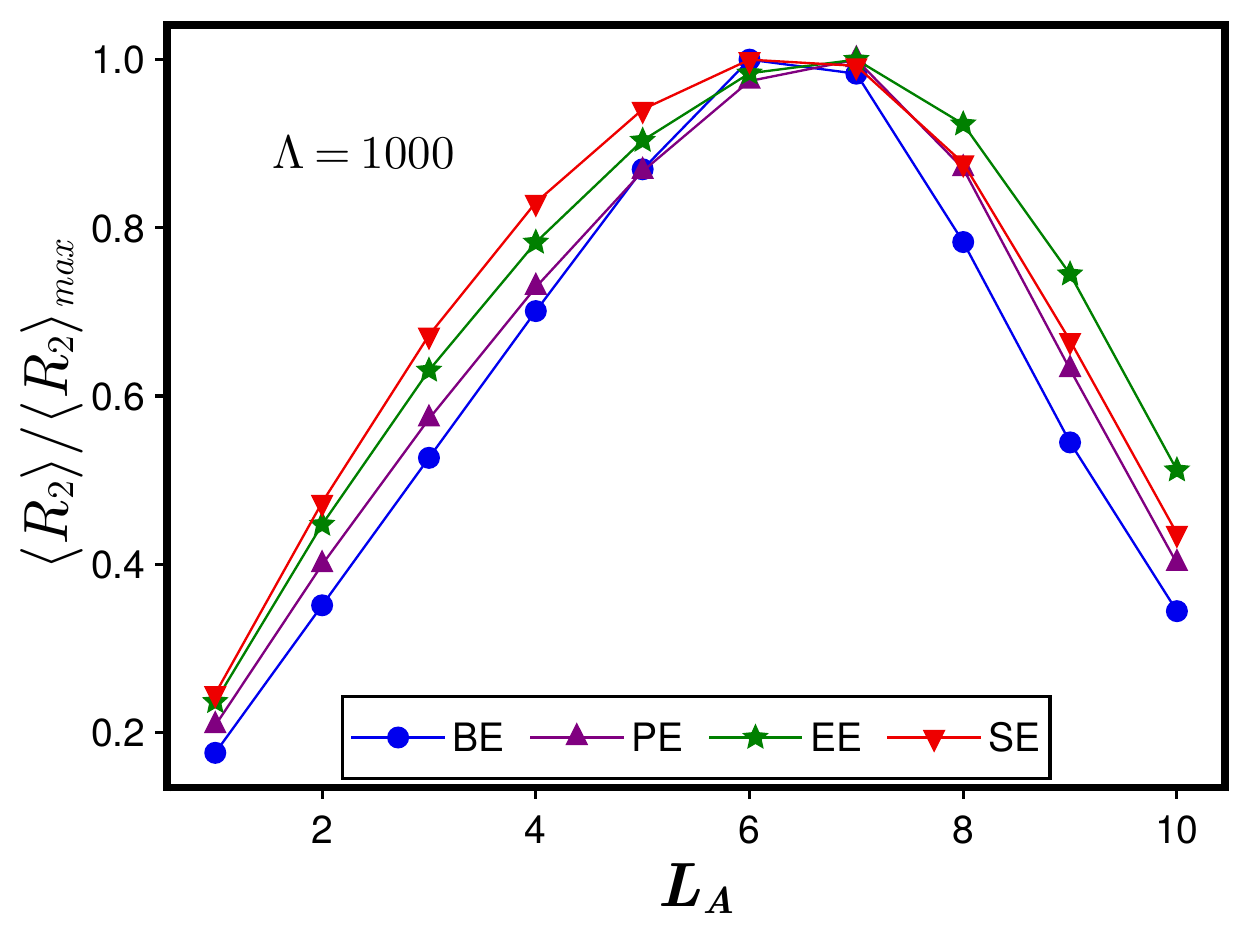}}
        \label{y3r2}
    \end{subfigure}
    \hfill
    \begin{subfigure}[b]{0.3\textwidth}
        {\includegraphics[width=\linewidth]{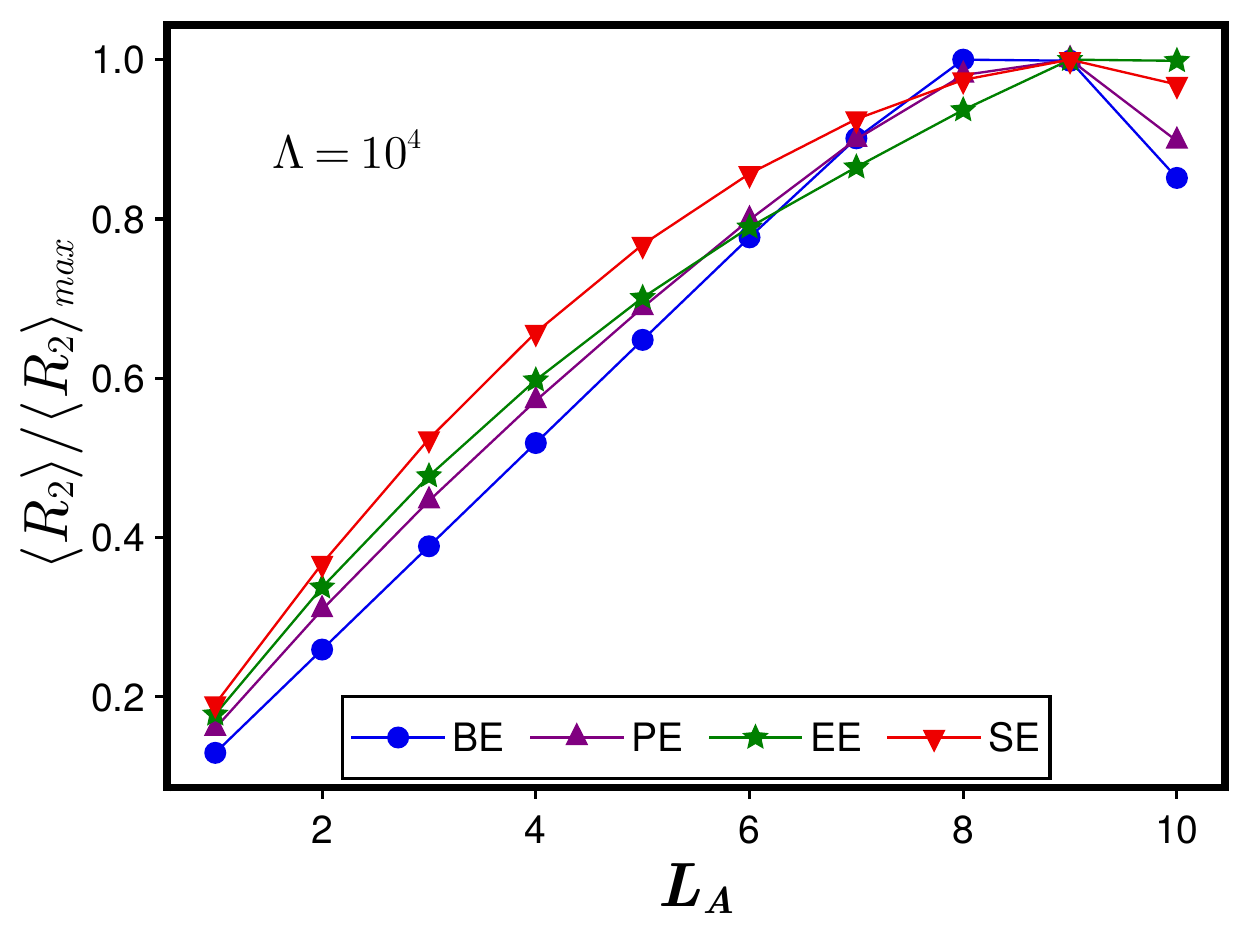}}
        \label{y4r2}
    \end{subfigure}
    \hfill
    \begin{subfigure}[b]{0.3\textwidth}
        {\includegraphics[width=\linewidth]{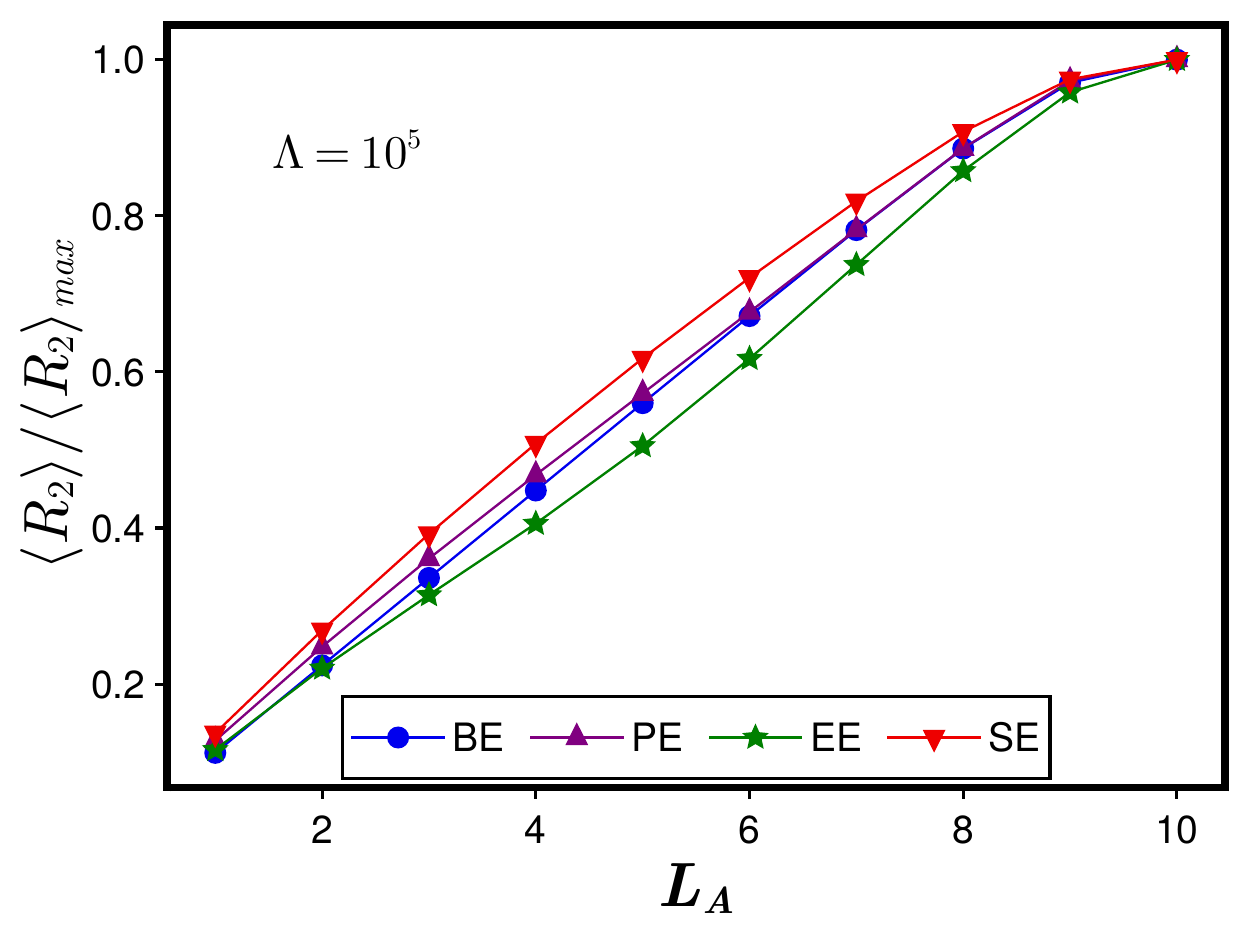}}
        \label{y5r2}
    \end{subfigure}
    \hfill
    \begin{subfigure}[b]{0.3\textwidth}
        {\includegraphics[width=\linewidth]{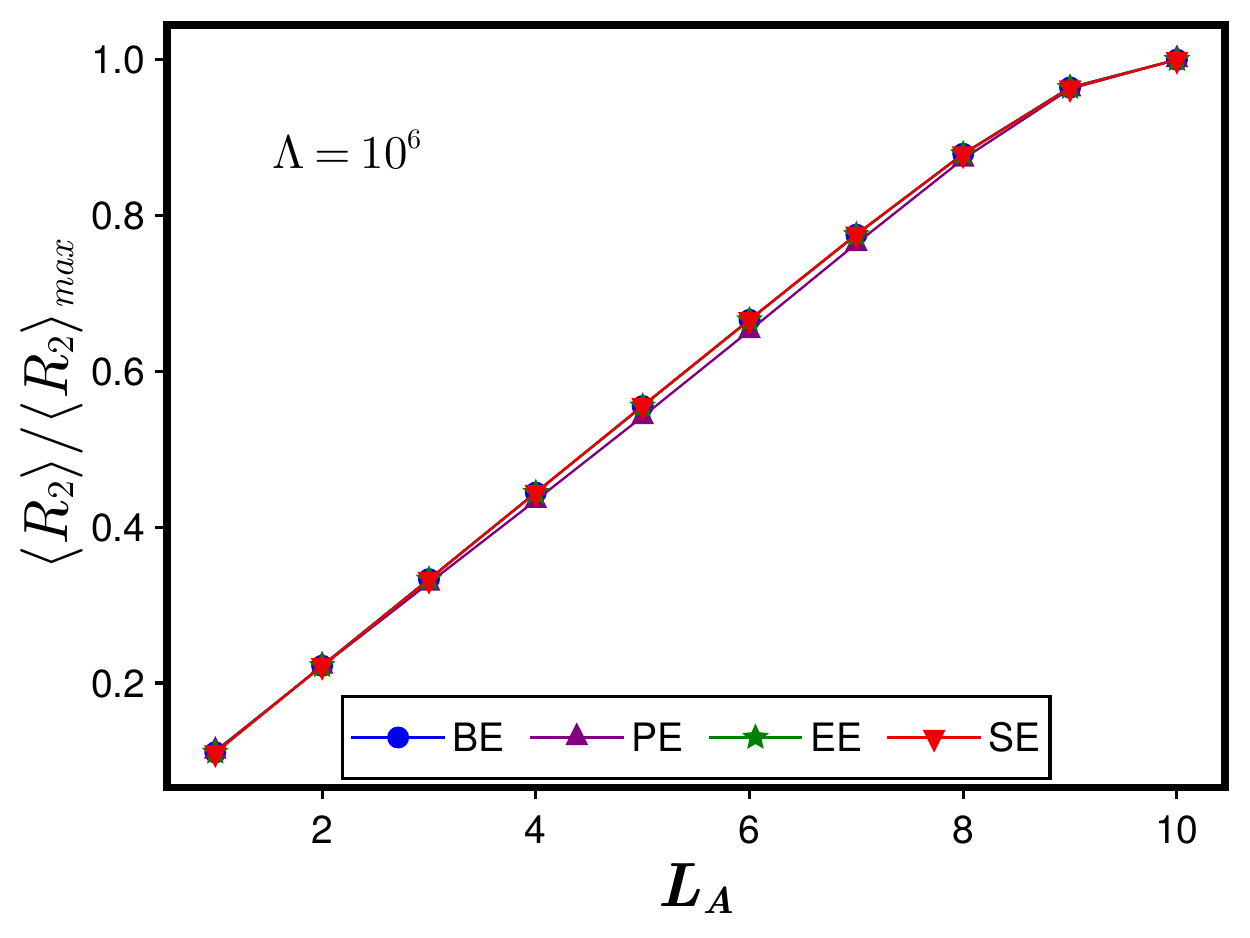}}
        \label{y6r2}
    \end{subfigure}
    
    \caption{The scaling of normalized $R_2$ with sub-system size $L_A$ is shown for the same cases as for $R_1$ in Figs. \ref{r1scaling}.}
    \label{r2scaling}
\end{figure}

Our analysis of these combinations reveals the following:   different ensembles indeed undergo almost similar evolutions qualitatively  if  the entropies are rescaled by their maximum value, referred here as  $\langle R_{1}(\Lambda) \rangle_{max}$ for a fixed $\Lambda$. This is indeed indicated by the illustrations in Figs. \ref{r1scaling} and \ref{r2scaling}, depicting $L_A$-dependence of  $\frac{\langle R_{1} \rangle}{\langle R_{1}\rangle_{max}}$  and $\frac{\langle R_{2}\rangle}{\langle R_2\rangle_{max}}$ for  six  values of $\Lambda$ ranging from $10 \to 10^6$.  Here the bi-partition, i.e., the ``cut" at which $\langle R_{n}\rangle$ is maximum, for a fixed $\Lambda$, can be derived from Eqn. (\ref{avgRnye1}) by invoking the condition $\frac{\partial \langle R_{n}\rangle}{\partial L_A}=0$; the corresponding value of $L_A$ is later referred as $L_m$. 
(We re-emphasize here the difference between  ${\langle R_n \rangle_{max}}$ and $R_{n, \infty}$:  while the former is the maximum entanglement entropy  for a fixed $\Lambda$, the latter is the limit reached at $\Lambda \to \infty$,  a maximum limit for the ensemble).   But as $\Lambda$ can depend on $L_A$ explicitly as well as through ensemble parameters (non-trivially as indicated by our critical point analysis discussed above),   in order to keep it fixed, its explicit variation with respect to $L_A$  must be balanced by other system parameters, e.g.,  $\alpha$  in BE, PE and EE.  we note that, in case of SE, balancing can occur due to different combinations of ensemble parameters.

As Figs. \ref{r1scaling}(a)-\ref{r1scaling}(d) and \ref{r2scaling}(a)-\ref{r2scaling}(d) indicate,  the size-dependence of  the entanglement entropy for small $\Lambda$ values is non-trivial,  first increasing and then decreasing. Based on our theoretical prediction Eqn. (\ref{avgRnye1}), this arises from a competition between the terms $g_n(\Lambda)$ and $L^{-{ \Lambda \over D_n}}$, former increasing and latter decreasing with $L_A$. 
Intuitively, a physical origin of small-$\Lambda$ behaviour  can  be explained as follows. A small $\Lambda$ corresponds to a state with lower participation ratio: the number of bases contributing significantly are restricted. As we change the subsystem size, $L_A$, the Hilbert space dimension increases exponentially, but the basis participation remains fixed and small. After a certain point, when $L_A = L_m$, the ``effective'' Hilbert space dimension becomes incommensurate to this exponential increase, leading to a fall in entropy. In fact, the Figs. \ref{r1scaling}(a)-\ref{r1scaling}(c) and \ref{r2scaling}(a)-\ref{r2scaling}(c) resembles the evolution of entanglement entropy (referred as Page-curve) in black hole physics \cite{kehrein2023page} ; a fermionic model proposed in this context, argues that the Page-curve behaviour amounts to the effective Hilbert space dimension of the sub-system becoming small.

On the contrary, for a large but fixed $\Lambda$, the term 
$L^{-{ \Lambda \over D_n}}$ remains small. As a consequence, the pre-factor $g_n(\Lambda)$ dominates the growth, leading to $\langle R_n\rangle(\Lambda) \approx \langle R_n\rangle_{max}$. We also note that, for $\Lambda$ fixed at large values, peak shifts to larger $L_A$ range. Typically, the entropy of an ergodic state is believed to be the maximum when the partition is balanced, \textit{i.e.,} at $L_A = \frac{L}{2}$ \cite{page1993average} and is expected to follow a volume-law scaling, \textit{i.e.}, Eqn. (\ref{stan}) with $q_1 \not=0, q_2=0$.  This is indeed confirmed by the displays in Figs. \ref{r1scaling}(e), \ref{r1scaling}(f), \ref{r2scaling}(e), and \ref{r2scaling}(f) in the large $\Lambda$-regime.

It is worth re-emphasizing the roles of $\Lambda_{ent}$ and $\Lambda$ in characterizing the  entanglement universality classes.  As Eqn. (\ref{avgRnye1}) indicates,  $\Lambda_{ent} = {\Lambda \over D_n}$ is the single functional of  the system parameters that  governs the variation of entanglement entropy for
any combination of system parameters including subsystem sizes.  Different ensembles with same $\Lambda_{ent}$, even though the latter may arise from  different combinations of system parameters, are predicted to have same entanglement entropy  if they are subjected to same constants of evolution.  More clearly, our theory predicts same entanglement entropy for two ensembles with same $\Lambda_{ent}$ values even if they correspond to different subsystem sizes.   This is however not the case if only their $\Lambda$ values are same; now the analogy is only qualitative and occurs for the ``ratio" of entanglement entropy and only if the ensembles have same subsystem sizes.  Clearly $\Lambda_{ent}$ characterizes more robust  universality class of the entanglement entropy and $\Lambda$ only the sub universality classes.

\section{Conclusion} \label{conclusion}

In the end, we summarize our main results and insights and open questions.
Growing potential relevance of multiparametric Gaussian states as an easy resource in quantum information motivated us in \cite{Shekhar_2023} to analyse their entanglement statistics. The study predicted that a broad range of non-ergodic bipartite pure states, within a same universality class of global constraints, can further be classified into sub-universality classes characterized by the complexity parameter $\Lambda$.  But a  detailed numerical analysis,  described in present study, reveals that a robust form of universality,  insensitive even to subsystem sizes, can be achieved only in terms of $\Lambda_{ent}$,   obtained by a  rescaling  of $\Lambda$ by another system-dependent parameter $D_n$.    As $\Lambda_{ent}$ can take continuous values between zero (initial state) and infinity (maximum entanglement),  this indicates,  for finite system sizes, the existence of an infinite number of universality classes lying between separability and maximal entanglement limits.

Besides  identification of the correct parameter $\Lambda_{ent}$ leading to  universality,  another  important contribution of the present work is to analyse the separate roles of two complexity scales, i.e., $\Lambda$ and $D_n$.  As discussed in main text,  a knowledge of $\Lambda$ is sufficient to predict the evolution of the entropy of a non-ergodic pure state represented by a multiparametric Gaussian ensemble (details of all ensemble parameters not required),  but its rescaling  by $D_n$ is necessary for a  comparison of the different states; $D_n$ seems to have it origin in a "rescaling" of the  Schmidt eigenvalues which enter in entanglement entropy. 
For deeper insights in the exact role of $\Lambda$, we have also analysed the variation of entropy with subsystem size while keeping $\Lambda$ as well as  full system size  fixed;  this in turn requires readjustment of other system parameters. For a large but fixed $\Lambda$,  the state approaches ergodic, equivalently,  maximum entanglement limit as the subsystem size increases and obeys the standard  volume-law.  For smaller $\Lambda$,  the entropy first increases for small system sizes to a maximum and then decreases; this behaviour seems to be arising from a competition between two different tendencies of  the quantum correlations, namely, the one pushing towards their homogeneity (indicated by increasing entanglement between two subsystems) and the noise, appearing through $D_n$, characterizing hindrance to homogeneity. We also find that the states with same $\Lambda$ undergo qualitatively analogous evolution of the rescaled entanglement entropy as $L_A$ varies; this reveals a hidden connection and a possible classification of the states based on complexity parameter $\Lambda$.

An important aspect of our analysis is the consideration of the state matrix ensembles represented by an ensemble density in a multiparametric Gaussian form:  the latter extends applicability of our results to a broad range of non-ergodic states with independent Gaussian components. The appearance of such states in wide-ranging areas, from black holes to nanophysics, renders a rich potential applicability to our results. 

The present study has focussed on non-ergodic pure Gaussian states with complex components in the bipartite basis. A straightforward extension of our analysis is  possible for the states with real coefficients; such bipartite states can appear in case of many-body  systems with time-reversal symmetry and integer angular momentum.  A derivation of  common mathematical formulation similar to the one discussed here as well as in \cite{Shekhar_2023}  for mixed states or multipartite states is also very desirable.
Further while our present analysis as well as in \cite{Shekhar_2023} is confined to multiparametric Gaussian ensembles of state matrices, we intuitively believe the results will be applicable for non-Gaussian state matrices too. This needs to be verified.

\section{Acknowledgement}

D.S. is supported by the MHRD under the PMRF scheme (ID 2402341) and P.S.  is grateful to SERB, DST, India for the financial support provided for the  research under Matrics grant scheme. 

\newpage

\bibliographystyle{ieeetr}
\bibliography{references}

\pagebreak

\appendix

\section{Derivation of Eqn.(\ref{avgRnye})} \label{appAvgRn}

The ensemble average for any arbitrary function $F(\lambda_1, \ldots, \lambda_N)$ can be defined as 
\begin{eqnarray}
\langle F(\Lambda, S_1)  \rangle = C_{hs} \; \int F_n(r) \;\delta \left(S_1-\sum_k \lambda_k \right) \; P_{\lambda} \; {\rm D} {\lambda}
\label{r1df1a}
\end{eqnarray}
where $J$ gives the normalization condition for the {\it jpdf} of the Schmidt eigenvalues under trace constraint $S_1 =\sum \lambda_k$:
\begin{eqnarray}
J = C_{hs} \; \int \delta \left(S_1-\sum_k \lambda_k \right) \; P_{\lambda} \; {\rm D} {\lambda}
\label{j}
\end{eqnarray}
Choosing $J=1$ subjects  $P_{\lambda}$ to following normalization condition:
$ \int \delta \left(S_1-\sum_k \lambda_k \right) \; P_{\lambda} \; {\rm D} {\lambda} =C_{hs}^{-1}$.

As discussed in detail in \cite{Shekhar_2023}, for any arbitrary function $F(\lambda_1, \ldots, \lambda_N)$ we have (choosing $J=1$)

\begin{eqnarray}
& &{\partial \langle F \rangle \over\partial Y} =  \alpha  \; \langle F \rangle  +  \frac{\partial   }{\partial S_1} \left(2  \; \gamma \; S_1 - {1\over 2} \beta N(N+ 2\nu-1) \right) \; \langle F \rangle 
+ \beta \; \left\langle \sum_{m, n=1}^N \frac{\partial F }{\partial \lambda_n}  \; \frac{ \lambda_n}{\lambda_n- \lambda_m} \right\rangle \nonumber \\
& &   -  \left\langle  \sum_{n=1}^N \frac{\partial F }{\partial \lambda_n}  \; \left(- {\beta \nu} + { 2\gamma} \lambda_n \right)  \right\rangle
+  \left\langle \sum_{n=1}^N \frac{\partial^2  F}{\partial \lambda_n^2} \; \lambda_n \right\rangle  -  2  \frac{\partial }{\partial S_1} \left\langle \sum_{n=1}^N  \frac{\partial  F}{\partial \lambda_n}  \; \lambda_n  \right\rangle +  \frac{\partial^2 }{\partial S_1^2}  (S_1 \; \langle F \rangle)  \nonumber \\
 \label{fg1b}
\end{eqnarray}
where $\alpha = {\partial \log C_{hs}\over \partial Y}$ with $C_{hs}$ as a normalization constant for the JPDF $P_c$\cite{Shekhar_2023}: $\int P_c(\Lambda) \; {\rm D}\Lambda =C_{hs} \; J$. 
As $\log C_{hs}$ varies  slowly with $Y$, we have ${\alpha\over N N_{\nu}} \approx 0$ and the contribution from corresponding term can be ignored for simplification. 


\vspace{0.1in}

\noindent{\bf Solution for $R_1$:}
Using $F =R_1$ where $R_1=- \sum_n \lambda_n \log \lambda_n$ in eq.(\ref{fg1b}), we have (details discussed in \cite{Shekhar_2023})


\begin{eqnarray}
    &&{1\over N N_{\nu}} \frac{\partial \langle R_1(S_1) \rangle}{\partial Y} =
    {{\beta\over 2 N}  } \; \langle R_0 \rangle  - {1\over 2} \left[\beta +\frac{\beta (N-1)}{N_{\nu}}- \frac{4  \; S_1}{N N_{\nu}} +\frac{2 (N-2)}{N N_{\nu}}\right] + \nonumber \\
    &- &{1\over 2} \left( \beta N - \frac{4  S_1}{N N_{\nu}} \right) \; \frac{\partial  \langle R_1\rangle }{\partial S_1}  + \frac{S_1}{N N_{\nu}} \; \frac{\partial^2 \langle R_1 \rangle }{\partial S_1^2}  
 \label{rvn2b}
\end{eqnarray}
%
with $N_{\nu}=N + 2 \nu-1$, and $ R_0  = - \sum_n \log \lambda_n$. 
 As our interest is in $S_1=1$ condition only, 
we consider the variation of $S_1$ only in the neighbourhood of $S_1 \sim 1$. 


For trace condition $S_1=\sum_k \lambda_k$ on $N$ eigenvalues $\lambda_k$, we have $S_1/N \le \lambda_k \le S_1$, thereby implying  $- S_1 \log S_1 \le R_1 \le  \log N - S_1 \log S_1$ and  $N \log (\frac{N}{S_1}) \le R_0 \le \infty$. Thus, for  $S_1 \sim 1$, the contribution from the terms with partial derivatives i.e., $\frac{S_1}{N^2} \; \frac{\partial^2 \langle R_1 \rangle }{\partial S_1^2}$ and $\frac{2}{N^2} \frac{\partial J }{\partial S_1}$ can be neglected as compared to other terms in eq.(\ref{rvn2b}). Further as $\alpha$ corresponds to a logarithmic response of the normalization factor $C_{hs}$  to a change in $Y$, this is also expected to be negligible in units of $N N_{\nu}$. The resulting differential equation using the rescaled parameter $\Lambda = N\, N_{\nu} \, (Y-Y_0)$ becomes
\begin{equation}
    {2\over \beta}  \frac{\partial \langle R_1 \rangle}{\partial \Lambda} = g_1(\Lambda, S_1) - \frac{\partial \langle R_1 \rangle}{\partial S_1},
    \label{vnd1b}
  \end{equation}
with  $g_1(\Lambda, S_1) =\left( {\langle R_0 \rangle\over N}- q_0 \right)$ with $q_0 =\frac{N_{\nu}+(N-1)}{N_{\nu}}$.

The above equation can be solved by using the standard method  of characteristics for partial differential equations. The characteristic equations are 
\begin{eqnarray}
  \frac{d \Lambda}{ 2/\beta} = \frac{d S_1}{1}  = \frac{d \langle R_1 \rangle}{g_1} \label{rt1b} 
 \end{eqnarray}
The general solution of eq.(\ref{rt1b}) is given by the function 
$F_1(\Phi, \Psi)=0$ or equivalently $\Phi= f(\Psi)$ where $F_1, f$ are arbitrary functions and  the functions $\Phi(\langle R_1\rangle, \Lambda, S_1)$  and $\Psi(\langle R_1\rangle, \Lambda, S_1)$ are obtained by solving any of the two pairs of eq.(\ref{rt1b}).
We have, from first pair, 
\begin{eqnarray}
{\rm d}\Phi \equiv  d \left(S_1 - \frac{\beta \Lambda}{2}\right) = 0  \label{rt2b} 
 \end{eqnarray}
The above gives 
\begin{eqnarray}
\Phi = \left(S_1 - \frac{\beta \Lambda}{2} \right)
\label{rt3b}
\end{eqnarray}
From second pair, we have,
\begin{eqnarray}
 d \langle R_1 \rangle - g_1 {\rm d}S_1 =0  \label{rt4b} 
\end{eqnarray}
The above in turn gives 
\begin{eqnarray}
 \Psi =\langle R_1 \rangle - \int g_1 {\rm d}S_1 \label{rt5b} 
\end{eqnarray}
The general solution of eq.(\ref{rt1b}) can now be given as $F(\Psi, \Phi)=0$  with $F$  as an arbitrary function.  Equivalently we can write $\Psi = f(\Phi)$ with $f(S_1, \Lambda) \equiv f\left(S_1 - \frac{\beta \Lambda}{2} \right)$ as an arbitrary function and $I_1 \equiv \int g_1 {\rm d}S_1 $. This in turn gives 
\begin{eqnarray}
\langle R_1 \rangle (\Lambda, S_1) = I_1 + f\left(S_1 - \frac{\beta \Lambda}{2} \right) \label{rt6b} 
\end{eqnarray}

\begin{figure}
\includegraphics[width = 0.6\textwidth]{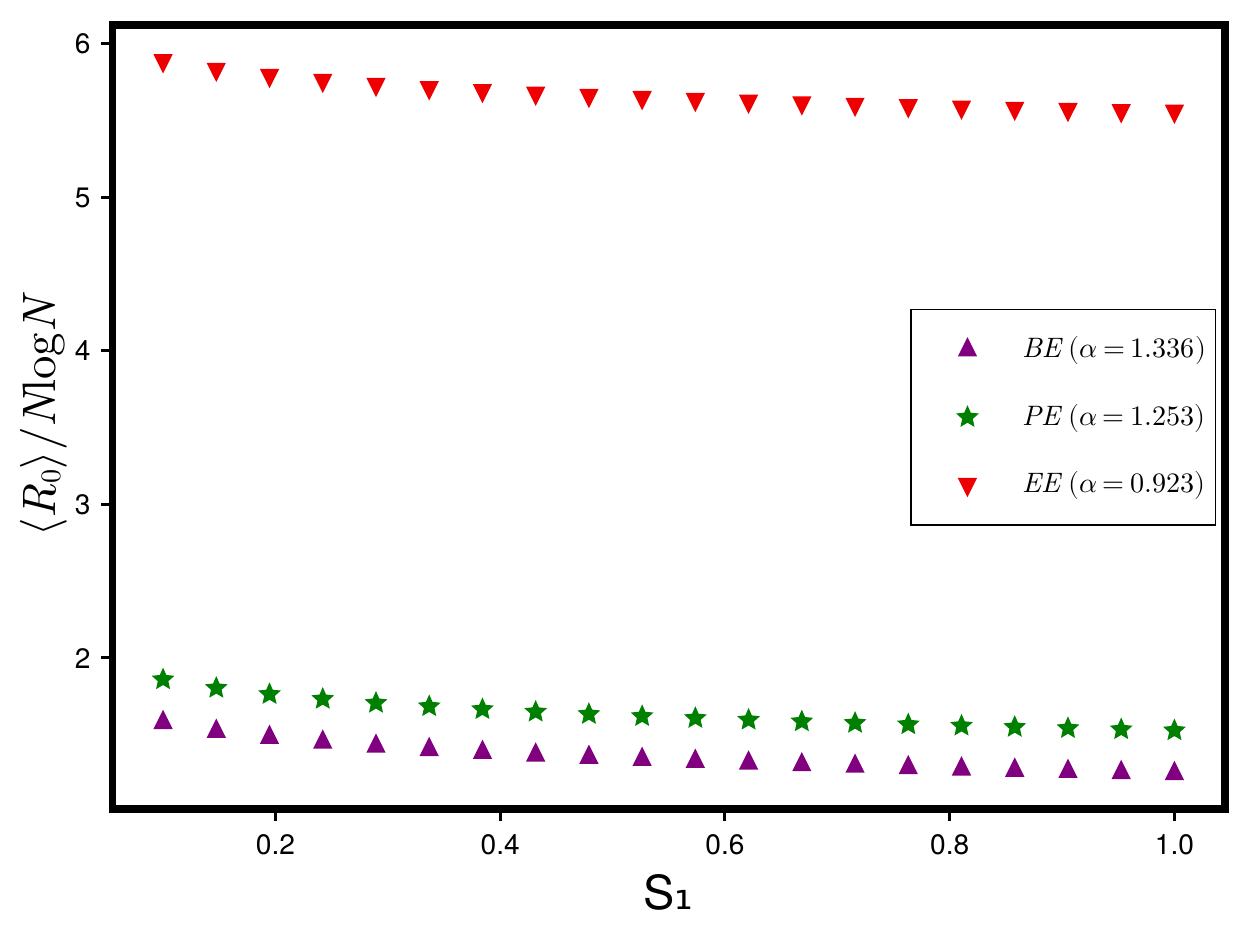}
\caption{{\bf $\langle R_0 \rangle$-dependence on $S_1$:} The figure displays the evolution of $\langle R_0 \rangle$ with $S_1$  for BE, PE and EE. As can be seen from the figure, $R_0$ is almost constant near $S_1 \sim 1$. }
\label{r0}
\end{figure}

The integral $I_1$ over $S_1$ can further be approximated by noting that  $\langle R_0 \rangle$ changes very slowly with $S_1$ i.e., $R_0 \sim -\log S_1 $. (This can be seen by writing $R_0 = - \sum_{n=1}^N \log \lambda_n$ in terms of constraint $S_1=\sum_{n=1}^N \lambda_n$. This leads to $R_0 =- \sum_{n=1}^{N-1} \log \lambda_n -\log\left(S_1-\sum_{n=1}^{N-1} \lambda_n \right) = -\log S_1  - \sum_{n=1}^{N-1} \left({\lambda_n\over S_1} +\log \lambda_n\right) $. This is also supported by figure(\ref{r0}) displaying the  $\langle R_0 \rangle$-dependence on $S_1$ for BE, PE and EE).  The above permits the approximation $\int g_1 {\rm d}S_1 \approx  g_1 \, S_1$. and leads to, for $S_1=1$,
\begin{eqnarray}
\langle R_1\rangle(\Lambda, 1) = g_1(\Lambda, 1) + f\left(1-\frac{\beta \Lambda}{2} \right)
\label{rt7b} 
\end{eqnarray}
As $\langle R_1 (0, 1) \rangle=0$  for a separable initial state chosen at $Y = Y_0$ i.e., $\Lambda=0$, we have $f\left(0,1 \right) = -g_1(0, 1)$. Further to satisfy the boundary condition  at $\Lambda \to \infty$,  we must have $f(\infty, 1) =\langle R_1 (\infty, 1 ) \rangle - g_1(\infty, 1)$. But as $g_1(\infty,1) \approx \frac{\langle R_0(\infty, 1) \rangle}{N} \approx \langle R_1 (\infty, 1 ) \rangle $, this implies $f(\infty, 1) \approx 0$. One possible form of $f$ that satisfies the above conditions on $f$ can be given as $f\left(\Lambda,1 \right) = - g_1(\Lambda, 1) \; L^{-\tau \,\left(1-(\beta \,\Lambda/2)\right) } $; the latter leads to 
$\langle R_1\rangle(\Lambda, 1) = g_1(\Lambda, 1) \left(1- \; L^{-\tau \,\left(1-(\beta \,\Lambda/2)\right)} \right)$.

\vspace{0.1in}

\noindent{\bf Solution for $R_2$:}
Similarly using $F =R_2$ where $R_2 = -\log \sum_{k=1}^N \lambda_k^2$ in eq.(\ref{fg1b}), we have

\begin{eqnarray}
    & &\frac{\partial \langle R_2(S_1) \rangle}{\partial Y} =
    2 \gamma \; \langle R_2 \rangle - (2\beta (N+\nu-1)+2) \; S_1 \left\langle {1\over S_2} \right\rangle + 4 \left\langle {S_3 \over S_2^2} \right\rangle  + 4   +  \nonumber \\
    &+&   S_1 \; \frac{\partial^2 \langle R_2 \rangle }{\partial S_1^2} + (2 S_1 - \frac{1}{2}\beta N(N + 2\nu - 1) + 2) \frac{\partial  \langle R_2\rangle }{\partial S_1}
    \label{rvn3b}
\end{eqnarray}

Using similar approximations as in $R_1$ case, the resulting differential equation using the rescaled parameter $\Lambda = N\, N_{\nu} \, (Y-Y_0)$ now becomes
  \begin{equation}
    {2\over \beta}  \frac{\partial \langle R_2 \rangle}{\partial \Lambda} = g_2(\Lambda, S_1) - \frac{\partial \langle R_2 \rangle}{\partial S_1},
    \label{vnd1c}
  \end{equation}
with  $g_2(\Lambda, S_1) = -\eta \, S_1 \, \left\langle {1\over S_2} \right\rangle $ (with  ${1\over S_1^2} \le {1\over S_2} \le {N\over S_1^2}$), where, $\eta = \frac{2 \beta (N+\nu-1)}{N (N+2\nu-1)}$. Proceeding as in previous case, the solution of the above equation  can be given as 
\begin{eqnarray}
\langle R_2\rangle(\Lambda, S_1) = I_2(\Lambda, S_1) + f\left(S_1-\frac{\beta \Lambda}{2} \right) 
\label{rt7c} 
\end{eqnarray}
Proceeding as in the case of $I_1$, here again  $I_2 = \int g_2 \; {\rm d}S_1$ can be approximated by noting that $\langle {1\over S_2} \rangle$ changes very slowly near $S_1=1$; this in turn permits  $I_2 \approx  - {\eta \over 2} \langle {1\over S_2} \rangle \, S_1^2$ and leads to, for $S_1=1$,
\begin{eqnarray}
\langle R_2\rangle(\Lambda, 1) = {1\over 2} \, g_2(\Lambda, 1) + f\left(1 -\frac{\beta \Lambda}{2} \right) 
\label{rt7c} 
\end{eqnarray}
The form of function $f$ in the above can be fixed by invoking boundary conditions: with $\langle R_2 (0, 1) \rangle=0$  for a separable initial state chosen at $Y = Y_0$ i.e., $\Lambda=0$, $f$ must satisfy the condition that $f\left(0,1 \right) = - {1\over 2}  g_2(0,1)$. 
Similarly with $\langle R_2 (\infty, 1 ) \rangle  = - {1\over 2} g_2(\infty, 1) $, we  have $f(\infty, 1) = \langle R_2 (\infty, 1 ) \rangle - {1\over 2}  g_2(\infty, 1) = - g_2(\infty, 1)$.  The above conditions on $f$ can again be satisfied for $f\left(\Lambda,1 \right) = - g_2(\Lambda,1) + {1\over 2} g_2(\Lambda,1)\; L^{-\tau \,\left(1- (\beta \,\Lambda/2)\right)} $ with $\tau$ as an arbitrary constant.  For $S_1=1$, the above leads to 
$\langle R_2\rangle(\Lambda, 1) = - {1\over 2} \, g_2(\Lambda, 1) \left(1- \; L^{-\tau \,\left(1-(\beta \,\Lambda/2) \right)}\right)$.

\section{Solution for $R_0$} \label{solR0}

Using $F =R_0$ where $R_0= -\sum_{k=1}^N \log\lambda_k$ in eq.(\ref{fg1b}), we have 

\begin{eqnarray}
& &\frac{\partial \langle R_0(S_1) \rangle}{\partial Y} =
2 \gamma  \; \langle R_0 \rangle 
+(2 S_1 - {1\over 2} \beta N(N+2 \nu-1))
\frac{\partial  \langle R_0\rangle }{\partial S_1}
+ (1-\beta \nu) \left\langle S_{-1} \right\rangle \nonumber \\
&+ & 2 N 
+\frac{\partial^2 S_1 \, \langle R_0 \rangle }{\partial S_1^2}
\label{rvn4d}
\end{eqnarray}
%
where $\nu = \frac{(N_B -N_A + 1)}{2}$. Further
$S_m$ is defined as (with $m$ as a positive/ negative integer)
\begin{eqnarray}
S_{m} \equiv \sum_{k=1}^N \lambda_k^m. 
\label{smd}
\end{eqnarray}  

Here $\left\langle S_{-1} \right\rangle$ ranges from $N^2/S_1 \le S_{-1} \le \infty$. We also note that $S_{-1}  \gg R_0$ and therefore the former can be treated as almost constant while  $\langle R_0 \rangle$ varies with $\Lambda$. 

For simplification here we consider the balanced case $N_A=N_B$ and the case $\beta=2$. This gets rid of the term  $S_{-1}$.
Proceeding again as in $R_1$-case, the above equation can be approximated as

\begin{eqnarray}
\frac{\partial \langle R_0(S_1) \rangle}{\partial \Lambda} &=& \frac{2 \gamma  \; \langle R_0 \rangle}{N^2} + \frac{2 }{N} - \frac{\partial  \langle R_0\rangle }{\partial S_1}. 
\label{rvn5d}
\end{eqnarray}

Further solving the equation as in $R_1$ case above, we have, for arbitrary $S_1$,
\begin{eqnarray}
\log \left({2 \gamma \, \langle R_0\rangle \over N^2}  + {2  \over N}  - \phi_0\right) = { 2 \gamma \over N^2} S_1 + f_0\left( \Lambda- S_1 \right) 
\end{eqnarray}
with $f_0(\Lambda, S_1) \equiv f_0\left(\Lambda- S_1 \right)$ as an arbitrary function and $\phi_0$ a constant of integration.  Inverting the above gives
\begin{eqnarray}
 \langle R_0\rangle(\Lambda,S_1) =   {N^2 \over 2 \gamma}\; {\rm e}^{f_0\left(\Lambda- S_1\right) + {2 \gamma S_1 \over N^2}}  - {2  N  \over 2 \gamma} +\phi_0
\label{rt7e} 
\end{eqnarray}

The unknown constant $\phi_0$ can now be determined by invoking the boundary condition at $\Lambda \to \infty$. Eq.(\ref{rt7e}) gives
$\phi_0 =\langle R_0 \rangle(\infty, S_1) -
{N^2 \over 2 \gamma}\; {\rm e}^{f_0\left(\infty- S_1\right) + {2 \gamma S_1 \over N^2}}  + {N }$.  This in turn gives 
\begin{eqnarray}
 \langle R_0\rangle(\Lambda,S_1) = \langle R_0 \rangle(\infty, S_1)  +   {N^2 \over 2 \gamma}\; {\rm e}^{{2 \gamma S_1 \over N^2}}  \left({\rm e}^{f_0\left(\Lambda- S_1\right)} - {\rm e}^{f_0\left(\infty- S_1\right)}\right)
\label{rt8e} 
\end{eqnarray}

The form of $f_0$ can now be determined by  substituting eq.(\ref{rt7e}) in eq.(\ref{rvn5d}). This leads to 
\begin{eqnarray}
\frac{\partial f_0}{\partial \Lambda} &=& - \frac{\partial f_0}{\partial S_1}. 
\label{rvmd}
\end{eqnarray}
A particular solution of the above, satisfying required boundary conditions at $\Lambda=0, \infty$, can be given as $f_0 = -\log (\Lambda-S_1) \tau$ with $\tau$ as an unknown constant. This on substitution in eq.(\ref{rt8e}) along with   $S_1=1$ gives 
\begin{eqnarray}
 \langle R_0\rangle(\Lambda,1) = \langle R_0 \rangle(\infty, 1)  +   {N^2 \over 2 \gamma (\Lambda-1) \tau}\; {\rm exp}\left[{2 \gamma \over N^2}\right]  \nonumber  \\
 \approx \langle R_0 \rangle(\infty, 1)  +   {N^2 \over 2 \gamma (\Lambda-1) \tau}  
\label{rt9e} 
\end{eqnarray}
where the $2nd$ equality is obtained by the approximation ${\rm e}^{ 2 \gamma/N^2} \approx  1 +{2 \gamma  \over N^2}$

\section{Solution for $\langle {1\over S_2 } \rangle$} 
\label{solQ}

Using $F = {1\over S_2 }$ where $S_2= \sum_{k=1}^N \lambda_n^2$ in eq.(\ref{fg1b}) and proceeding as in previous cases, the dependence of ${1\over S_2 }$ can be derived. Using notation $Q \equiv {1\over S_2 }$, we have

\begin{eqnarray}
    & &\frac{\partial \langle Q \rangle}{\partial Y} =
    6 \gamma \; \langle Q \rangle
    +(4 + 2 S_1 - {1\over 2} \beta N(N+2 \nu-1))
    \frac{\partial \langle Q \rangle}{\partial S_1}
    - (2 \beta (N-1) + 2 \beta \nu +2) S_1 \left\langle Q^2 \right\rangle \nonumber \\
    &+ & 8  \left\langle S_3 \, Q^3\right\rangle  
    +\frac{\partial^2 S_1 \, \langle Q \rangle}{\partial S_1^2}
    \label{run4d}
\end{eqnarray}

with all symbols same meaning as in previous appendix. 


For simplification here we consider the balanced case $N_A=N_B$. Further noting that typically ${1\over N^k} \le S_{k+1} \le 1$ and $1 \le Q \le N$, the terms with $\left\langle S_3 \, Q^3\right\rangle$ and $6 \gamma \; \langle Q \rangle$ can be neglected with respect to $\left\langle Q^2 \right\rangle$.
The above equation can then be approximated as

\begin{eqnarray}
\frac{\partial \langle Q \rangle}{\partial \Lambda} & \approx &   g_q
 - \frac{\partial \langle Q \rangle}{\partial S_1}. 
\label{run5d}
\end{eqnarray}
with $g_q \equiv {2 S_1\beta \over N}  \left\langle Q^2\right\rangle $.

Further solving the equation as in $R_1$ case above, we have, for arbitrary $S_1$,

\begin{eqnarray}
\left \langle Q \right \rangle(\Lambda, S_1) =  I_q(\Lambda, S_1) + f_q(\Lambda, S_1) 
\label{run6d}
\end{eqnarray}
where $I_q(\Lambda, S_1) \equiv \int g_q \, {\rm d}S_1$ with $f_q(\Lambda, S_1) \equiv f_q\left(\Lambda- S_1 \right)$ as an arbitrary function.

As $\langle Q (0, 1) \rangle=1$  for a separable initial state at  $\Lambda=0$, we have $f_q\left(0,1 \right) =1-I_q(0, 1)$. Further as $ \langle Q (\infty, 1) \rangle $ i.e.,  we must have $f_q(\infty, 1) = \langle Q (\infty, 1) \rangle -I_q(\infty,1)$.  The form of $f_q$ that satisfies the above conditions can be given as $f\left(\Lambda,1 \right) = (\langle Q (\Lambda, 1) \rangle - I_q(\Lambda, 1) )\; {\rm e}^{\tau \,\left(1-(\beta \,\Lambda/2)\right) } $; the latter leads to

\begin{eqnarray}
\langle Q\rangle(\Lambda, 1) = {\rm e}^{\tau \,\left(1-(\beta \,\Lambda/2)\ \right)} + I_q(\Lambda, 1) \left(1- \; {\rm e}^{\tau \,\left(1-(\beta \,\Lambda/2) \right)} \right)
\label{run10} 
\end{eqnarray}
with $\tau$ as an arbitrary function independent of $\Lambda$, subjected to conditions that in large $N$ limit, $\tau \to 0$ but $\Lambda \ge {1\over \tau}$. Here again, with $Q^2$ varying very slowly near $S_1=1$,  $I_q(\Lambda, S_1)$ near $S_1=1$ can be approximated as $I_q(\Lambda,1)= g_q(\Lambda,1)/2 = {\beta \over 2 N}  \left\langle Q^2\right\rangle$.

\section{$\Lambda_{ent}$ dependence of $R_{1,2}$} \label{lentR12}

\begin{eqnarray}
 \langle {\mathcal R}_n \rangle = 
 {\mathcal C}_{hs} \int  {\mathcal R}_n(r) \;\delta \left({\mathcal S}_1 - \sum_k r_k \right) 
\;   P_{r} \; {\rm D} r
\label{r1df2a}
\end{eqnarray}

Differentiating the above equation  with respect to $Y_1$, 
substitution of eq.(\ref{pdl1a}) in the right side of the above equation and proceeding as in \cite{Shekhar_2023}, we have 
\begin{eqnarray}
 {1\over u_n} \; \frac{\partial \langle {\mathcal R}_n \rangle}{\partial \Lambda} 
= \tilde{g}_n(\Lambda, {\mathcal S}_1) -  {1\over v_n} \;\frac{\partial \langle {\mathcal R}_1 \rangle}{\partial {\mathcal S}_1}, 
 \label{r1df3a}
\end{eqnarray}
 
where $\tilde{g}_1(\Lambda, {\mathcal S}_1) =  {\langle {\mathcal R}_0 \rangle\over N} $, $\tilde{g}_2(\Lambda, {\mathcal S}_1) = - \frac{ 4 \, {\mathcal S}_1}{N} \, \langle {1\over S_2 }\rangle$, $\tilde{g}_0(\Lambda, {\mathcal S}_1) = {2 \gamma\over N^2} (\langle {\mathcal R}_0 \rangle - N \, D)$ and $v_1 =1$, $v_2 = {\chi_2\over \chi_1}$, $v_0 \approx {2 \, D \over \beta}$ and $u_n={\chi_n \; \beta \over 2 \, D}$, $u_0=1$, $\chi_1= \frac{N+2\nu D-1}{N+2 \nu-1}$, $\chi_2= \frac{N+\nu D-1}{N+2 \nu-1}$.

{\bf Solutions for $\langle {\mathcal R}_1 \rangle$ and  $\langle {\mathcal R}_2 \rangle$ :}
Proceeding as in previous {\it appendices}, the general solution of eq.(\ref{r1df3a}) for $n=1,2$ can be given as 
\begin{eqnarray}
\langle {\mathcal R}_n \rangle (\Lambda, {\mathcal S}_1) =  f\left(v_n \, {\mathcal S}_1 - u_n \, \Lambda \right) +  v_n \, {\mathcal I}_n   \hspace{0.5in} n=1,2
\label{r1df6a} 
\end{eqnarray}
where ${\mathcal I}_n(\Lambda, {\mathcal S}_1) \equiv \int \tilde{g}_n \, {\rm d}{\mathcal S}_1  $.

Further using the rescaled constraint ${\mathcal S}_1=\sum_n r_n={1\over D}$, we have

\begin{eqnarray}
\langle {\mathcal R}_n \rangle \left(\Lambda, {1\over D} \right)  =  f\left({v_n\over D}- u_n \,\Lambda \right) +  v_n \, {\mathcal I}_n 
\label{r1df7a} 
\end{eqnarray}

Following from the relations in eqs.(\ref{ax1}, \ref{ax2}), and using the boundary conditions on $\langle R_n \rangle$ i,e $\langle R_n \rangle(0,1)= 0$ and $\langle R_n \rangle(\infty,1)= L_A$,  
the boundary conditions for $\langle {\mathcal R}_n \rangle$ can be given as  
$\langle {\mathcal R}_n \rangle \left(0, {1\over D} \right)  = {n \, \log D\over D}$
 and $\langle {\mathcal R}_n \rangle \left(\infty, {1\over D} \right)  = {L_A\over D} + {n \, \log D\over D}$. These on substitution in eq.(\ref{r1df7a}) give the boundary condition on $f$: 
 $\lim_{\Lambda \to 0}f\left({v_n\over D}- u_n \,\Lambda \right) = {n \, \log D\over D} -  v_n \, {\mathcal I}_n\left(0, {1\over D} \right)$ 
 and
 $\lim_{\Lambda \to \infty}f\left({v_n\over D}- u_n \,\Lambda \right) = {n \, \log D\over D} + {L_A\over D} -  v_n \, {\mathcal I}_n \left(\infty, {1\over D} \right)$.
 
Further, following from the relations in eqs.(\ref{ax1}, \ref{ax2}),
we have ${\mathcal I}_1(\Lambda, {\mathcal S}_1) \equiv \int {{\mathcal R_0}\over N} \, {\rm d}{\mathcal S}_1  = 
{1\over D} \int \left({R_0\over N} +\log D\right) \, {\rm d}S_1$. For 
$S_1=1$, this can again be approximated as (as in {\it appendix A})
${\mathcal I}_1(\Lambda, {\mathcal S}_1) \approx {1\over D} \left({R_0\over N} +\log D\right) ={{\mathcal R}_0\over N D} $. Proceeding similarly, it can be shown that ${\mathcal I}_2(\Lambda, {1\over D}) \approx - {\eta \over 2} \langle Q \rangle = I_2$.

The particular solution for $f$ satisfying boundary conditions can now be given as 
 $f\left({v_n\over D}- u_n \,\Lambda \right) = 
   {n \, \log D\over D} - v_n \, {\mathcal I}_n \; L_A^{\frac{ (2 v_n -\chi_n  \beta \Lambda)}{2 D}}$ (neglecting $L_A/D$ in large $N$ limit) . This in turn leads to 

\begin{eqnarray}
\langle {\mathcal R}_n \rangle \left(\Lambda, {1\over D} \right)  =   v_n \, {\mathcal I}_n \, \left(1- L_A^{\frac{ (2 v_n -\chi_n  \beta \Lambda)}{2 D}} \right) + {n \, \log D\over D}
\label{r1df8aa} 
\end{eqnarray}
with ${\mathcal I}_1 \approx {\langle{\mathcal R}_0 \rangle\over N D}$ and ${\mathcal I}_2 \approx - {\eta \over 2} {\langle {\mathcal Q} \rangle\over D^2}$.

{\bf Solution for $\langle {\mathcal R}_0 \rangle$:}
For $n=0$ case of eq.(\ref{r1df3a}), the boundary conditions are different from those for $n=1,2$ and we need to proceed as in case of eq.(\ref{rvn5d}).
The solution of eq.(\ref{r1df3a}) for $n=0$ can be given as

\begin{eqnarray}
 \langle {\mathcal R}_0\rangle\left(\Lambda, {1\over D} \right) = \langle {\mathcal R}_0 \rangle(\infty, 1)  +   {N^2 D \over 2 \gamma \Lambda}\; {\rm e}^{{ 2 \gamma \over N^2}} +  {N \, \log D\over D}
\label{raa2} 
\end{eqnarray}

{\bf Solution for $\langle{\mathcal Q}\rangle$:}
Using ${\mathcal Q}(r)=(\sum_k  r_k^2)^{-1}$ and proceeding as before, we have 

\begin{eqnarray}
\langle {\mathcal Q}\rangle (\Lambda,{\mathcal S}_1) = 
{\mathcal C}_{hs} \int \left({1\over \sum_k r_k^2}\right) \;\delta \left({\mathcal S}_1-\sum_k r_k \right) \; P_r \; {\rm D}r.
\label{qt1}
\end{eqnarray}
Differentiation of the above equation with respect to $Y_1$, 
using eq.(\ref{pdl1a}) and proceeding as in {\it appendix C}, we have 
\begin{eqnarray} 
 \frac{\partial \langle {\mathcal Q} \rangle}{\partial Y_1} 
= \tilde{g}_s(\Lambda, {\mathcal S}_1) -  v_s \;\frac{\partial \langle {\mathcal Q} \rangle}{\partial {\mathcal S}_1}, 
 \label{r0a1}
\end{eqnarray}
where $\tilde{g}_s(\Lambda, {\mathcal S}_1) = -2 \gamma \langle {\mathcal Q} \rangle + {2 \beta (N-1) {\mathcal S_1} \langle {\mathcal Q}^2 \rangle  \over D}$ and $v_s = {\beta N^2 \over 2}-2 \gamma {\mathcal S}_1$.   Proceeding as in {\it appendix C}, the general solution of eq.(\ref{r0a1}) can be given as

\begin{eqnarray}
\langle {\mathcal Q}\rangle(\Lambda, 1) = {\rm e}^{\frac{\tau \, (1-\beta \,\Lambda)}{2}\,} + g_q(\Lambda, 1) \left(1- \; {\rm e}^{\frac{\tau \, (1-\beta \,\Lambda)}{2}}\right)
\label{qun} 
\end{eqnarray}

\section{Complexity parameter formulation} \label{comparamApp}

A perturbation of the state by a change of the parameters $h_{kl;s} \rightarrow h_{kl;s}+\delta h_{kl;s}$ and $b_{kl} \rightarrow b_{kl}+\delta b_{kl}$  over time causes the matrix elements $C_{kl;}$ undergo a dynamics in the matrix space. In \cite{Shekhar_2023}, we considered a combination of multiparametric variations defined as  $ T \; \rho \equiv \sum_{k,l;s}\left[f_{kl;s}{\partial \rho\over\partial h_{kl;s}} - \gamma b_{kl;s} {\partial \rho \over\partial b_{kl;s}}\right]$. Using Gaussian nature of the density, it can be shown that $T \rho = L \rho$ where $L = \sum_{k,l,s} \frac{\partial}{\partial C_{kl;s}}\left[ \frac{\partial  \rho_c}{\partial C_{kl;s}}+\gamma \; C_{kl;s} \; \rho \right]$ and $\rho_c \propto \rho$.

As the equation   $  T \rho =  L \rho$
is difficult to solve for generic parametric values,  we seek a transformation from the set of $M$ parameters $\{h_{kl;s}, b_{kl;s} \}$ to another set $\{y_1,\ldots, y_M \}$ such that only $Y_1$ varies under the evolution governed by the operator $T$ and rest of them i.e., $Y_2, \ldots, Y_M$ remain constant:  $T \rho_c \equiv \frac{\partial \rho_c}{\partial y_1}$.  This in turn requires 

\begin{eqnarray}
T Y_1=1, \qquad T Y_{\alpha}=0  \qquad \forall \; \;  \alpha >1. 
\label{hbya}
\end{eqnarray}

As can be seen from the definition of $T$ given above, the contribution from the term with $ b_{kl;s} \; {\partial \over\partial b_{kl;s}} =0$  for $b_{kl;s}=0$ and similarly the term with $(1-\gamma h_{kl;s}) \; {\partial \over\partial h_{kl;s}}=0$  as $h_{kl;s} \to \gamma^{-1}$; (we note, from eq.(\ref{jpdfMulti}), that ${\partial \rho\over\partial b_{kl;s}}$ and ${\partial \rho\over\partial h_{kl;s}}$  remain well-defined in the neighbourhood of  $b_{kl;s}=0$ and $h_{kl;s} \to \gamma^{-1}$  respectively). Further $T \rho=0$ as all $b_{kl;s}=0$ and $h_{kl;s} \to \gamma^{-1}$; this in turn implies $\frac{\partial \rho}{\partial Y_1} =0$ and thereby equilibrium limit of a stationary Wishart ensemble. 


Using $T$ given above, the set of equations (\ref{hbya}) can be solved by the standard method of characteristics, leading to 

\begin{eqnarray}
    \frac{d h_{kk;s}}{ f_{kk;s}} &=& \ldots = \frac{db_{kl;s}}{b_{kl;s}} = {dY_1} \label{lt1} \\
    \frac{d h_{kk;s}}{ f_{kk;s}} &=& \ldots = \frac{db_{kl;s}}{b_{kl;s}} =\frac {dY_n}{0}  \qquad  n>1
  \label{ltna}
\end{eqnarray}
where equality relations include all $\{kl;s \}$ pairs. The solution $Y_1$ of eq.(\ref{lt1}) is  given by eq.(\ref{Yparam}) and rest of the equations give $Y_{\alpha}=c_k$ with $c_2, \ldots,c_M$ as the constants of evolution.

We note that a most general solution of eq.(\ref{lt1}) is $Y_1=-\frac{1}{2 M \, \gamma} \; \sum_{k,l}' \sum_{s=1}^{\beta} a_{kl;s} \; f_{kl's} + q_{kl;s} \ln |b_{kl;s}|$ with $a_{kl;s}, q_{kl;s}$ as arbitrary constants of integrations. Assuming all $f_{kl;s}$ and $b_{kl;s}$ as non-zero, we have $2 N^2$ analogous differential equations to solve and resulting constants can be chosen equal i.e., $a_{kl;s}=q_{kl;s}=1$ thereby leading to eq.(\ref{Yparam}). In case one of the $b_{kl;s}=0$, we have $2 N^2-1$ analogous differential equations to solve, with general solution now corresponds to $q_{kl;s}=0$.  Similarly many solution  of eq.(\ref{ltna}) can also be found leading to different sets of basis constants.

\end{document}